\documentclass[11pt,a4paper]{article}
\usepackage{cite,mathtools,amsmath,amssymb,color,comment,graphicx}
\usepackage[usenames]{xcolor}
\usepackage[bookmarksopen,colorlinks=true,linkcolor=dark_green,citecolor=dark_red,urlcolor=dark_red,linktocpage=false]{hyperref}
\usepackage[height=22.5cm,width=16.5cm,centering]{geometry}

\definecolor{dark_blue}{rgb}{0,0,0.6}
\definecolor{dark_green}{rgb}{0,0.4,0}
\definecolor{dark_red}{rgb}{0.6,0,0}

\newcommand{\GRIDSIZE}{N} 
\newcommand{\BOXSIZE}{L} 

\newcommand{\nn}{\nonumber} 
\newcommand{\be}{\begin{equation}} 
\newcommand{\ee}{\end{equation}} 
\newcommand{\bea}{\begin{eqnarray}} 
\newcommand{\eea}{\end{eqnarray}} 
\newcommand{\wvv}{{\langle w \gamma^2 v^2 \rangle}}

\begin{document}

\begin{titlepage}

\begin{center}

\hfill DESY 20-170 \\

\vskip 2cm

{\fontsize{16pt}{0pt} \bf
A hybrid simulation of gravitational wave production 
}
\vskip 0.5cm
{\fontsize{16pt}{0pt} \bf 
in first-order phase transitions
}

\vskip 1.2cm

{\Large
Ryusuke Jinno, Thomas Konstandin and Henrique Rubira
}

\vskip 0.4cm

\end{center}
\begin{center}

\textsl{Deutsches Elektronen-Synchrotron DESY, 22607 Hamburg, Germany}
\vskip 8pt

\end{center}

\vskip 1.2cm

\begin{abstract}
The LISA telescope will provide the first opportunity to probe the scenario of a first-order phase transition happening close to the electroweak scale. 
In thermal transitions, the main contribution to the GW spectrum comes from the sound waves propagating through the plasma. 
Current estimates of the GW spectrum are based on numerical simulations of a scalar field interacting with the plasma or on analytical approximations -- the so-called sound shell model. 
In this work we present a novel setup to calculate the GW spectra from sound waves. 
We use a hybrid method that uses a 1d simulation (with spherical symmetry) to evolve the velocity and enthalpy profiles of a single bubble after collision and embed it in a 3d realization of multiple bubble collisions, assuming linear superposition of the velocity and enthalpy. 
The main advantage of our method compared to 3d hydrodynamic simulations is that it does not require to resolve the scale of bubble wall thickness. 
This makes our simulations more economical and the only two relevant physical length scales that enter are the bubble size and the fluid shell thickness (that are in turn enclosed by the box size and the grid spacing). 
The reduced costs allow for extensive parameter studies and we provide a parametrization of the final GW spectrum as a function of the wall velocity and the fluid kinetic energy.
\end{abstract}

\end{titlepage}

\tableofcontents
\thispagestyle{empty}

\renewcommand{\thepage}{\arabic{page}}

\newpage
\setcounter{page}{1}

\section{Introduction}
\label{sec:Introduction}

LIGO's milestone discovery of gravitational waves (GW) from Black Hole and Neutron Star \cite{Abbott:2016blz,Abbott:2016nmj,TheLIGOScientific:2017qsa} binaries has opened a new avenue for testing and eventually finding new physics, using GW science to probe phenomena from microscopic to cosmological scales. Moreover, the upcoming LISA launch \cite{Seoane:2013qna, amaroseoane2017laser,LISA}, planned to happen in the mid-2030s is especially interesting for scrutinizing GWs sourced by first-order phase transitions around the TeV scale. 

Even though within the Standard Model (SM) the Higgs field performs a cross-over from the false to the true vacuum \cite{Kajantie:1996mn, Gurtler:1997hr, Csikor:1998eu}, in many extensions of the SM this scenario is modified to be a first-order phase transition. Further motivation for this scenario is the possibility of explaining the baryon asymmetry through sphaleron processes in electroweak baryogenesis \cite{Kuzmin:1985mm, Cohen:1993nk, Rubakov:1996vz, Riotto:1999yt,Morrissey:2012db}.
When the Universe cools down and an energetic barrier in the Higgs potential separating the true and the false vacua arises, there is a chance that disjoint regions of the space perform the transition from the symmetric vacuum to the broken one either via quantum or thermal processes \cite{Coleman:1977py, Linde:1980tt, Steinhardt:1981ct}. Those true vacuum regions grow as spherical bubbles and eventually collide, breaking up spherical symmetry and generating a quadrupole moment and therefore GWs \cite{Witten:1984rs}.

In recent years, a large effort has been made to better understand the bubble expansion dynamics. This depends on a balance between a driving term proportional to the energy difference between the two vacua and a friction term proportional to the interactions of the scalar field with the surrounding plasma \cite{Ellis:2019oqb}. Next-to-leading order calculations of the friction term have shown that it is hard for bubbles to runaway \cite{Bodeker:2017cim}, as initially argued by Ref.~\cite{Bodeker:2009qy} (see Ref.~\cite{Hoeche:2020rsg} for recent developments). Macroscopically, the parameter $\alpha$, the ratio between the energy released in the phase transition and the initial plasma energy density before transition, characterize the bubble dynamics~\footnote{For a recent analysis on the energy budget of phase transitions see Ref.~\cite{Giese:2020rtr}.}. It basically quantifies how much energy is available to be converted into shear stress through the interaction of the scalar field and the fluid. In the regime of weak phase transitions ($ \alpha \ll 1$), less energy is released to the fluid sector, such that the system is rather linear (meaning smaller velocities) and the fluid is well-characterized by overlapping sound-waves.

Currently, the only simulations of a system including a relativistic fluid and a scalar field are 
performed by the Helsinki-Sussex group~\cite{Hindmarsh:2013xza,Hindmarsh:2015qta,Hindmarsh:2017gnf,Cutting:2019zws}. These simulations require on the one hand a simulation volume
large enough to fit a sizable number of bubbles (at least 100s). On the other hand, the grid spacing must be small enough
to resolve the Higgs bubble wall thickness. The wall velocity in these simulations is adjusted by adding a phenomenological 
friction term to the Higgs equation. These results are then extrapolated to the physical point, where the Higgs wall thickness is many orders smaller than the bubble size. 

\begin{figure}[h]
\centering
\includegraphics[width=0.32\textwidth]{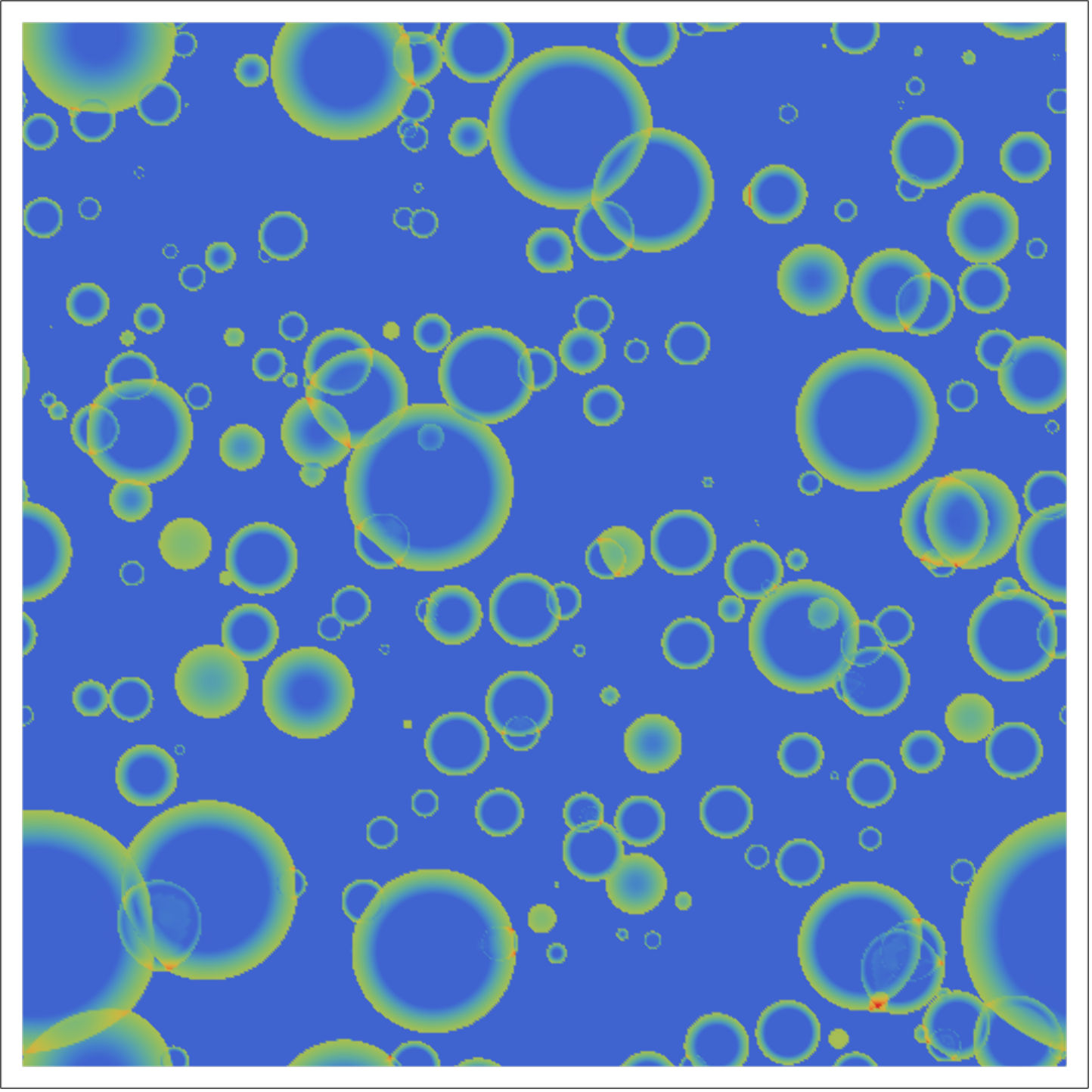} 
\includegraphics[width=0.32\textwidth]{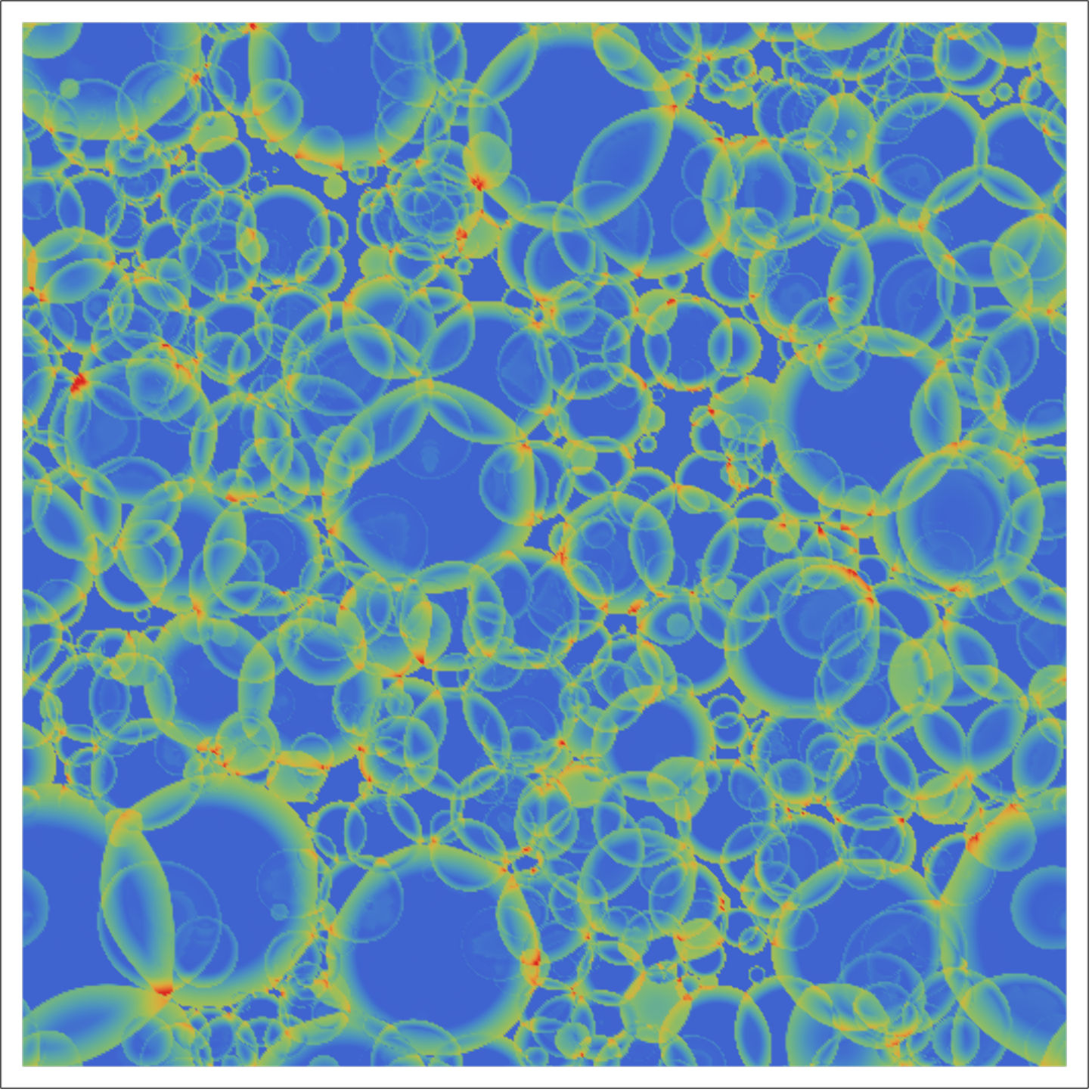} 
\includegraphics[width=0.32\textwidth]{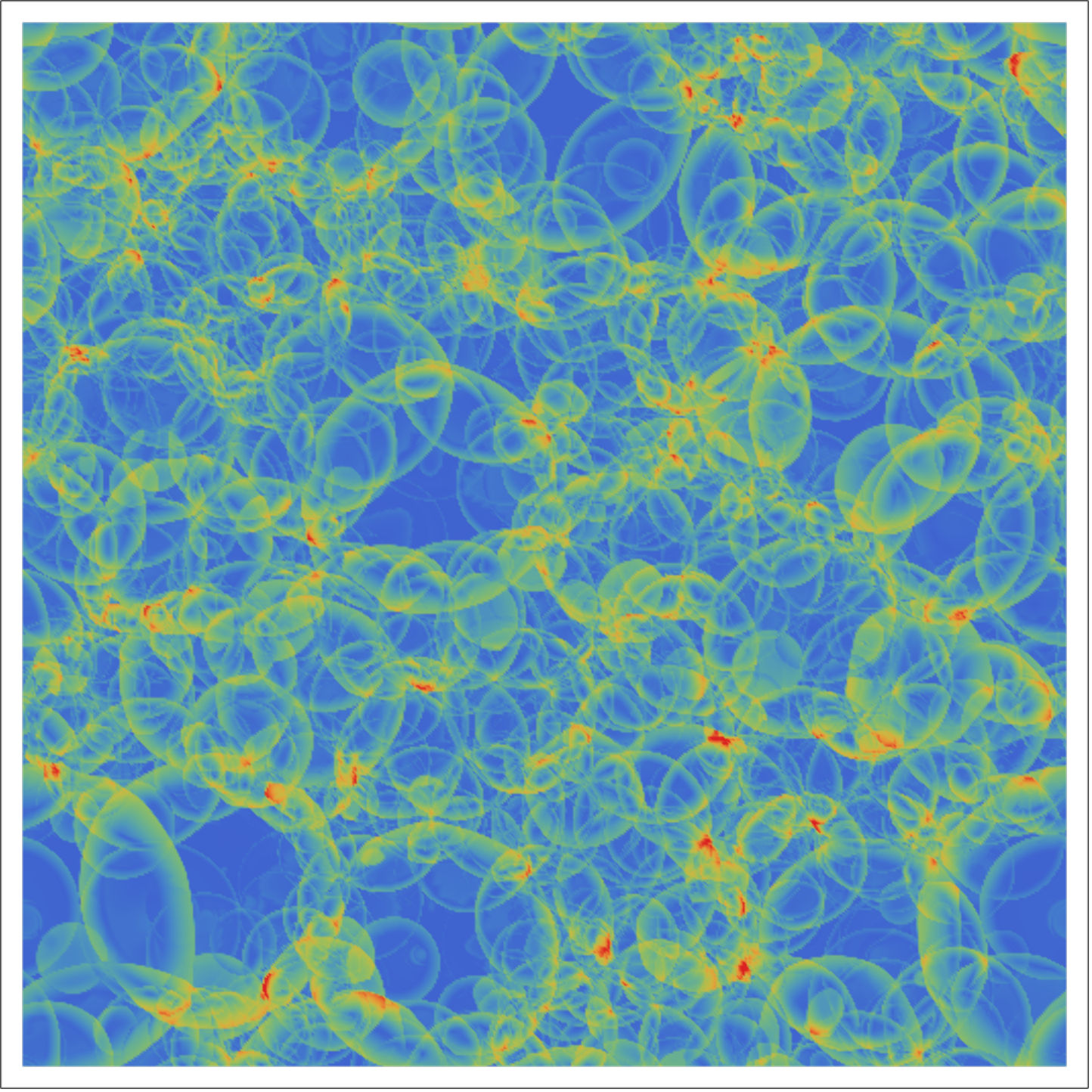} 
\caption{
An example slice of the numerical simulation. 
In this figure we used $\xi_w = 0.8$ and $v_{\rm max} = 0.1$ and the box size $V = L^3$ with $L = 80 \xi_w / \beta = 64 / \beta$ and the grid size $\GRIDSIZE^3 = 512^3$. 
}
\label{fig:simulation}
\end{figure}

In this work we present a new method to calculate the sound shell contribution of GWs. The main intent of our approach is to remove the Higgs bubble wall thickness as a relevant scale from the simulation. 
The only physically relevant scales in our simulation are then the bubble size and the fluid shell thickness that depending on the
wall velocity and the strength of the phase transition is somewhat smaller than the bubble size.
Our simulations then require to have a volume large enough to fit a sizable number of bubbles (for enough statistics and to resolve the IR tail of the spectrum) and a grid spacing that is fine enough to resolve the fluid shell thickness. Unlike the full hydrodynamic simulations, a moderate grid size of $\GRIDSIZE^3 = 256^3$ typically suffices to meet these demands. 
Besides, our approach can cope with rather large time steps in the simulation. 
In essence, this allows us to run a large suit of simulations and provide the spectrum for a large range of wall velocities and different phase transition strengths. We can also use realistic bubble nucleation histories with an exponential increase in nucleation probability. In Fig.~\ref{fig:simulation} we show an example slice of the fluid in one of our numerical simulations. We explain the setup of our simulation in Section~\ref{sec:Strategy}. The main results are shown in Section~\ref{sec:Result} and we conclude in Section~\ref{sec:DC}.

\begin{figure}[h]
\centering
\includegraphics[width=0.75\textwidth]{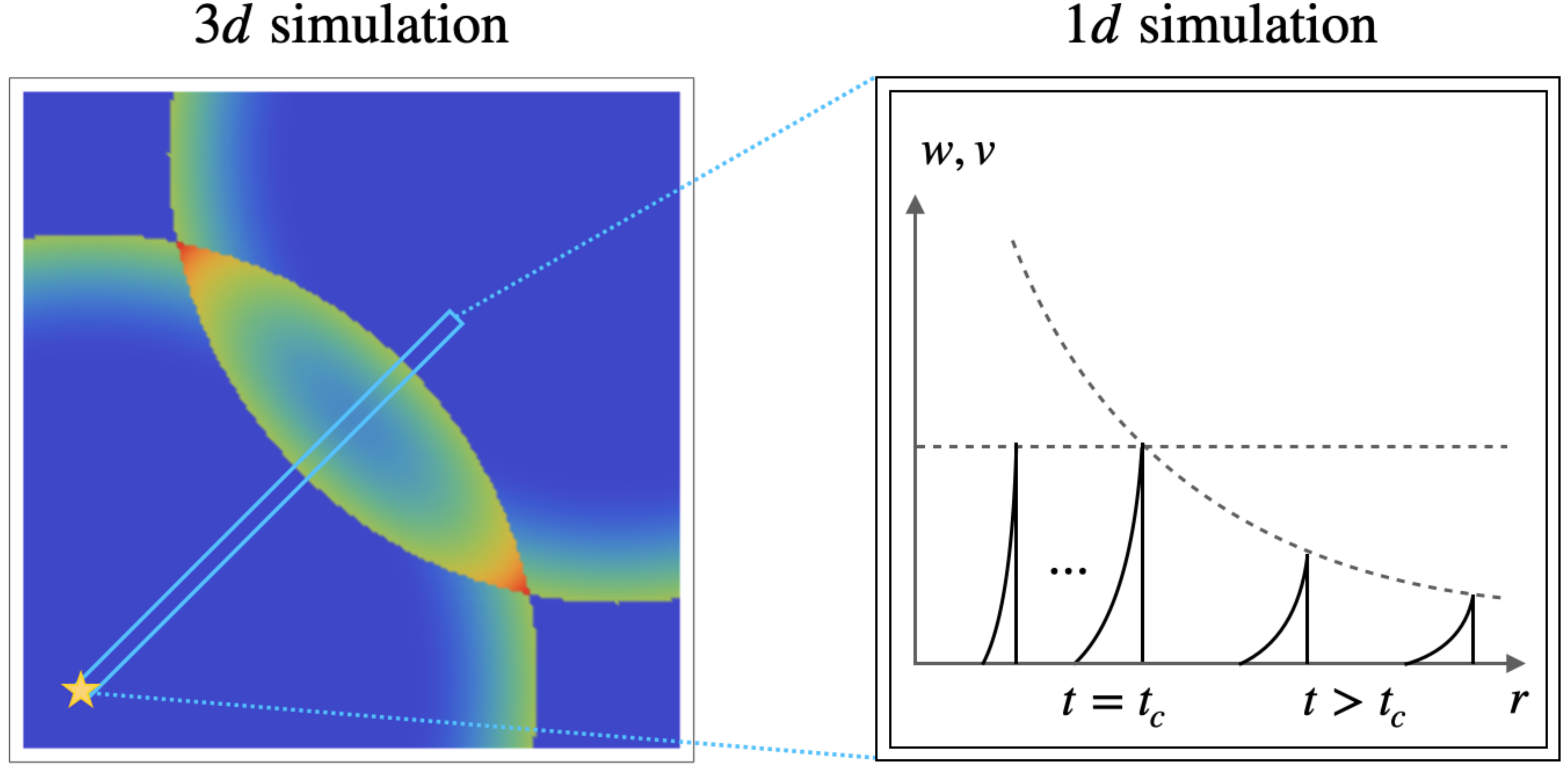} 
\caption{
Schematic illustration for the numerical simulation.
In the 3d simulation we generate bubble nucleation points (denoted by the star) numerically.
For each direction we embed the 1d fluid profile with direction-dependent collision time $t_c$.
The 1d profile before collision can be obtained from the literature~\cite{Espinosa:2010hh}, while after collision it is obtained by solving the 1d evolution equation (i.e.~3d evolution with spherical symmetry): see Fig.~\ref{fig:1d_Evolution} and Ref.~\cite{Jinno:2019jhi}.
The 1d fluid profile generally develops discontinuities (i.e.~shocks), which are dealt with using the Kurganov-Tadmor scheme~\cite{KURGANOV2000241} (see Appendix~\ref{app:1d_numerics}). 
}
\label{fig:embedding}
\end{figure}

\section{Strategy}
\label{sec:Strategy}

In order to remove the Higgs field from the simulation, we model the system in the following way: first, consider a single bubble with spherical symmetry. Before colliding with surrounding bubbles, the fluid adheres to the conventional self-similar solutions. After the collision, the fluid follows the hydrodynamic equations. After collision it is reasonable to neglect the Higgs field, since it is quickly damped to the broken phase~\cite{Hindmarsh:2013xza}. These 1d solutions are then embedded into the simulation volume (a similar technique has been used for the scalar field evolution~\cite{Lewicki:2020jiv}). Every grid point potentially obtains contributions from different bubbles that depend on when the corresponding surface element of the bubble collided with the neighboring bubbles. In this section, we flesh out the details of this approach.

\subsection{Overview}
\label{subsec:Overview}

The GW spectrum is determined from fluid dynamics and the GWs are stored in the tensor components $h_{ij}$ of the metric 
\begin{align}
ds^2
&=
- dt^2 + a^2 (\delta_{ij} + h_{ij}) dx^i dx^j.
\end{align}
Neglecting cosmic expansion during the transition,\footnote{
Throughout the paper we neglect the effect of cosmic expansion.
See Ref.~\cite{Guo:2020grp} for this effect.
}
the time evolution of $h_{ij}$ for each Fourier component is given by
\begin{align} \label{eq:heq}
\ddot{h}_{ij} + k^2 h_{ij}
&=
\frac{2}{M_P^2} \Lambda_{ij,kl} T_{kl},
\end{align}
where $M_P = 1/\sqrt{8 \pi G}$ is the reduced Planck mass, $\Lambda_{ij,kl} = P_{ik} P_{jl} - P_{ij} P_{kl} / 2$ with $P_{ij} = \delta_{ij} - \hat{k}_i \hat{k}_j$ is the projection tensor for the transverse-traceless (TT) part, and $T_{ij}$ is the energy-momentum tensor of the system. In our setup the energy-momentum tensor $T_{ij}$ stems from the fluid dynamics during and after transition\footnote{Since the plasma friction makes it difficult for bubbles to run away, the plasma energy dominates over the scalar contribution. The energy contained in the plasma scales with the bubble radius to the third power, while the energy in the scalar field grows with the bubble radius square~\cite{Caprini:2019egz}.}.
Throughout the paper we assume an equation of state for radiation and a perfect fluid 
\begin{align}
T_{\mu\nu}
&=
w u_\mu u_\nu + p g_{\mu \nu},
\end{align}
where $w$, $p$, $u^\mu$, and $g_{\mu \nu}$ are the enthalpy density, pressure, fluid four-velocity, and metric, respectively.
Notice that only the first term contributes to the GW production.
The GW spectrum is then written in terms of the source term using Weinberg's formula \cite{Weinberg:1972kfs}
\begin{align}
\Omega_{\rm GW} (q)
&\equiv 
\frac{1}{\rho_{\rm tot}} \frac{d \rho_{\rm GW}}{d \ln q}
=
\frac{q^3}{4 \pi^2 \rho_{\rm tot} M_P^2 V}
\int \frac{d\Omega_k}{4\pi} 
\left[ \Lambda_{ij,kl} T_{ij}(q, \vec{k}) T_{kl}^*(q, \vec{k}) \right]_{q = k}.
\label{eq:OmegaGW}
\end{align}
Here $\rho_{\rm tot}$ is the total energy density of the Universe,
$q$ and $\vec{k}$ are GW frequency and wavenumber, respectively, with $k \equiv |\vec{k}|$.
Also, $V$ is the simulation volume and our convention for the Fourier transform is
\begin{align}
T_{ij} (q, \vec{k})
&=
\int dt~e^{i \, q \, t} \int d^3x~e^{-i \vec{k} \cdot \vec{x}}~T_{ij} (t, \vec{x}).
\label{eq:FT}
\end{align}
The Fourier transform in the time direction is performed over the simulation time $T$.

In this paper we propose modeling the 3d energy-momentum tensor field $T_{ij}$ from 1d (more precisely 3d with spherical symmetry) hydrodynamic simulations.
The main idea is illustrated in Fig.~\ref{fig:embedding}.
For every bubble we construct the fluid along radially outgoing rays. Depending on whether the corresponding surface element of the 
bubble already collided with surrounding bubbles or not, we either embed the self-similar solution or the fluid profile obtained from 
1d hydro simulations. 

Unless the transition is strong ($\alpha \gtrsim 1$), the enthalpy change $\Delta w = w - w_0$ and velocity $\vec{v}$ can be treated perturbatively ($w_0$ being the enthalpy deep inside the broken phase, see Fig.~\ref{fig:IC}).
Since we expect the system to behave perturbatively, the solution of the full simulation is given by the superposition of contributions from different bubbles
\begin{align}
\frac{\Delta w}{w_0} (t, \vec{x})
&\simeq
\sum_{i:{\rm bubbles}}
 \frac{\Delta w^{(i)}}{w_0} (t, \vec{x}),
~~~~~~
\vec{v} (t, \vec{x})
\simeq
\sum_{i:{\rm bubbles}}
\vec{v}^{(i)} (t, \vec{x}),
\end{align}
where the superscript $(i)$ denotes the contribution from bubble $i$.
Approximating each surface element to be spherically symmetric, the radial profile of the fluid can be obtained from the solution of 1d simulations (i.e.~3d simulations with spherical symmetry)
\bea
\frac{\Delta w^{(i)}}{w_0} (t, \vec{x})
&\simeq&
\frac{\Delta w^{\rm (1d)}}{w_0} (t -t_n^{(i)}, t_c^{(i)}-t_n^{(i)}, r^{(i)}), \\
\vec{v}^{(i)} (t, \vec{x})
&\simeq&
\hat{n}^{(i)} v^{\rm (1d)} (t -t_n^{(i)}, t_c^{(i)}-t_n^{(i)}, r^{(i)}).
\eea
Here $\Delta w^{\rm (1d)} / w_0$ and $v^{\rm (1d)}$ are the enthalpy and radial velocity obtained from the 1d simulation.
The collision time $t_c^{(i)} = t_c^{(i)} (\hat{n}^{(i)})$ depends on the index $i$ for the bubble and the direction $\hat{n}^{(i)}$ of the surface element measured from the bubble nucleation point $\vec{x}_n^{(i)}$: $\hat{n}^{(i)} \equiv \vec{r}^{(i)} / |\vec{r}^{(i)}|$ with $\vec{r}^{(i)} \equiv \vec{x} - \vec{x}_n^{(i)}$.
Since the 1d solution is common to all the surface elements and all the bubbles, we have to solve the 1d evolution only once and then embed the solution into the 3d lattice with a proper rescaling with the bubble-dependent nucleation time $t^{(i)}_n$ and direction-dependent collision time $t^{(i)}_c$. The embedding of the profiles in the grid therefore dominates the execution time of the simulation.
In the following subsections, we explain the 1d and 3d simulations in detail.

\subsection{1d profile}
\label{subsec:1d}

\subsubsection{Equations, initial conditions, and shocks}

Before bubble collision, the fluid profile is calculated from energy-momentum conservation, $\partial_\mu T^{\mu \nu} = 0$, together with the energy injection at the wall position~\cite{Espinosa:2010hh}.
Because of the spherical symmetry of the bubble, the profile becomes self-similar and depends only on $\xi \equiv r / (t - t_n)$.
The resulting expansion mode of the bubble has three different types: deflagration, hybrid, and detonation~\cite{Espinosa:2010hh}.

After bubble collision, the Higgs wall is damped into the broken phase and the fluid is launched into free propagation. 
This damping occurs on particle physics time scales while the bubble dynamics follows cosmological times scales, and we neglect the Higgs
field right after collision.
Assuming $d$-dimensional spherical symmetry
(i.e.~$d = 1$: planar, $d = 2$: cylindrical, $d = 3$: spherical)
and a relativistic ideal gas $T_{\mu \nu} = w u_\mu u_\nu + p g_{\mu \nu}$ with $w = \rho + p$ and $p = \rho / 3$ the evolution equations become (see Refs.~\cite{McKee, Jinno:2019jhi})
\begin{align} \label{eq:1d_system}
\partial_t u + \partial_r f + g
&= 0,
\end{align}
where $r$ denotes the radial coordinate, and
\begin{align}
u
&= 
\left(
\begin{matrix}
u_1 \\
u_2
\end{matrix}
\right)
=
\left(
\begin{matrix}
w \gamma^2 - p \\
w \gamma^2 v
\end{matrix}
\right),
~~
f
= 
\left(
\begin{matrix}
w \gamma^2 v \\
w \gamma^2 v^2 + p
\end{matrix}
\right),
~~
g
= 
\frac{d-1}{r}
\left(
\begin{matrix}
w \gamma^2 v \\
w \gamma^2 v^2
\end{matrix}
\right).
\end{align}
Here $v$ is the fluid velocity and $\gamma \equiv 1 / \sqrt{1 - v^2}$.
We solve these equations starting just after the bubble collision with the initial condition set in the following way:
we expect that the first crossing of the two profiles do not change their shapes significantly as long as the system is in the linear regime (see Fig.~\ref{fig:IC};
this assumption is discussed in more detail in Appendix~\ref{app:first_collision}).
Then the initial condition is given by the self-similar profile of Ref.~\cite{Espinosa:2010hh}
with the enthalpy in front of the wall shifted by an appropriate amount (such that the enthalpy in the broken and symmetric phases coincide).
This is a consequence of the linearity of the system. Figure~\ref{fig:IC} illustrates how the fluid propagates inside the other bubble (black line) when the profile takes the value in the broken phase.

It is well known that the system (\ref{eq:1d_system}) can develop discontinuities (shocks).
With our initial condition the discontinuity is already established from the beginning at the bubble wall.
Solving the hydrodynamic equations numerically requires special attention and we deal with this issue using the method described by Kurganov and Tadmor~\cite{KURGANOV2000241}. The numerical precision of this scheme is discussed in Appendix~\ref{app:1d_numerics}. As mentioned before, the execution time is dominated by the embedding into the 3d grid. The 1d simulation only takes a few minutes and is not a limiting factor of the accuracy.

\begin{figure}
\centering
\includegraphics[width=0.7\textwidth]{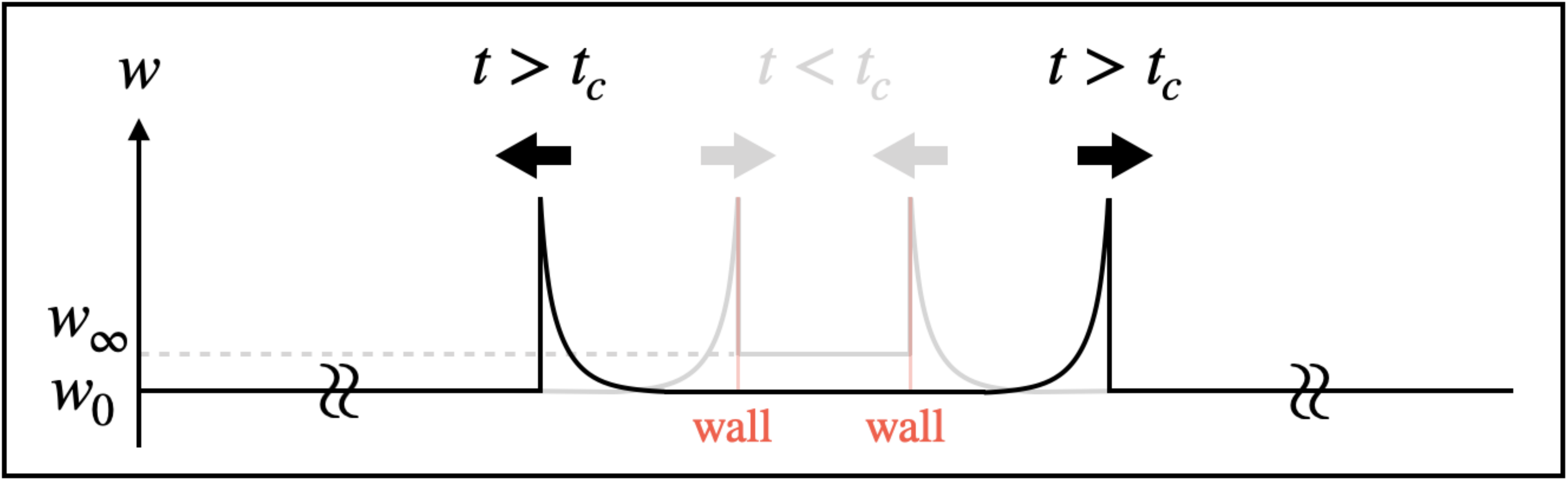} 
\caption{
Illustration for the initial condition of the 1d simulation.
The figure shows the collision of two fluid profiles.
Before collision (gray), the enthalpy takes different values in front ($w_\infty$, symmetric phase) and behind ($w_0$, broken phase) the profile.
Assuming that the first collision does not change the profile significantly, the profile just after collision (black) is given by the same shape as the gray line with the front enthalpy replaced by $w_0$.
We use the black line as the initial condition for the 1d simulation.}
\label{fig:IC}
\end{figure}

\subsubsection{Numerical examples}

\begin{figure}
\centering
\includegraphics[width=\textwidth]{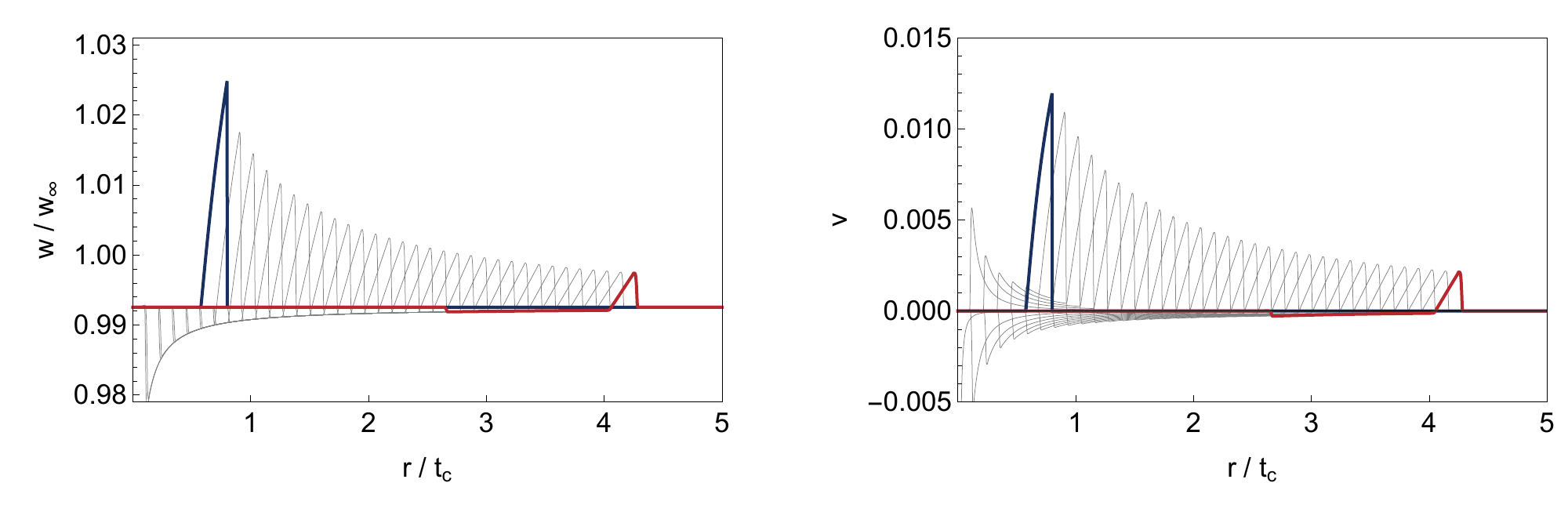} 
\caption{
Example of our 1d simulation, showing how the initial condition (blue line) evolves in time.
In this plot we used $\alpha = 0.0046$ and $\xi_w = 0.8$ (corresponding to $v_{\rm max} \simeq 0.012$), and evolved the system from $t = t_c$ to $7 t_c$ (with the nucleation time $t_n$ taken to be $0$).
}
\label{fig:1d_Evolution}
\end{figure}

In Fig.~\ref{fig:1d_Evolution} we show a typical result for the time evolution of the 1d profile.
The initial condition is taken from the profile in Ref.~\cite{Espinosa:2010hh} with $(\alpha, \xi_w) = (0.0046, 0.8)$ (corresponding to the maximal fluid velocity at the wall position $v_{\rm max} \simeq 0.012$) with $w$ and $v$ outside replaced with their values inside as explained above. 
The simulation time is from $t = t_c$ (blue line) to $t = 7 t_c$ (red line).

Here, several comments are in order. 
First, a sharp discontinuity remains at the front end of the profile even though this simulation is {\it after} collision and therefore no bubble wall or energy injection is present in this simulation\footnote{In the deflagration case, there are two discontinuities: one at the sound shell front and a steeper one at the bubble wall. Evolving the deflagration 1d simulation makes the bubble wall discontinuity eventually catch up with the sound shell front such that it resembles a detonation at late times.}.
As mentioned previously, special care is required for the numerical scheme to retain this feature.
We use the Kurganov-Tadmor scheme~\cite{KURGANOV2000241} for this.
A more detailed discussion is found in Appendix~\ref{app:1d_numerics}.
Second, the plot seems to diverge around the origin.
This is because of the term $g \propto 1/r$ in Eq.~(\ref{eq:1d_system}) and depending on the boundary conditions imposed at the origin, the incoming wave will reflect. 
This is an artifact from the spherical symmetry that would not affect the 3d simulation,
in which reflected waves do not converge at the origin thanks to the different collision time for different directions.
Indeed, only a very small volume is affected by the assumptions how this singularity is treated and we have checked that this issue does not affect the final GW spectrum in any major ways.
Notice that in our model the time dependence is explicit. 
Once the 1d profile is established, the 3d embedding can be performed for arbitrary times. 
Hence, one can perform the simulation with rather large time steps without sacrificing accuracy, unlike real simulations that rely on evolving the equations of motion in time. 
Finally, when embedding the 1d solution in the 3d box we sometimes need the 1d profile beyond $t = 7t_c$ (see also Appendix~\ref{app:1d_numerics} for a discussion about the choice of the maximal time of the 1d simulation). We extrapolate the 1d solution at the last time slice using $v^{\rm (1d)} \propto r^{-1}$, i.e.
\bea
\left. v^{\rm (1d)} (t, t_c, r)\right|_{t> 7 t_c}
 = \frac{\bar r}{r} 
\left. \times v^{\rm (1d)} (t, t_c, \bar r) \right|_{t = 7 t_c} \, ,
\eea
with $\bar r = r - c_s (t - 7t_c)$ and accordingly for $w-w_0$~(we actually extrapolate $\ln (w / w_0)$ in the simulation which is equivalent in leading order).

\subsection{3d embedding}
\label{subsec:3d}

Next we discuss how to embed our 1d result in the 3d simulation.
The procedure is illustrated in Fig.~\ref{fig:embedding}.
We generate nucleation points in a 3d box with size $\BOXSIZE$.
Each bubble has direction-dependent first collision time, and we rescale our 1d simulation result for each direction to embed the 1d fluid evolution into the 3d lattice.
From the 3d lattice thus constructed, we generate the transverse-traceless energy-momentum tensor field as
\begin{align}
\Lambda_{ij, kl} T_{kl}
&=
\Lambda_{ij, kl} \times w \gamma^2 v_k v_l.
\end{align}
Our final goal is to calculate the GW spectrum in Eq.~(\ref{eq:OmegaGW}).
However, we can factor out some trivial dependencies from $\Omega_{\rm GW}$
\begin{align}
\Omega_{\rm GW}
&=
\frac{w^2 \, \tau}{4 \pi^2 \rho_{\rm tot} M_P^2 \beta}
\times Q',
\label{eq:Q1}
\end{align}
where from Eq.~(\ref{eq:OmegaGW}) the dimensionless growth rate of the GW spectrum $Q'$ becomes 
\begin{align}
Q'
&\equiv
\frac{q^3 \beta}{w^2 \, V \, T}
\int \frac{d\Omega_k}{4\pi} 
\left[ \Lambda_{ij,kl} T_{ij}(q, \vec{k}) T_{kl}^*(q, \vec{k}) \right]_{q = k}.
\label{eq:Q2}
\end{align}
Here $\tau$ denotes the lifetime of the sound waves (typically determined by the onset of turbulence), and the prime means that $Q'$ is the growth rate (with respect to $\beta$-normalized time $\beta t$) of the dimensionless quantity $Q$.\footnote{
The definition is 
\begin{align}
Q
&\equiv
\frac{q^3 \beta^2}{w^2 \, V}
\int \frac{d\Omega_k}{4\pi} 
\left[ \Lambda_{ij,kl} T_{ij}(q, \vec{k}) T_{kl}^*(q, \vec{k}) \right]_{q = k}. \nn
\label{eq:Q2}
\end{align}
} 
We describe with more details how to obtain the GW spectrum from the velocity embedded grid in Appendix~\ref{app:3d_scheme}.
Notice that ideally the simulation is large enough that the correlations in the integrals already scale linearly in the integration time and volume, $T\gg 1/\beta$ and $V\gg 1/\beta^3$. Especially for the soft momentum modes in the IR tail, this is a nontrivial requirement that 
we checked explicitly. In this limit, $Q$ is a function of $k/\beta$ that depends only on the 1d profile that is embedded and the wall velocity. Any dependence on the simulation volume $V$ and simulation time $T$ drops out, while the concrete nucleation history becomes statistically irrelevant (or encoded in the parameter $\beta$, $ \xi_{w}$ and $\alpha$). 

Notice that the relative factor between $\Omega_{\rm GW}$ and $Q'$ can be recast in terms of the Hubble parameter $H$ as
\be
\frac{\Omega_{\rm GW}}{Q^\prime} = 
\frac{w^2 \, \tau}{4 \pi^2 \rho_{\rm tot} M_P^2 \beta} \simeq 
\frac{4 \rho_{\rm tot} \, \tau}{9 \pi^2 M_P^2 \beta} = 
\frac{4 H \, \tau}{3 \pi^2} \frac{H}{\beta} \, ,
\ee
meaning $\Omega_{\rm GW} \sim Q' \times (H / \beta)$ for the sound wave source lasting for the whole Hubble time.
Notice that this assumes a weak phase transition and a radiation equation of state, $\Gamma \equiv w/\rho_{\rm tot} \simeq 4/3$.

\section{Main results}
\label{sec:Result}

\subsection{Example of the GW spectrum}
\label{subsec:Structures}

\begin{figure}
\centering
\includegraphics[width=0.6\textwidth]{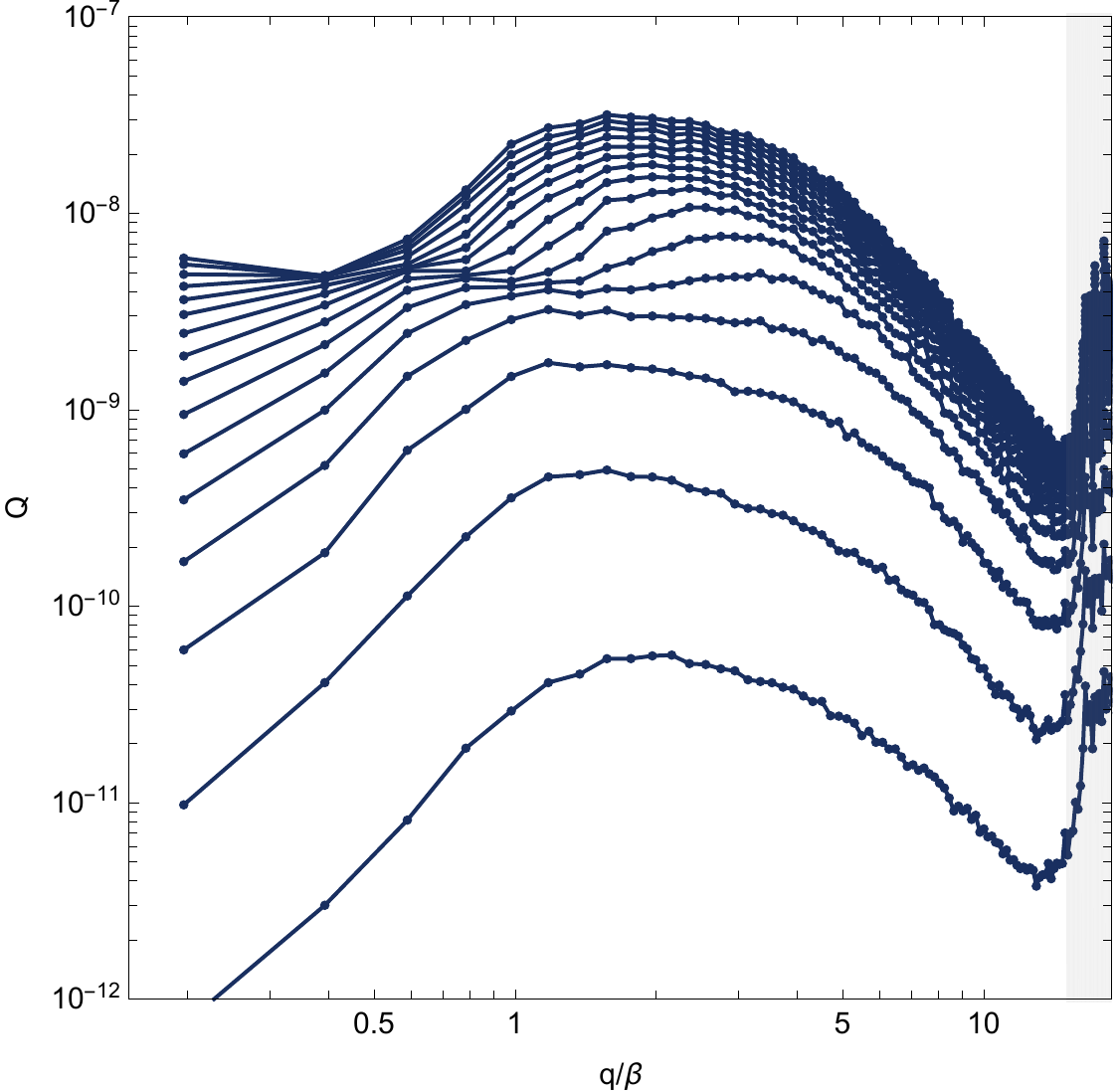} 
\caption{
GW spectrum $Q$ for $\alpha = 0.0046$ and $\xi_w = 0.8$ (corresponding to a detonation with $v_{\rm max} \simeq 0.012$).
The box size is $V = \BOXSIZE^3 = (40 \xi_w / \beta)^3 = (32 / \beta)^3$, the number of grids is $N^3 = 256^3$, and the integration range is from $t = 6 / \beta$ to $22 / \beta$ (the different lines change the upper integration limit in steps of $1/ \beta$) with the typical bubble nucleation time being $t \simeq 7 / \beta$. 
There is a sizable contribution in the IR mostly from the early times (up to percolation, around $0.5 \lesssim q / \beta \lesssim 1$) that we suppress by integrating only over rather late times $14/\beta < t < 22/\beta$ in the following.
}
\label{fig:3d_Detonation_vw=08_vmax=01}
\end{figure}

We first show an example of our 3d simulation in Fig.~\ref{fig:3d_Detonation_vw=08_vmax=01}.
The phase transition strength, wall velocity and fluid maximal velocity are $\alpha = 0.0046$, $\xi_w = 0.8$, and $v_{\rm max} \simeq 0.012$.
The expansion mode is a detonation for this parameter choice.
The box size is $V = \BOXSIZE^3 = (40 \xi_w / \beta)^3 = (32 / \beta)^3$,\footnote{
One can use the same nucleation history for different wall velocities.
In this case one needs to rescale the box size by the wall velocity to keep the nucleation history to be physical.
This is why naturally a factor $\xi_w$ comes with the box size.
}
the grids size is $N^3 = 256^3$ with periodic boundary conditions, and the GW spectrum is calculated with the integration range from $t = 6 / \beta$ to $22 / \beta$ (see Eq.~(\ref{eq:FT})). 
This typically results in simulations with about 2500 bubbles.
These are the default parameters we use in all simulations if not explicitly stated otherwise.

We notice several features in the spectrum.
The spectrum has a single peak at early times ($t \lesssim 12 / \beta$) while it breaks up into two peaks after that.
The peak at early times corresponds to the typical bubble size at the collision time.
This is a contribution to the GW spectrum akin to the envelope and bulk flow models that
seize after percolation completed. 
This peak moves to the IR without significant growth in the amplitude, which is qualitatively consistent with Refs.~\cite{Jinno:2017fby,Konstandin:2017sat}.
At later time, a second peak develops. 
This peak is the UV structure peaked at the scale of shell thickness.
Ideally, we would like to isolate the late time structure of the simulation that is dominated by the contributions from the sound waves.
We come back to this issue in Sec.~\ref{subsec:DC}.

\begin{figure}
\centering
\includegraphics[width=0.7\textwidth]{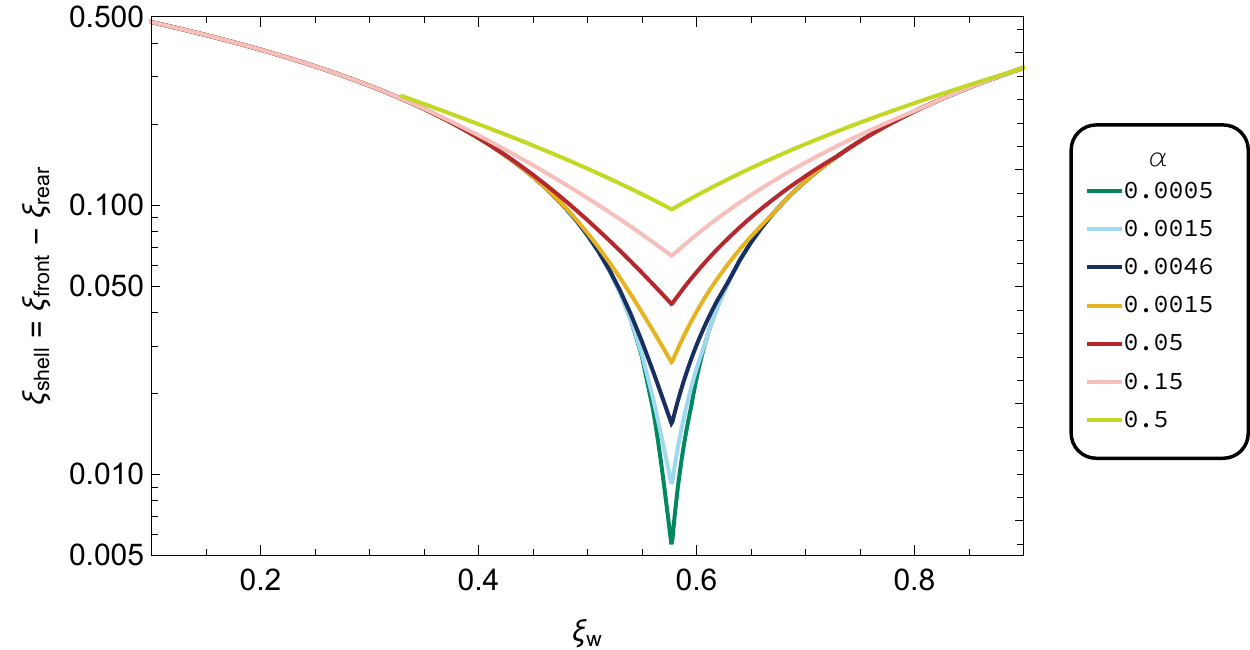}
\caption{
Shell thickness $\xi_{\rm shell} \equiv \xi_{\rm front} - \xi_{\rm rear}$ for different values of $\alpha$ and $\xi_w$.
}
\label{fig:xi_shell}
\end{figure}

We define the dimensionless shell thickness as (see Fig.~\ref{fig:xi_shell})
\begin{align}
\xi_{\rm shell}
&\equiv
 \xi_{\rm front} - \xi_{\rm rear},
\end{align}
where
\begin{align}
\xi_{\rm front}
&=
\left\{
\begin{matrix}
\xi_{\rm shock}
&~
{\rm (deflagration,~hybrid)}
\\[0.2cm]
\xi_w
&~
{\rm (detonation)}
\end{matrix}
\right. ,
~~~~
\xi_{\rm rear}
=
\left\{
\begin{matrix}
\xi_w
&~
{\rm (deflagration)}
\\[0.2cm]
c_s
&~
{\rm (hybrid,~detonation)}
\end{matrix}
\right. .
\end{align}
Note that the physical shell thickness is proportional to $\xi_{\rm shell} / \beta$.
As we discuss in the next section, the GW spectrum stretches from the (inverse of) typical bubble size at the collision time to the shell thickness.

\begin{figure}
\centering
\includegraphics[width=0.48\textwidth]{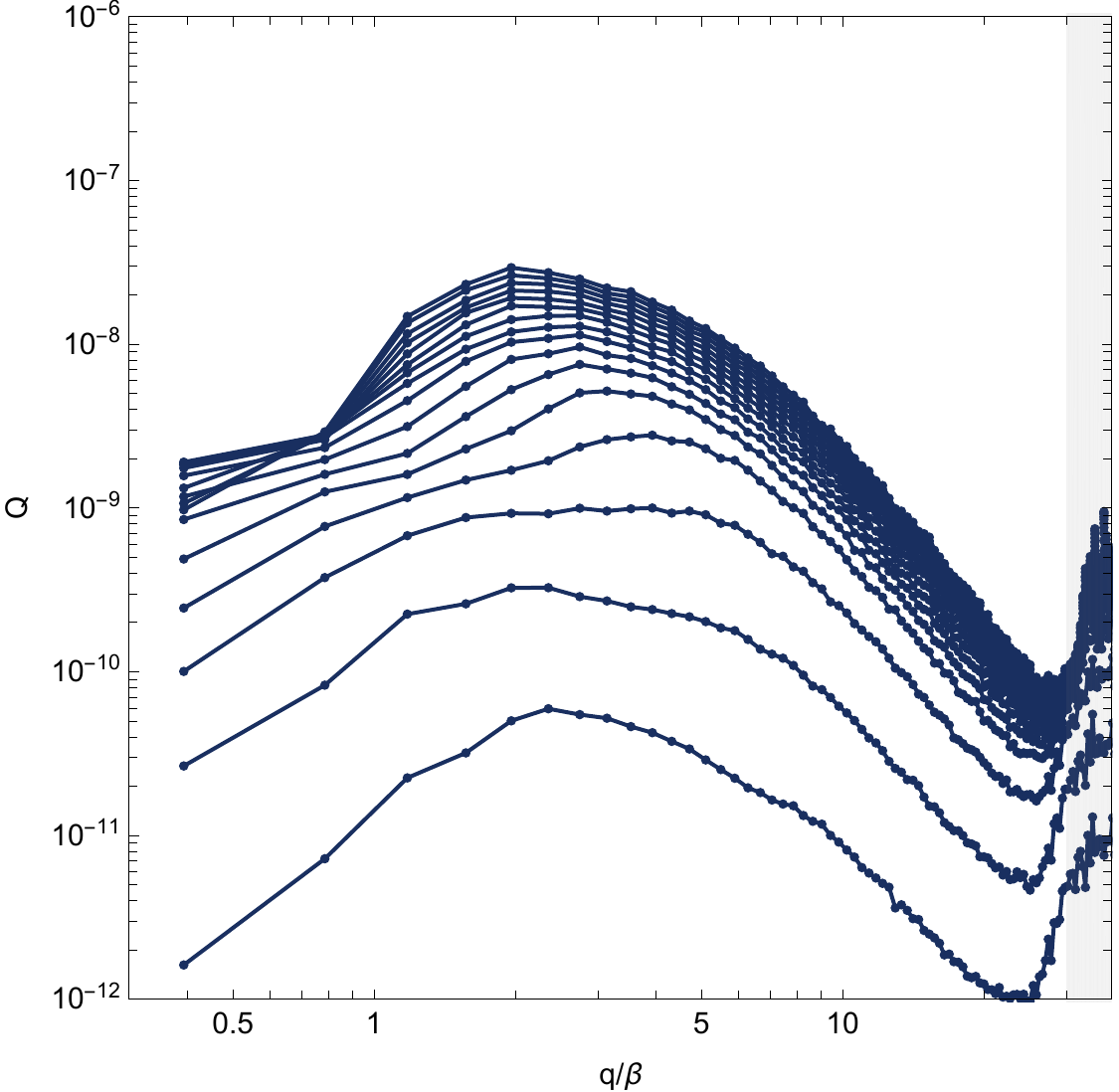} 
\includegraphics[width=0.48\textwidth]{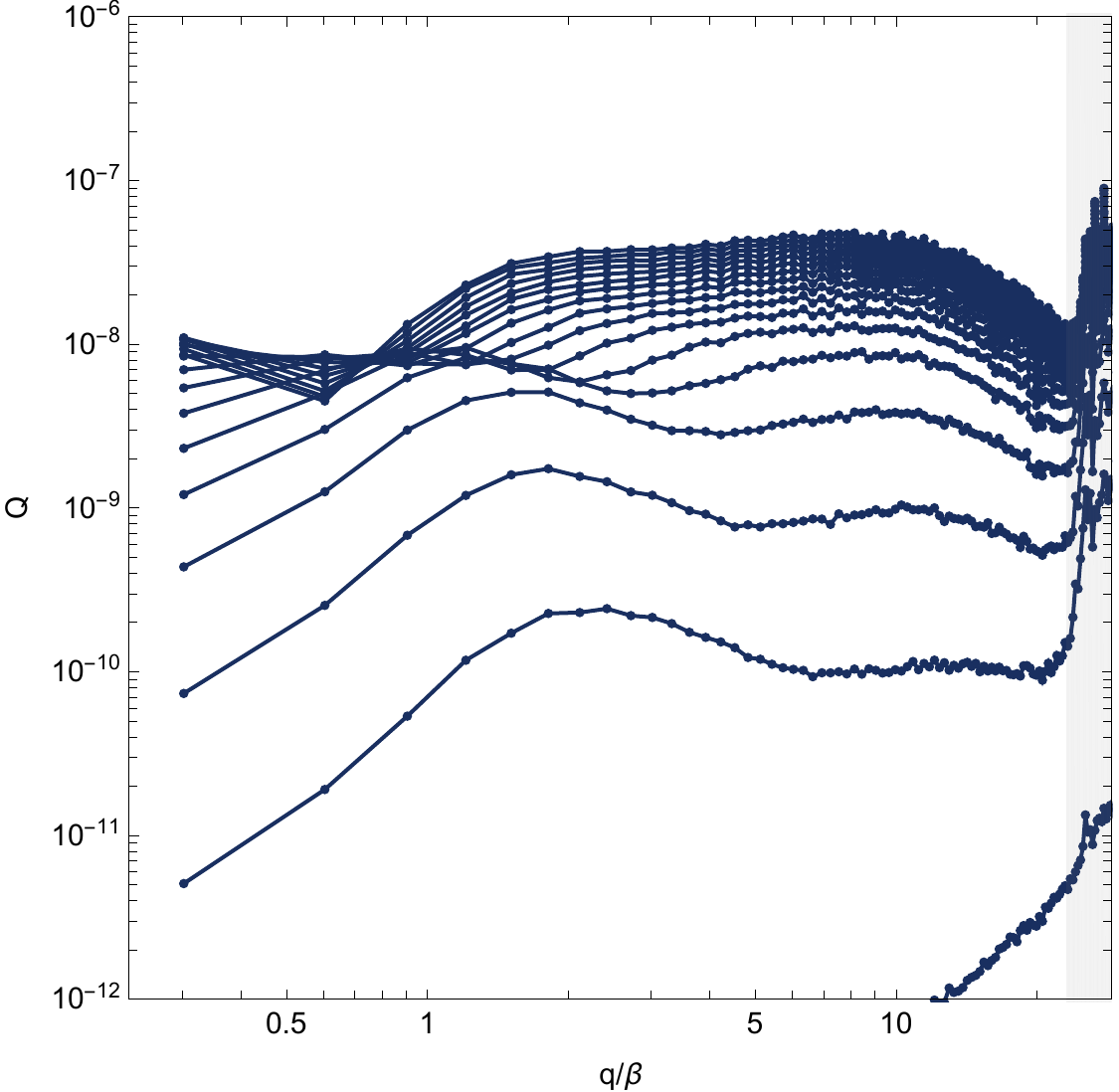} 
\vskip 0.5cm
\includegraphics[width=0.48\textwidth]{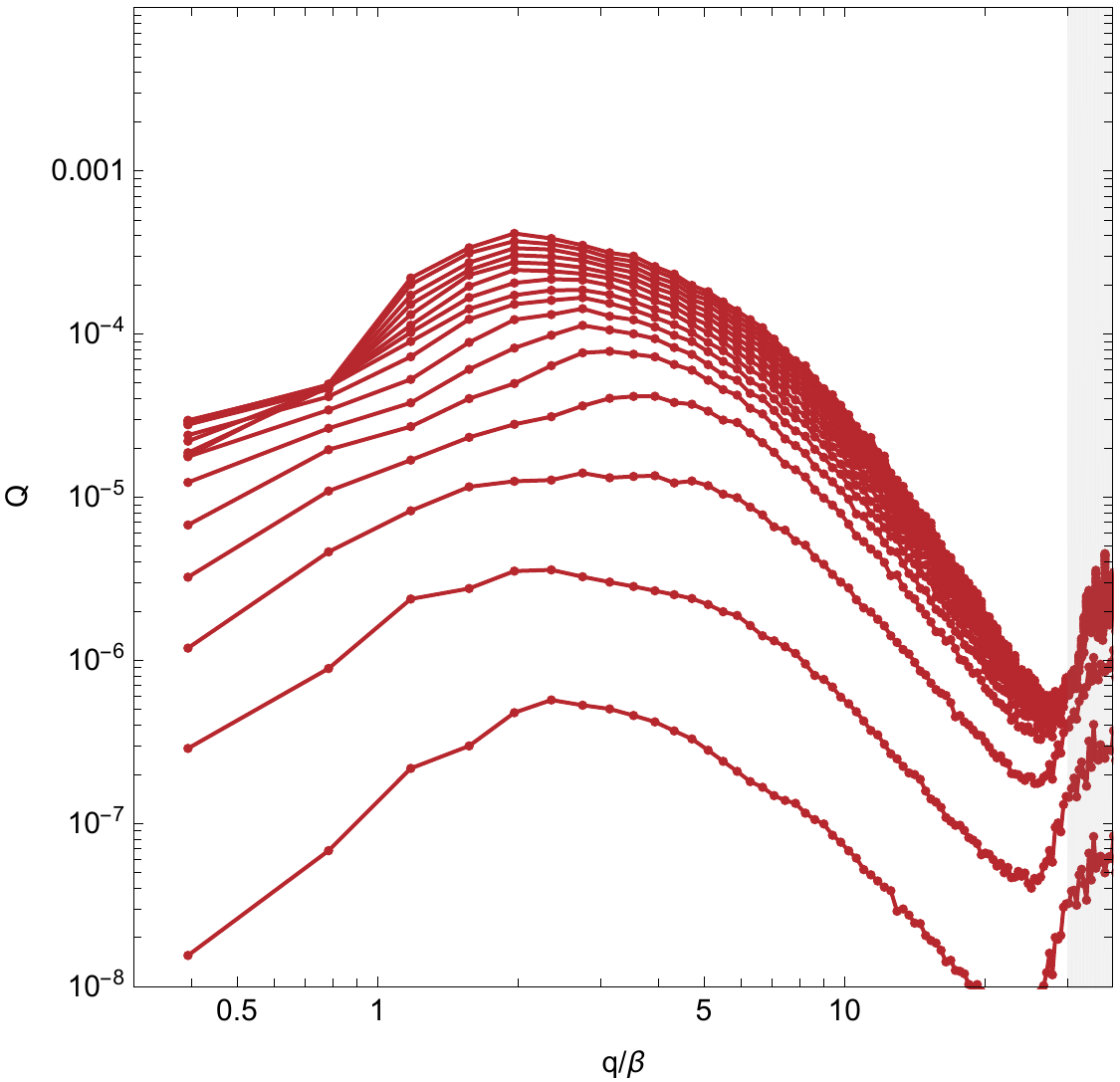} 
\includegraphics[width=0.48\textwidth]{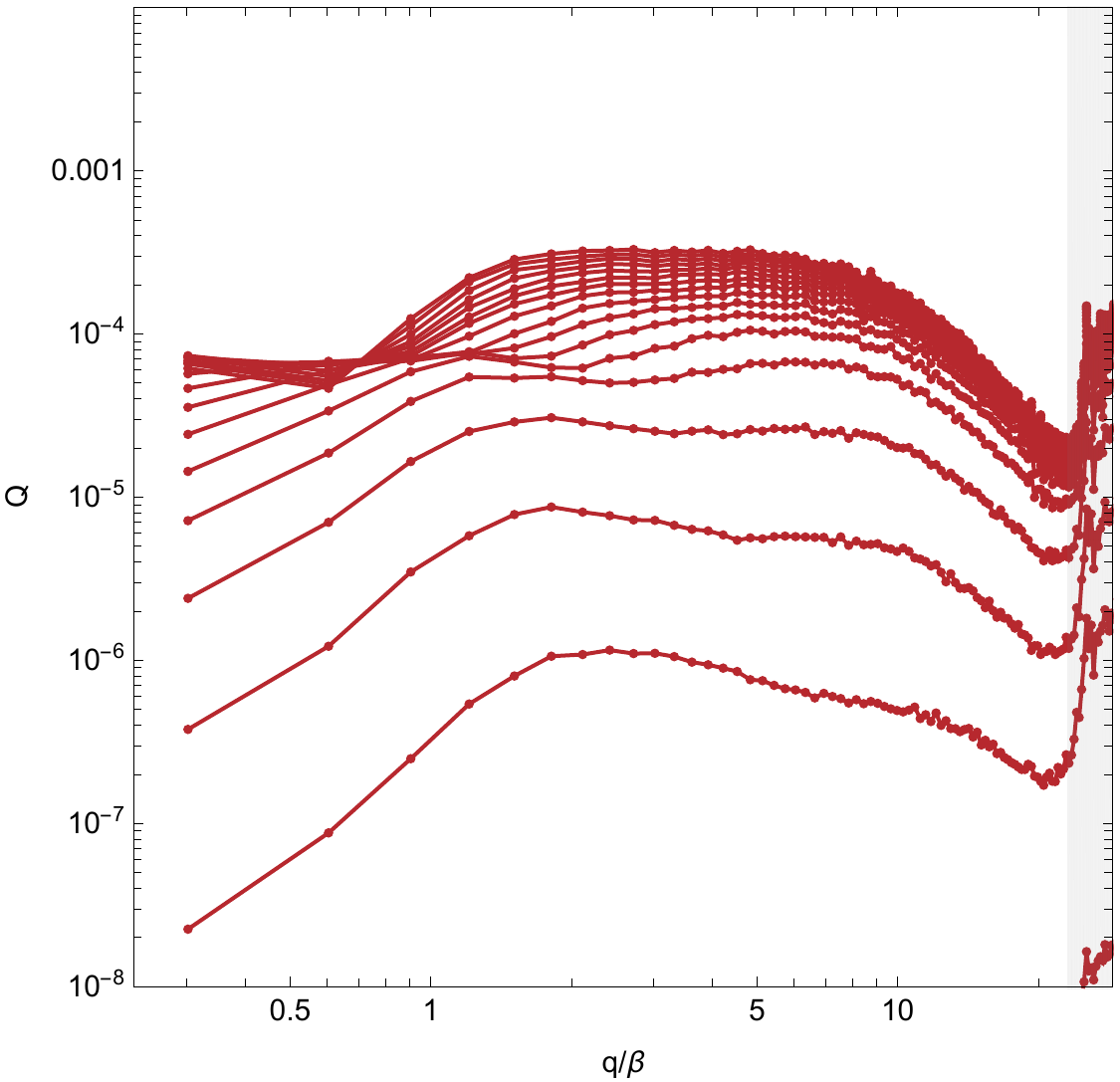} 
\caption{
GW spectrum in weak ($\alpha = 0.0046$, top panels) and intermediate ($\alpha = 0.05$, bottom panels) transitions.
The wall velocity is $\xi_w = 0.4$ and $\xi_w = 0.52$ for the left and right panels, respectively. In the right plot a clear separation of the shell thickness and bubble size length scales can be observed. 
}
\label{fig:3d_weak_intermediate}
\end{figure}

Following Ref.~\cite{Hindmarsh:2017gnf}, we consider two types of transitions
\begin{align}
\alpha
&=
\left\{
\begin{matrix}
0.0046
&~
{\rm (weak)}
\\[0.2cm]
0.05
&~
{\rm (intermediate)}
\end{matrix}
\right. ,
\end{align}
and take wall velocities from $\xi_w = 0.32$ to $\xi_w = 0.8$.
Fig.~\ref{fig:3d_weak_intermediate} shows examples of the GW spectrum for $\xi_w = 0.4$ (left columns) and $\xi_w = 0.52$ (right columns).
In the left panels, the two scales in the UV structure are hard to distinguish because the shell thickness $\xi_{\rm shell} \simeq 0.18$ (for both weak and intermediate phase transitions) is not much different from the bubble size at the collision time (see Sec.~\ref{subsec:DC} for a discussion).
However, in the right panels the shell thickness is much smaller than the bubble size: $\xi_{\rm shell} \simeq 0.06$ (weak) and $\xi_{\rm shell} \simeq 0.07$ (intermediate).

This is why the UV structure stretches to much higher frequencies for the right panels.
In Fig.~\ref{fig:xi_shell} we plot $\xi_{\rm shell}$ for the weak and intermediate transitions for different wall velocities.
Clearly the relation $\xi_{\rm shell} \simeq |\xi_w - c_s|$ holds for weaker transitions while it breaks down (for deflagrations and hybrids) as the transition becomes stronger.

\subsection{Extraction of the linear growth of the spectrum}
\label{subsec:DC}

\begin{figure}
\centering
\includegraphics[width=0.48\textwidth]{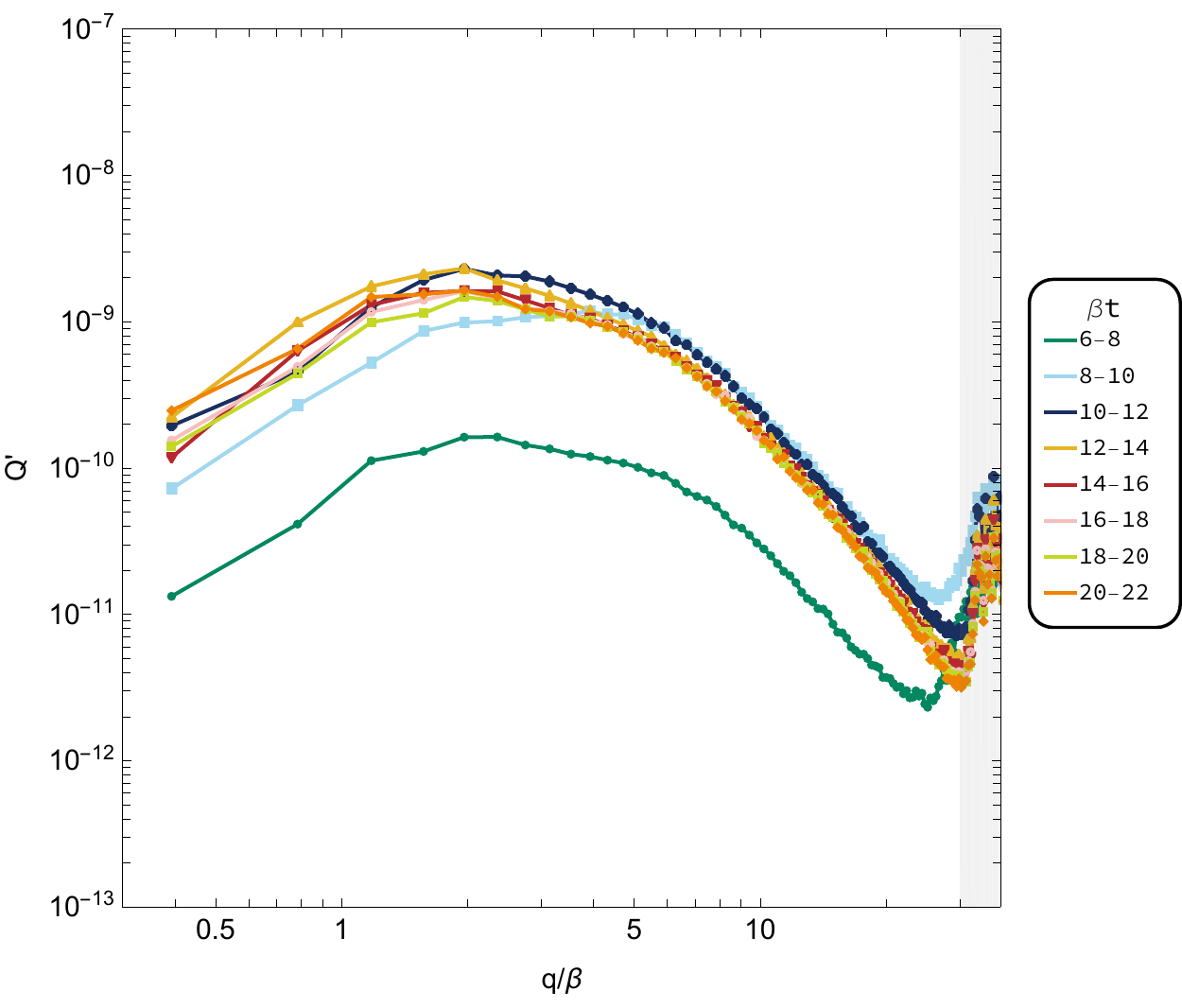} 
\hskip 0.1cm
\includegraphics[width=0.48\textwidth]{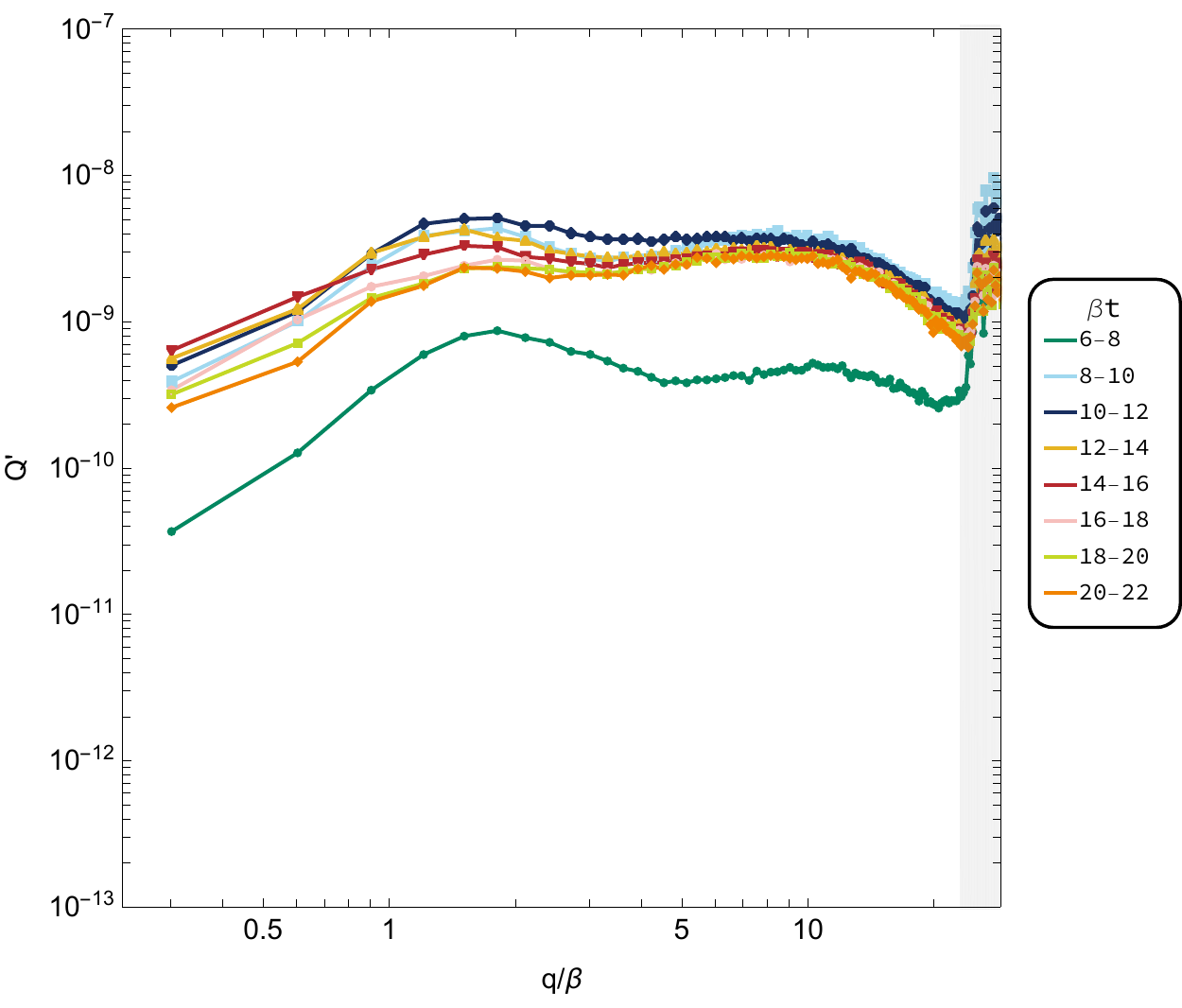} 
\includegraphics[width=0.48\textwidth]{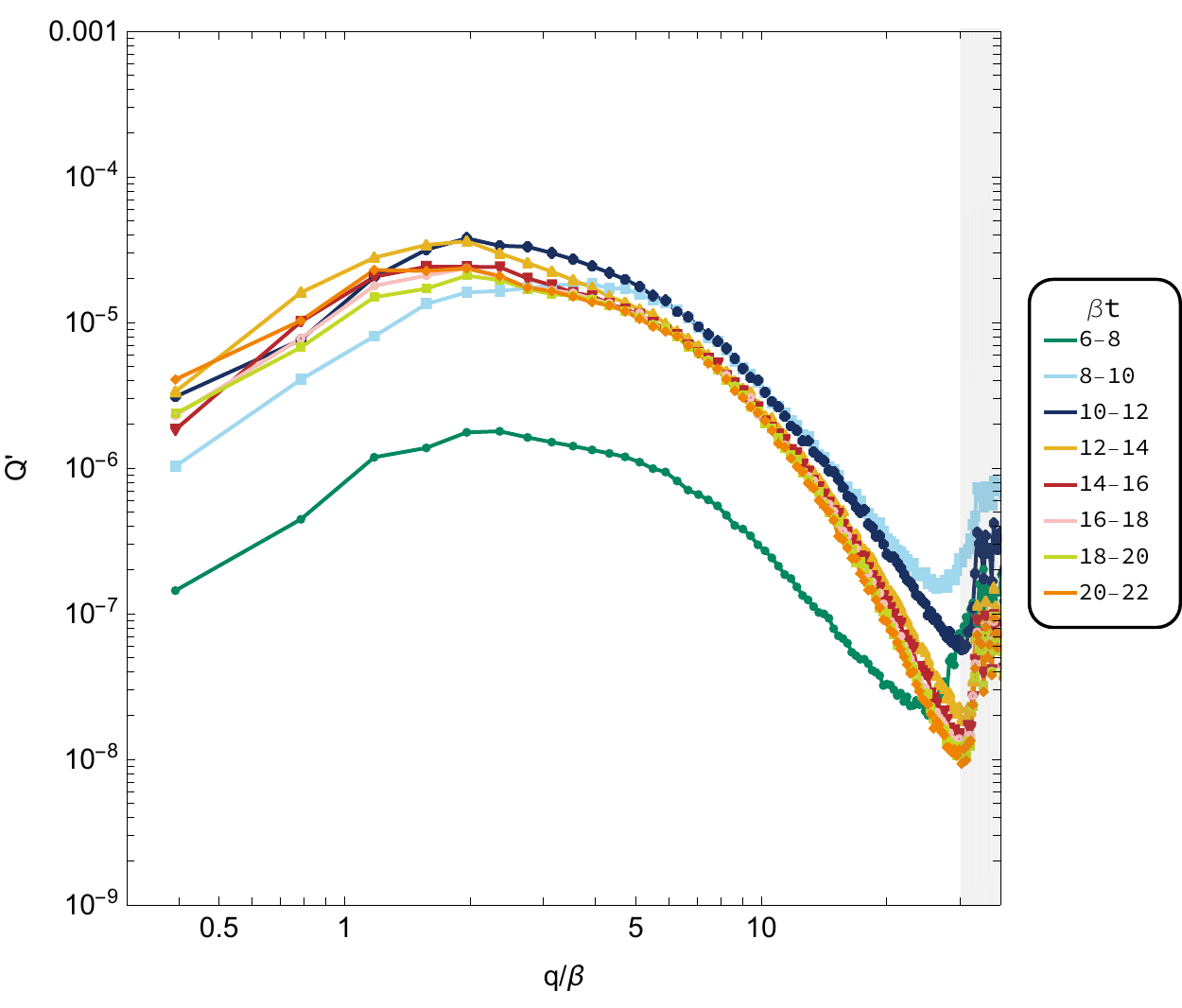} 
\hskip 0.1cm
\includegraphics[width=0.48\textwidth]{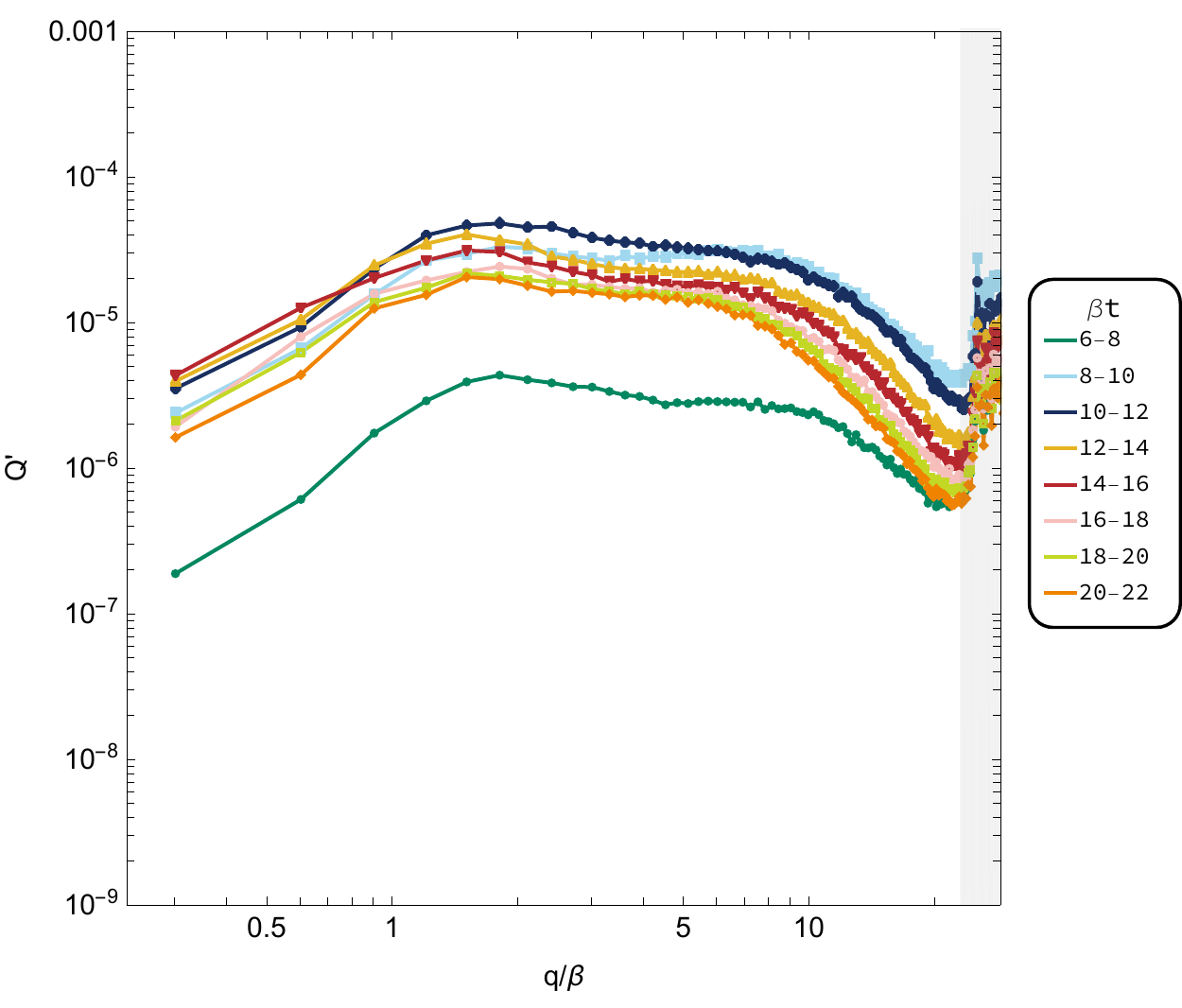} 
\caption{
Growth rate of the GW spectrum $Q'$ calculated for each short period $\Delta T = 2 / \beta$.
The four panels correspond to the parameter points in Fig.~\ref{fig:3d_weak_intermediate}
(top-left: $\alpha = 0.0046$, $\xi_w = 0.4$,
top-right: $\alpha = 0.0046$, $\xi_w = 0.52$,
bottom-left: $\alpha = 0.05$, $\xi_w = 0.4$,
bottom-right: $\alpha = 0.05$, $\xi_w = 0.52$).
In evaluating $Q'$ we take the lower and upper limit for the time integration in Eq.~(\ref{eq:FT}) to be the range shown in the legend.
For $t = 14 / \beta - 22 / \beta$ the UV part is stable. 
}
\label{fig:linear}
\end{figure}

Once the typical number of overlapping sound shells at an arbitrary spatial point exceeds unity, the root-mean-square of the fluid velocity remains almost constant in time (until the onset of turbulence)~\cite{Hindmarsh:2013xza,Hindmarsh:2015qta,Hindmarsh:2017gnf, Hindmarsh:2016lnk,Hindmarsh:2019phv,Ellis:2019oqb}. 
The GW spectrum grows linearly in time after that.
In other words, the correlation of the energy-momentum tensor at different times dies off if the two times are too dissimilar. 
Hence, even though the Weinberg formula involves integration over two times, the final result should only be linear in the 
simulation time. The early dynamics is dominated by the first few bubbles (which leads to a large enhancement in the IR) and we are rather interested in the late time behavior of the simulations.
We only integrate over rather late times (from $t = 14 / \beta$ to $22 / \beta$) so that the IR part of the spectrum is reduced.
Fig.~\ref{fig:linear} illustrates this point.
For each panel ($\alpha = 0.0046$ or $0.05$, $\xi_w = 0.4$ or $0.52$) we integrate $T_{ij}$ for a short period $\Delta T = 2 / \beta$ (and thus replacing the integration range $T$ in Eq.~(\ref{eq:FT}) with $\Delta T$).
The resulting linear growth $Q'$ is almost constant for $t = 14 / \beta$ -- $22 / \beta$.

Ideally, one would like to express the final GW spectrum in quantities that are easily calculated. The GW spectrum is proportional to the energy-momentum tensor squared and only the enthalpy part of the energy-momentum tensor contributes to the anisotropic stress. Hence one expects a naive scaling $Q^\prime \propto (\wvv / w_\infty)^2$. However, $\wvv$ can be evaluated at different stages of our simulation leading to different results. 

The first representation of $\left< w \gamma^2 v^2 \right>$ is the kinetic energy contained in the sound shell of the expanding bubble before collision. This is thus calculated from the self-similar profile by
\begin{align}
\kappa \alpha
&=
\frac{4}{\xi_w^3 w_\infty} \int d\xi~
w \gamma^2 v^2 \xi^2,
\end{align}
where $w_\infty$ is the enthalpy in the symmetric phase.
The second and third are average of $w \gamma^2 v^2$ in the 1d simulation after collision and the 3d value evaluated on the grid of the box, respectively. For $\wvv_{\rm 1d}$ we average $w \gamma^2 v^2$ over the spherical volume at the last time slice of the 1d simulation $t = 7t_c$. For $\wvv_{\rm 3d}$ we spatially average $w \gamma^2 v^2$ in the 3d box and further average for the period from $t = 14 / \beta$ to $22 / \beta$ (which is our integration range for $Q^\prime$).

The second question is what observable is most easily related to the model parameters. We studied several possibilities and found that the integrated GW spectrum allows for the most straightforward relation to the model parameters (compared to e.g.~the peak of the GW spectrum).
We hence use 
\begin{align}
Q'_{\rm int}
&\equiv
\int d\ln q~Q'(q) \, ,
\end{align}
to study the model dependence. We also found the final GW spectrum scales with the shell thickness, which is due to the fact that the 
anisotropic stress stems from the overlap of different sound shells rather than the sound shells individually, see Ref.~\cite{Hindmarsh:2019phv}. 

\begin{figure}
\centering
\includegraphics[width=0.49\textwidth]{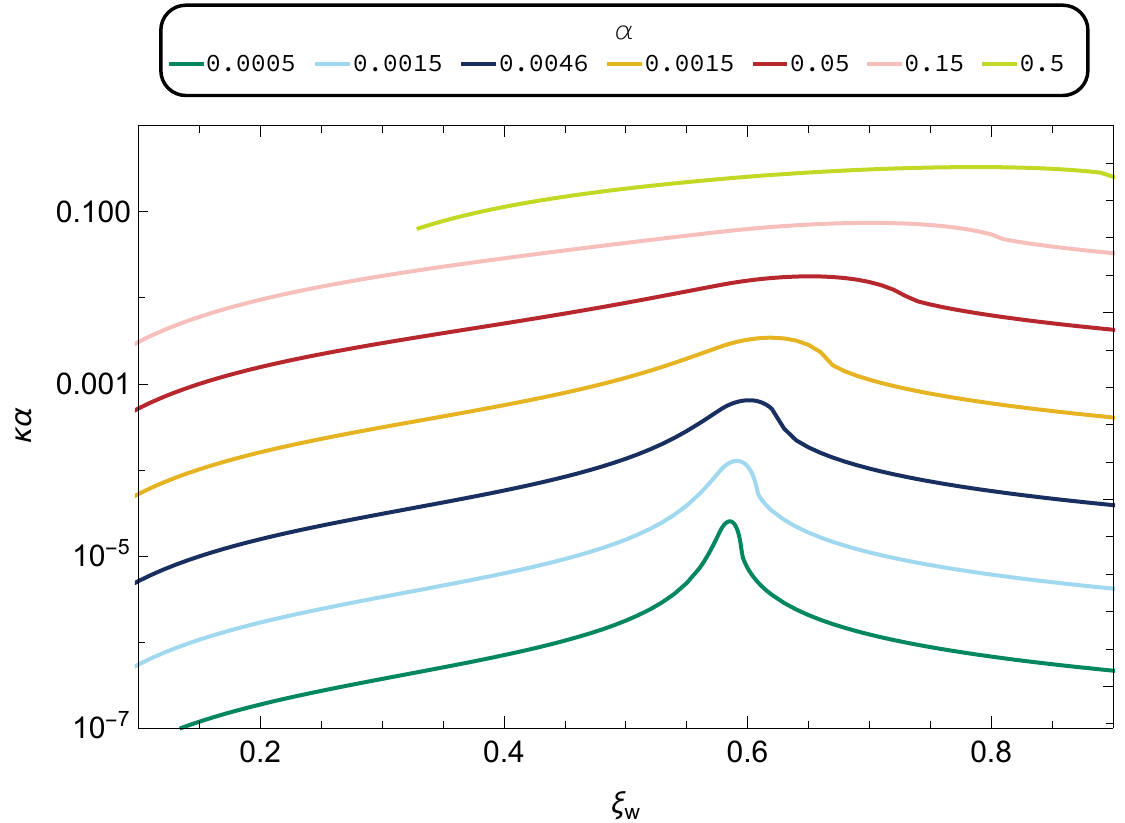}
\includegraphics[width=0.49\textwidth]{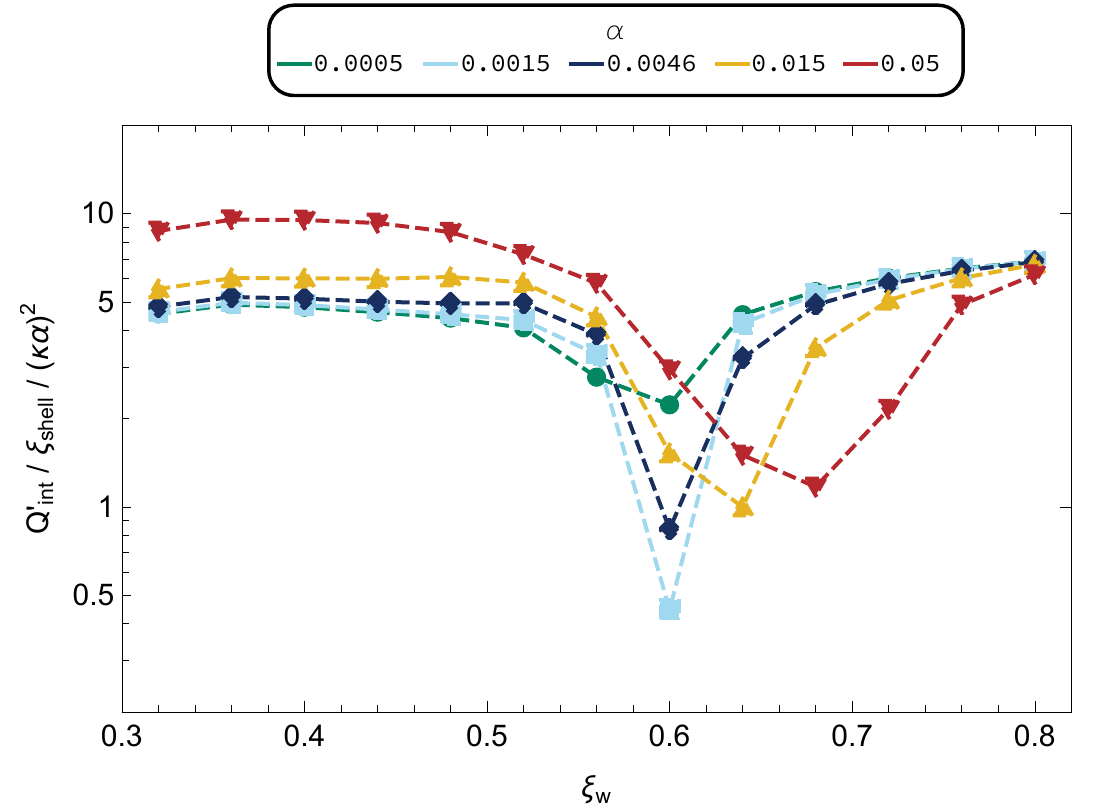}
\vskip 0.5cm
\includegraphics[width=0.49\textwidth]{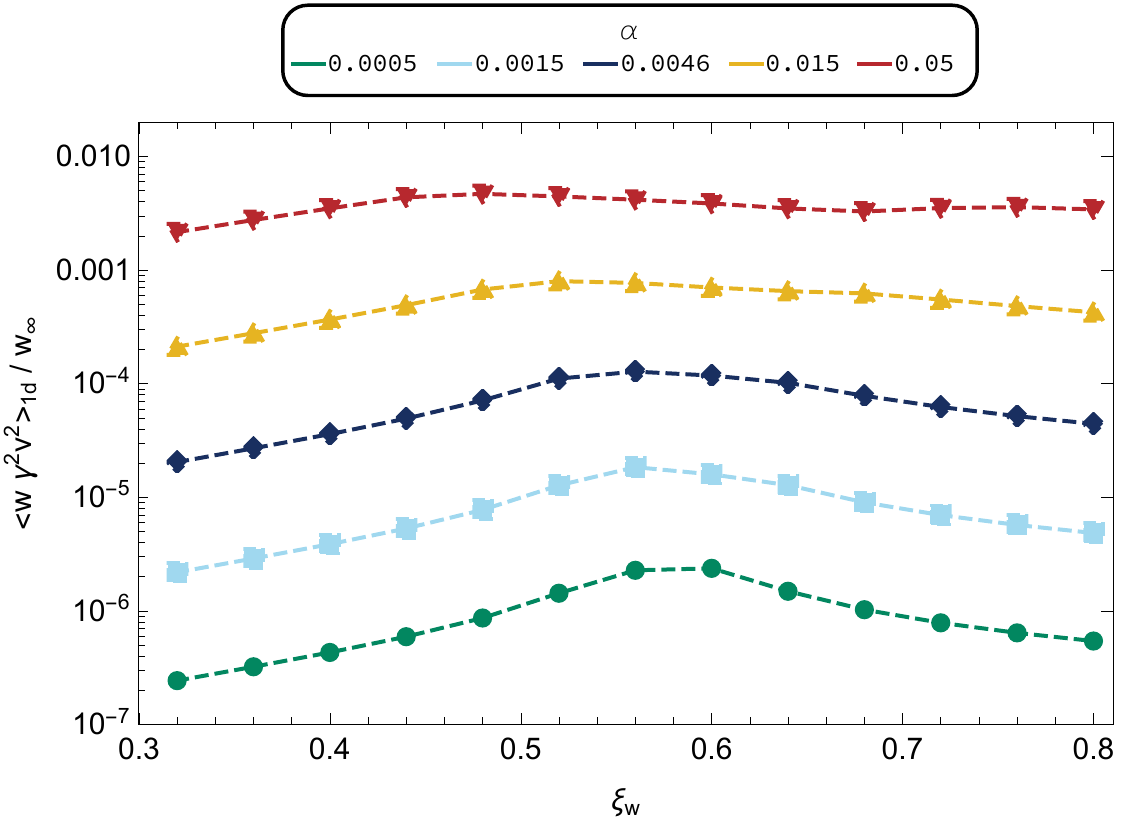}
\includegraphics[width=0.5\textwidth]{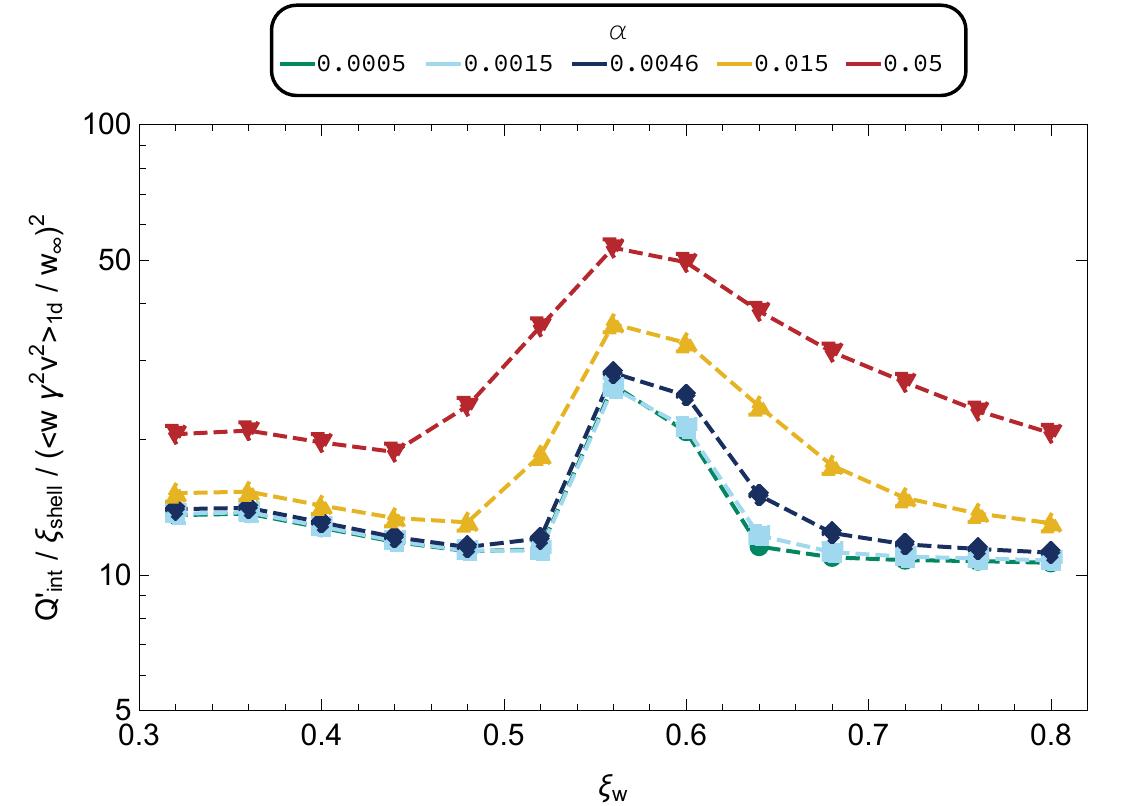}
\vskip 0.5cm
\includegraphics[width=0.49\textwidth]{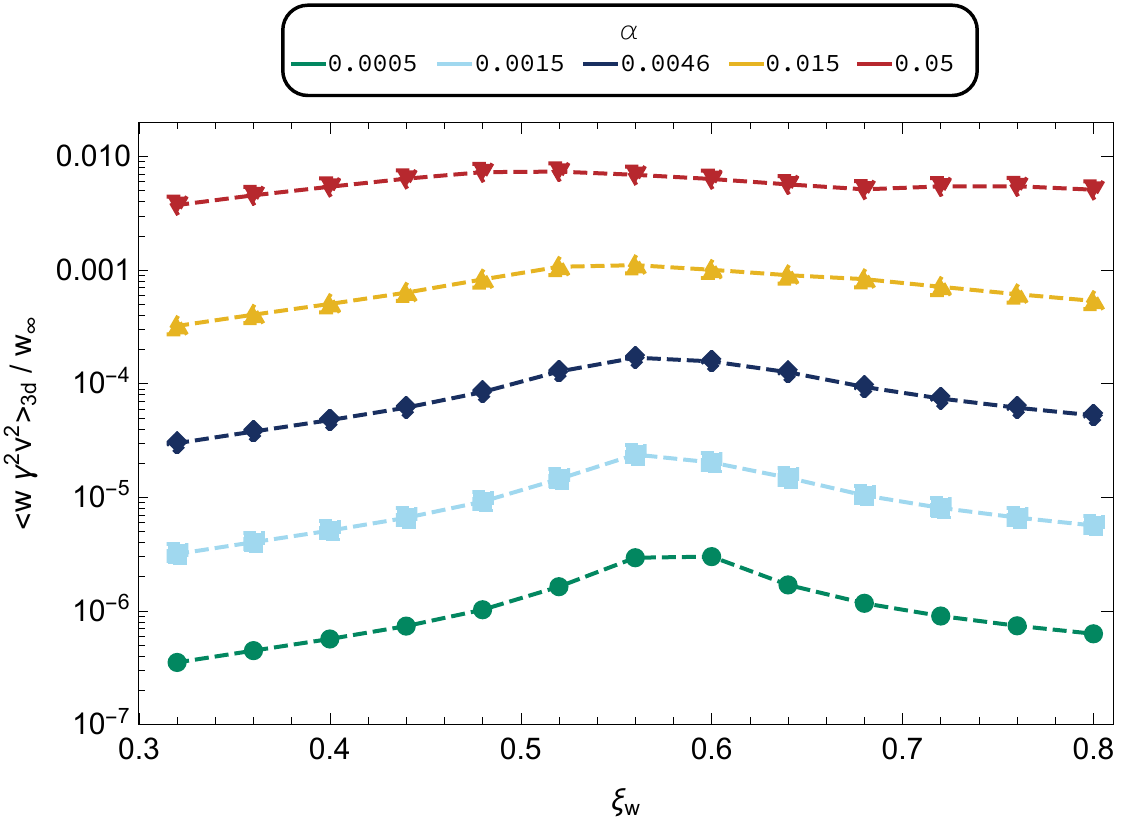}
\includegraphics[width=0.5\textwidth]{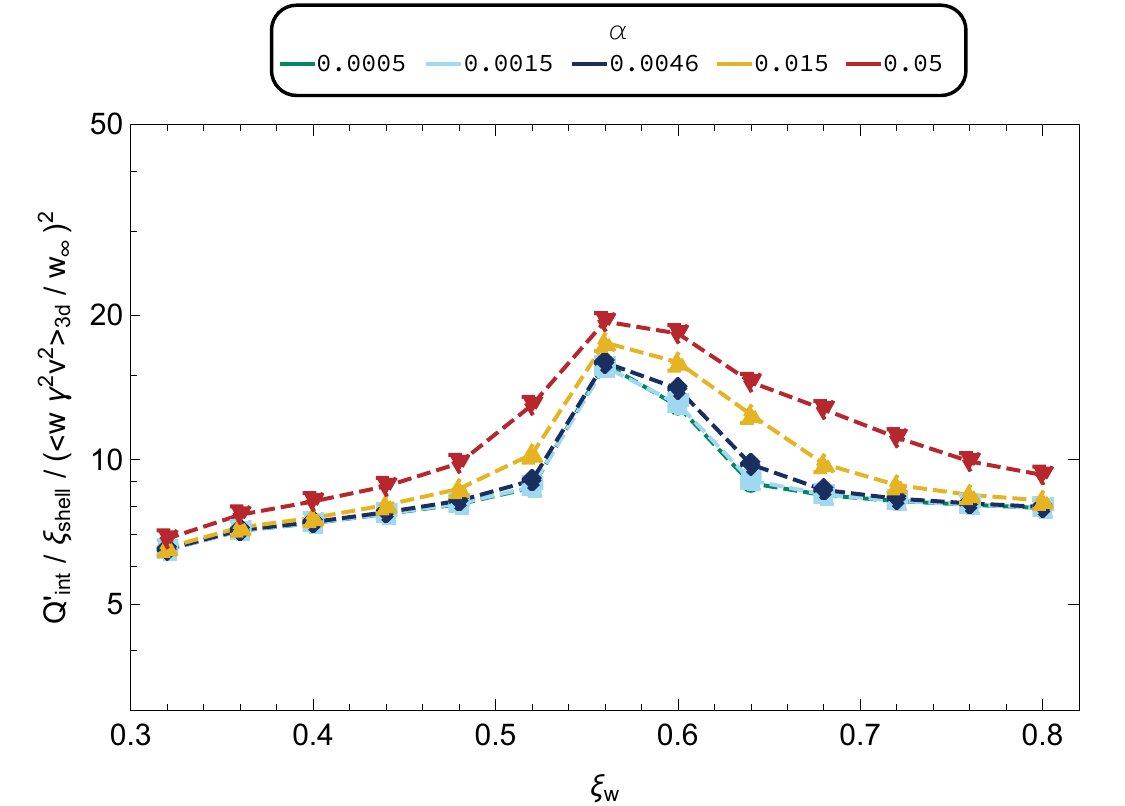}
\caption{
The left plots show the possible normalization factors $\kappa \alpha$, $\wvv_{\rm 1d} / w_\infty$, and $\wvv_{\rm 3d} / w_\infty$ for different values of $\alpha$ and $\xi_w$. The right plot shows the integrated growth factor $Q'_{\rm int}$ in relations to the normalization factors. 
}
\label{fig:normalization_factor}
\end{figure}

Figure~\ref{fig:normalization_factor} shows $Q'_{\rm int}$ for various values of $\alpha$ and $\xi_w$ and the three parameters quantifying the strength of the phase transition. 
While $\wvv_{\rm 1d}$ and $\wvv_{\rm 3d}$ show similar dependence, $\kappa \alpha$ behaves somewhat differently from the others. 
In particular, for weak phase transitions with a wall velocity close to the speed of sound, $\kappa \alpha$ shows 
a significant enhancement that is not seen in our simulations. 
As a result, the GW spectrum estimated from $\kappa \alpha$ overestimates the GW spectrum compared to our simulations.
As we discuss below and in Appendix~\ref{app:normalization}, this comes from the rearrangement of the fluid profile after collision, and this effect is stronger in this regime.
The strength parameter measured in the 3d simulation tracks the GW spectrum excellently, only leading to small variations for wall velocities close to the speed of sound. The strength parameter measured in the 1d simulation performs somewhat worse, especially for stronger phase transitions. This is probably due to non-linear effects arising in the simulations when many shells overlap in the same location.

\paragraph{Fitting the integrated GW spectrum}

As shown in the last section, the GW spectrum falls within a factor of $\sim 2$ -- $3$ for all the wall velocities for different expansion modes of transitions when normalized by $\wvv_{\rm 3d}^2$.
Hence a reasonable fit to the data is given by
\begin{align}
Q'_{\rm int}
&\simeq
9 \times \xi_{\rm shell} \times ( \wvv_{\rm 3d} /w_\infty )^2 .
\label{eq:Qint_fit_3d}
\end{align}
In this expression the normalization coefficient $\wvv_{\rm 3d}$ is the 3d strength measured in the simulation. 

More readily obtained is the 1d strength parameter, since this only requires a 1d simulation which is not as demanding.
As seen from Fig.~\ref{fig:normalization_factor}, $\wvv_{\rm 1d}$ also gives a reasonable normalization of the spectrum in the weak transition regime.
The data points for $\alpha = 0.05$ start to deviate from others because of nonlinear effects:
while $\Delta w^{(i)} / w_0$ and $\vec{v}^{(i)}$ from each bubble is well within the perturbative regime, the velocity sum can be ${\cal O}(0.1)$ on some grid points.
This makes $w \gamma^2 v^2$ on the 3d grids larger than $w \gamma^2 v^2$ estimated from a single bubble. 
From our data we obtain
\begin{align}
Q'_{\rm int}
&\simeq
12 \times \xi_{\rm shell} \times ( \wvv_{\rm 1d} /w_\infty )^2 .
\label{eq:Qint_fit_1d}
\end{align}
This estimate is conservative and underestimates the GW signal for the case of strong phase transitions or wall velocities close to the speed of sound.

Finally, normalization by $\kappa \alpha$ is not too bad either as seen from Fig.~\ref{fig:normalization_factor}.
As mentioned before, $\kappa \alpha$ tends to overestimate the GW signal in the regime of weak phase transitions when the wall
velocity is close to the speed of sound. At the same time, $\left< w \gamma^2 v^2 \right>_{\rm 1d,3d}$ tend to underestimate the GW production for these 
parameter choices, albeit to a lesser degree. 
Of course, to what extent these trends persist in the non-linear regime is debatable, since we assume linearity when the fluid is embedded into the grid and have to model the first collision,\footnote{
We thank D.~Cutting and M.~Hindmarsh for pointing out our improper choice of the initial condition for the simulation.
In an earlier version of the draft, the shift of enthalpy was poorly implemented for deflagrations.
} see Appendix~\ref{app:first_collision}.

In Fig.~\ref{fig:wv2_kappaalpha} we plot the relation between $\kappa \alpha$ and $\wvv_{\rm 1d}$.
The colored data points are for $\alpha = 0.0005, 0.015, \cdots, 0.05$ while the gray line is drawn from $\alpha = 0.0001$.
Interestingly, the relation seems universal and not to depend strongly on $\alpha$ for small wall velocities. Hence, this figure can be readily used to translate $\kappa \alpha$ into $\wvv_{\rm 1d}$. Special care has to be used for wall velocities close to the speed of sound, since the GW spectrum from the simulation does not show a significant increase for hybrids, unlike $\kappa \alpha$.

\begin{figure}
\centering
\includegraphics[width=0.7\textwidth]{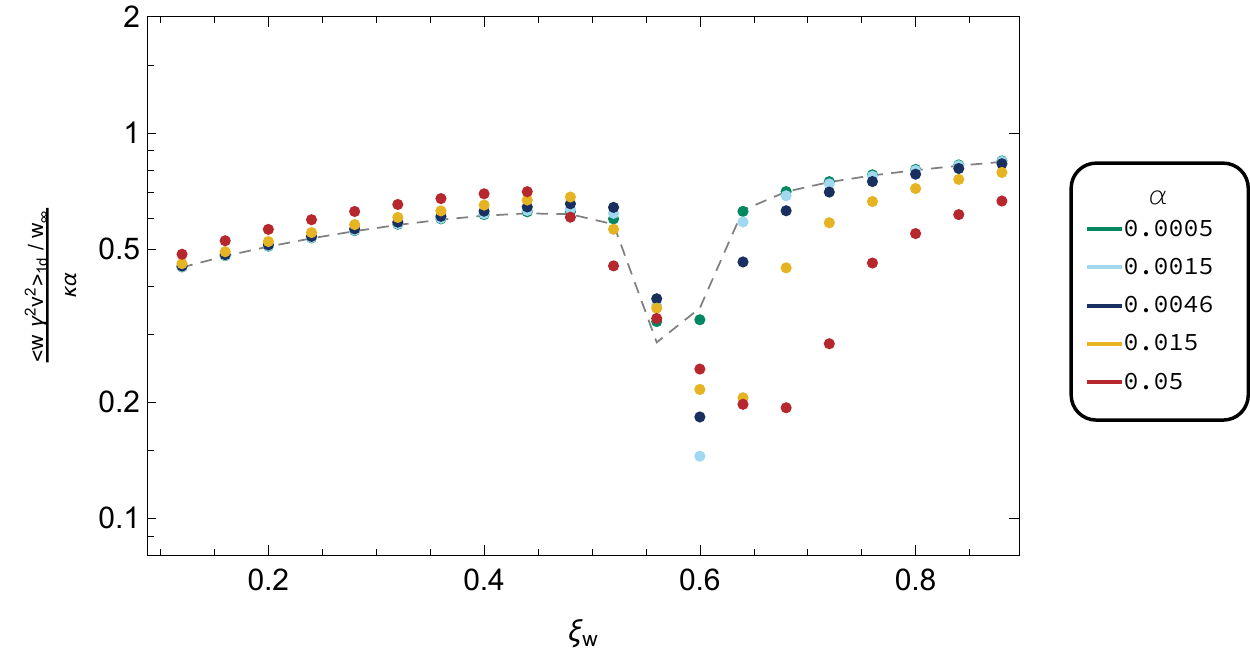}
\caption{
Ratio between $\kappa \alpha$ and $\wvv_{\rm 1d} / w_\infty$ for different values of $\alpha$ and $\xi_w$.
The gray dashed line is drawn from data for $\alpha = 0.0001$.
}
\label{fig:wv2_kappaalpha}
\end{figure}

\paragraph{Fitting the spectral shape}

\begin{figure}
\centering
\includegraphics[width=0.49\textwidth]{./figs/Q_wv2_3d.pdf}
\includegraphics[width=0.49\textwidth]{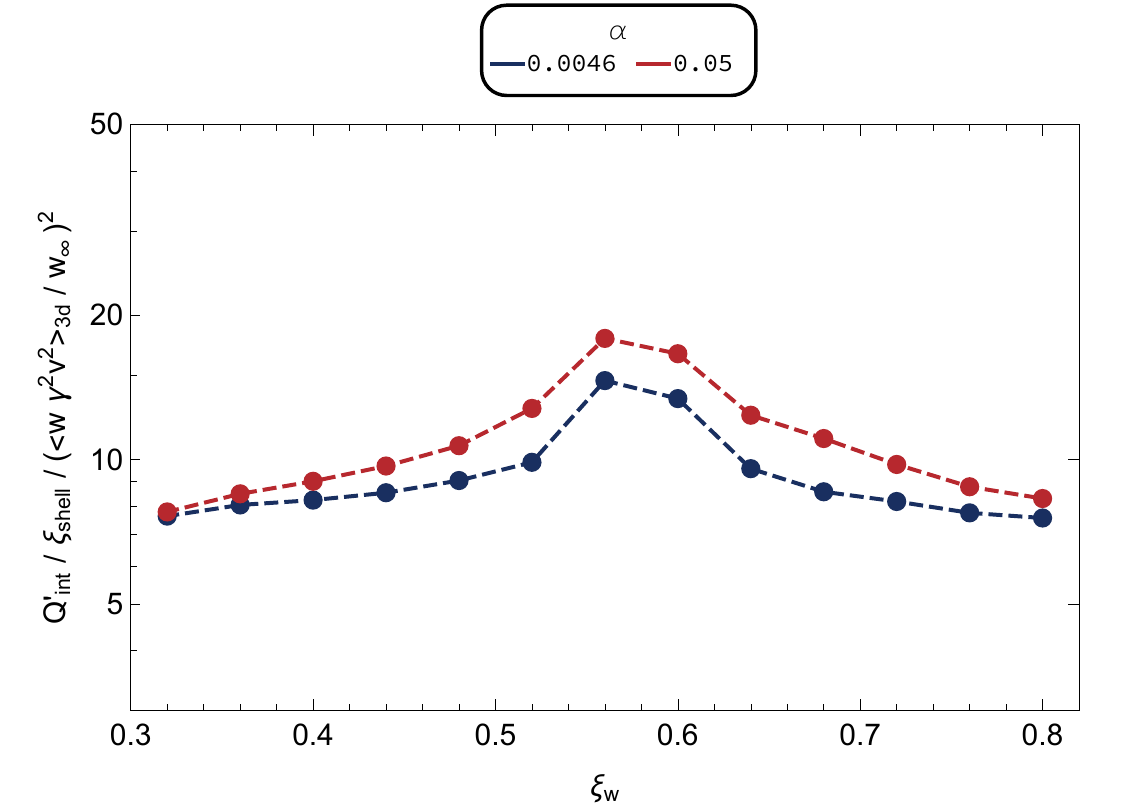}
\vskip 0.5cm
\includegraphics[width=0.49\textwidth]{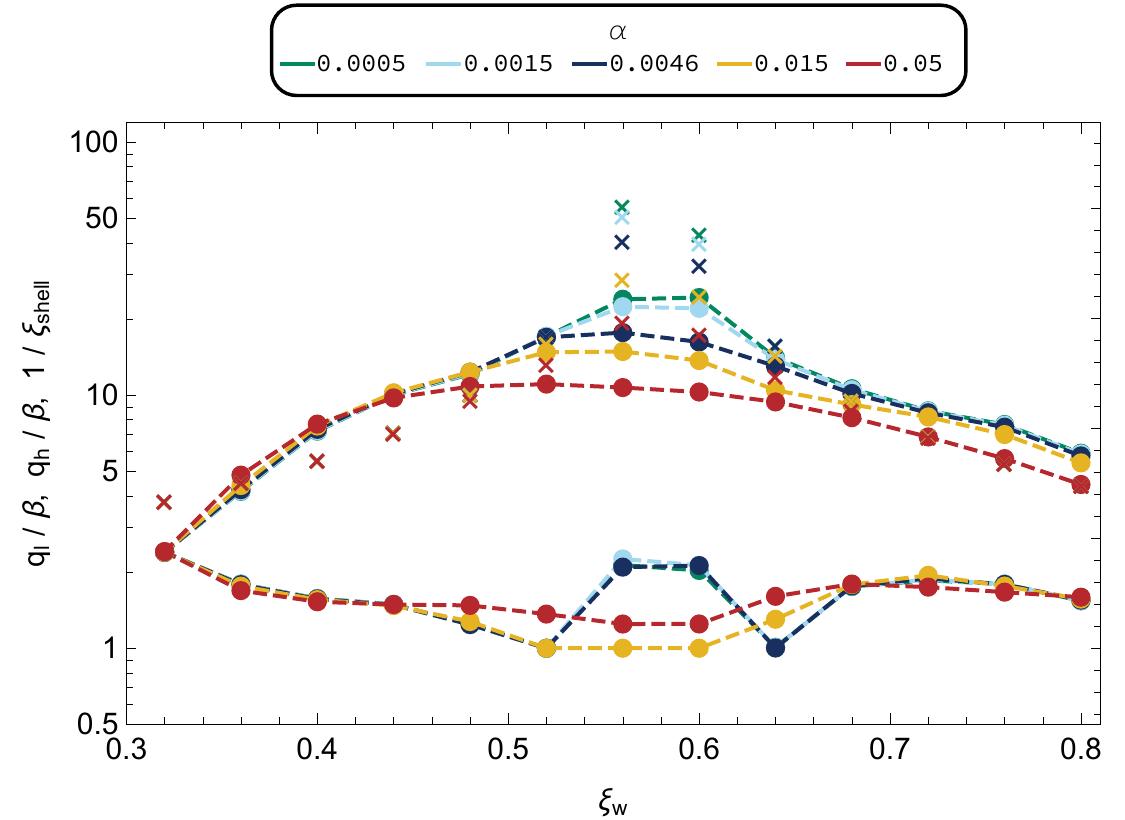}
\includegraphics[width=0.49\textwidth]{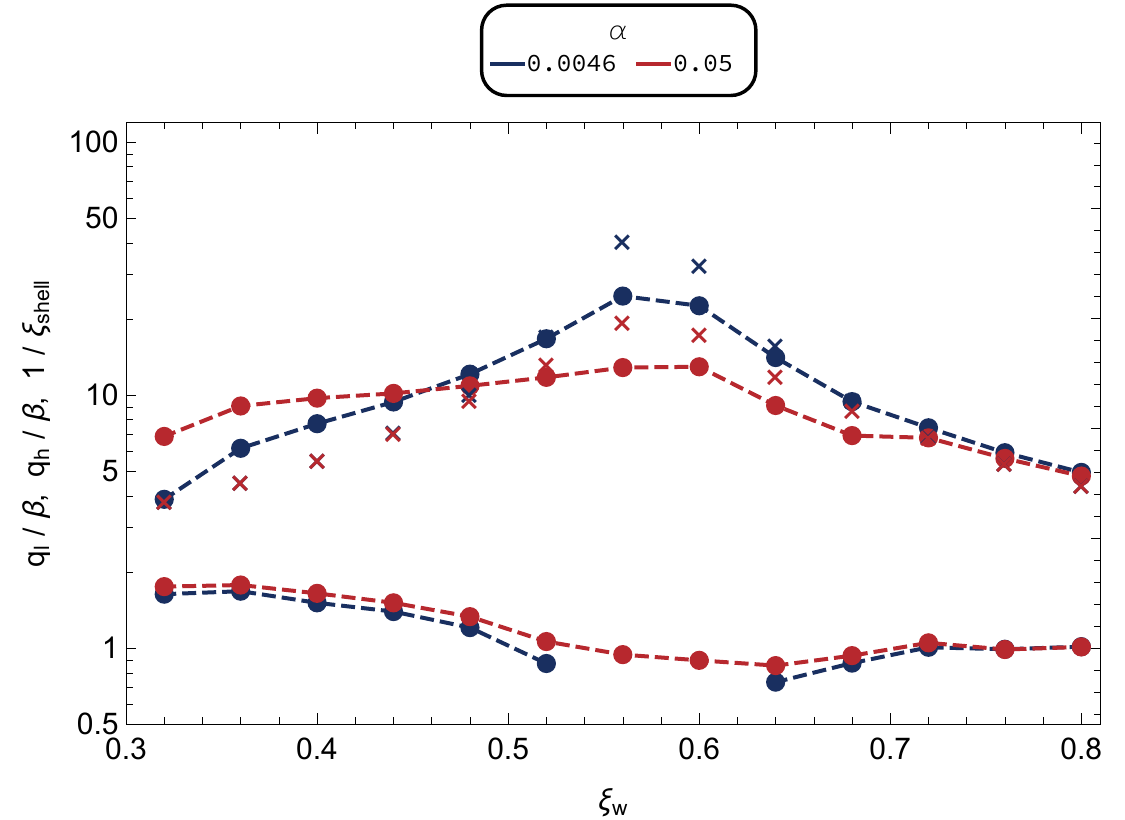}
\vskip 0.5cm
\includegraphics[width=0.49\textwidth]{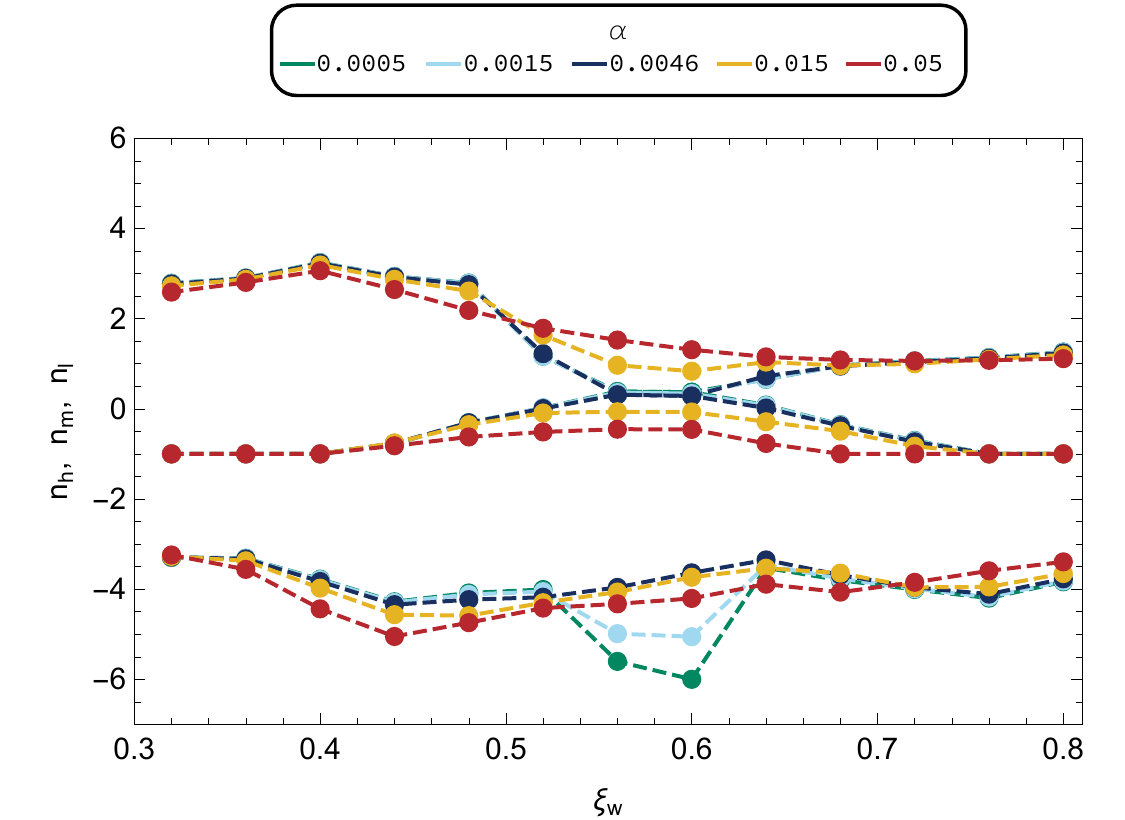}
\includegraphics[width=0.5\textwidth]{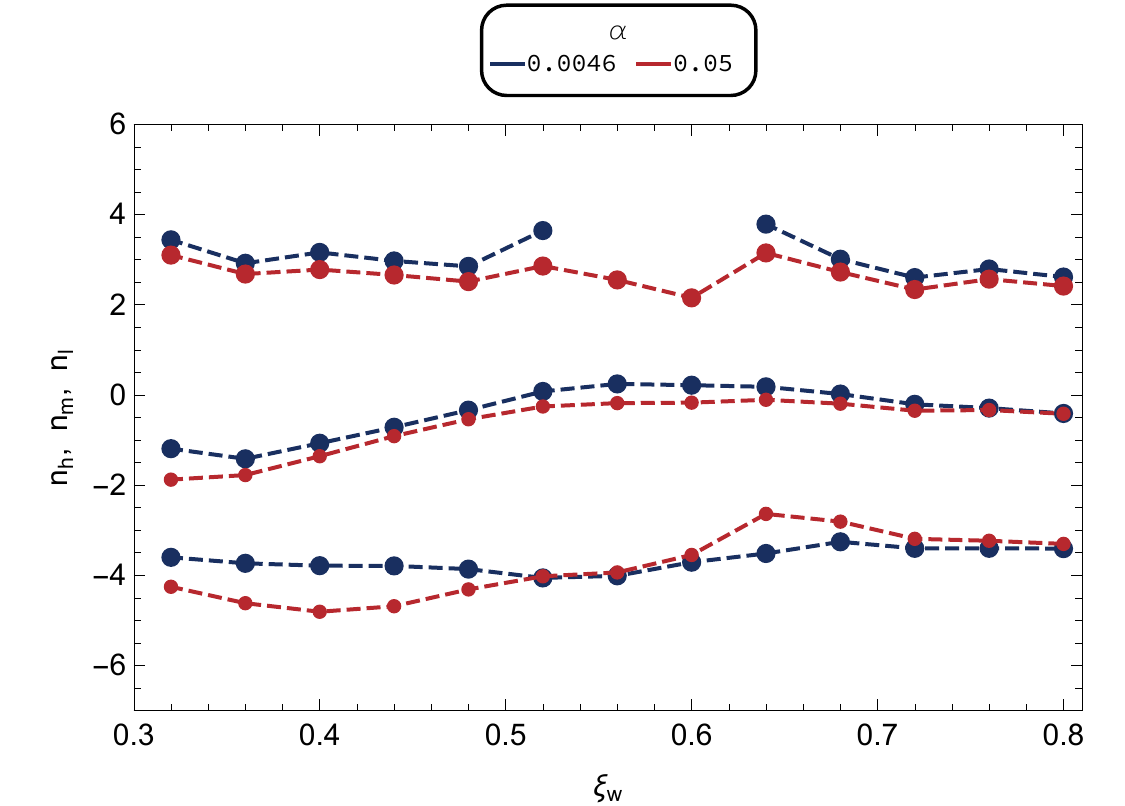}
\caption{
The plots show the fitting parameters in Eq.~(\ref{eq:Q_fit_1d}) of the GW spectrum for $N=256$ and $L = 40 \xi_w / \beta$ (left) and $N=512$ and $L = 80 \xi_w / \beta$ (right). Missing points indicate that the corresponding parameters could not be measured reasonably due to
different issues (dominance by GWs from collisions or a lack of resolution). We can see that some features for thin sound shells are better resolved in the high-resolution simulations. The crosses indicate the inverse shell thickness $1 / \xi_{\rm shell}$ that should be compared to the position of the second peak $q_h / \beta$ in the spectrum. 
}
\label{fig:fitting_n_bigbox}
\end{figure}

A detailed comparison with experimental sensitivity curves also requires the knowledge of the peak frequency (or frequencies) of the spectrum and 
to certain extent the asymptotic behavior. We fitted all spectra to a double-broken power law
\begin{align}
\frac{Q'}{
\xi_{\rm shell} \times ( \wvv_{\rm 3d} /w_\infty )^2}
&\propto
\frac{1}{(q / q_l)^{-n_l} + (q / q_l)^{-n_m} + (q_h / q_l)^{-n_m} (q / q_h)^{-n_h}}
\label{eq:Q_fit_1d}
\\[0.2cm]
&\simeq
 \left\{
\begin{matrix}
(q / q_l)^{n_l} &~~~~(q \ll q_l) \\[0.2cm]
(q / q_l)^{n_m} &~~~~(q_l \ll q \ll q_h) \\[0.2cm]
(q_h / q_l)^{n_m} (q / q_h)^{n_h} &~~~~(q_h \ll q)
\end{matrix}
\right. .
\label{eq:Qpr_fitting}
\end{align}
Note that the template is proportional to $q^{n_l}$, $q^{n_m}$, and $q^{n_h}$ for the three frequency ranges, respectively.
Notice also that these two peaks should not be confused with the features from the early evolution seen e.g.~in the top-right panel of Fig.~\ref{fig:3d_weak_intermediate}.
Both of the peaks come from the late-time behavior of the spectrum.
Fig.~\ref{fig:fitting_n_bigbox} shows the fitting result. The amplitude can be determined via the fit of the integrated 
GW spectrum, c.f. Eq.~(\ref{eq:Qint_fit_3d}).

One sees that the bending at the low frequency is almost constant in $\xi_w$ while the one at high frequency correlates with the shell thickness.
This is somewhat unexpected, since one would expect that the (physical) bubble size scales with the wall velocity. However, the fluid profile for (subsonic) deflagrations actually ends with the shock front that propagates at least as fast as the speed of sound. Also for detonations no decrease in the momentum scale coming from the bubble size can be observed. Even though we simulate phase transition with wall velocities up to $\xi_w = 0.8$, there is no clear trend in the data.

We also performed simulations with a bigger box in order to investigate the box size effect (especially on the IR exponent). The right panels of Fig.~\ref{fig:fitting_n_bigbox} use $L = 80 \xi_w / \beta$ and $N=512$, while the left plots are smaller simulations with $L = 40 \xi_w / \beta$ and $N=256$. For most parts, this allows to measure the IR behavior with better accuracy. Our data is consistent with
\begin{align}
q_l
&\simeq 1,
~~~~~~
q_h
\simeq 1 / \xi_{\rm shell},
\label{eq:q_final}
\end{align}
and
\be
n_l
\in \, [2,4] \, , \quad
n_m
\in \, [-1,0] \, , \quad
n_h
\in \, [-4,-3] \, ,
\label{eq:n_final}
\ee
This combines the steep rise of the envelope approximation with the fast fall-off from the bulk flow model. 

We finally summarize how to use our result.
For given $\alpha$ and $\xi_w$ one can calculate the kinetic energy fraction before collision $\kappa \alpha$ and the shell thickness $\xi_{\rm shell}$ from the fluid profile of an expanding bubble using Refs.~\cite{Espinosa:2010hh,Giese:2020rtr,Giese:2020znk}.
It can be converted to the 1d kinetic energy long after collision $\wvv_{\rm 1d} / w_\infty$ using Fig.~\ref{fig:wv2_kappaalpha}.
Then the linear growth of the GW spectrum $Q'$ is calculated from Eq.~(\ref{eq:Qpr_fitting}) with the overall normalization given by Eq.~(\ref{eq:Qint_fit_1d}).
The parameters characterizing the spectral shape can be read off from Fig.~\ref{fig:fitting_n_bigbox} (left: small box simulation, right: big box simulation) or Eqs.~(\ref{eq:q_final}) and (\ref{eq:n_final}).
Finally $Q'$ can be converted to the GW spectrum $\Omega_{\rm GW}$ using Eq.~(\ref{eq:Q1}).

\section{Discussion and conclusions}
\label{sec:DC}

In this paper we proposed a novel way to calculate the sound wave contribution to the GW spectrum
in first-order phase transitions. 
The main idea is illustrated in Fig.~\ref{fig:embedding}: assuming linearized fluid equations of motion after the transition, both the enthalpy and velocity field are well described by the superposition of contributions from different bubbles.
The contribution from each bubble can be calculated from 1d simulation (i.e.~3d with spherical symmetry) with a proper rescaling of the collision time depending on the direction of the surface element of each bubble.
One of the advantages of our method is that we can incorporate the shock front (discontinuities in fluid profile) relatively easily. Hence, our scheme only contains the scalar field as a boundary condition in the 1d simulation. The dynamical range of the simulation has to resolve only the bubble size and the sound shell thickness but not the bubble wall thickness. This allows for economic simulations and more extensive parameter scans. 

In our setup, it is possible to separate the mean bubble separation from the simulation volume (we always spawn more than 2500 bubbles in our simulation) and to separate the shell thickness from the grid spacing more easily. For lattice simulations including the scalar field this is prohibitively expensive in the regime of thin sound shells when the wall velocity is close to the speed of sound.

We discussed the dependence of the resulting GW spectrum on the fluid kinetic energy and on the shell thickness, and successfully scaled out the proportionality factor. We provide expressions for the contributions of the GW spectrum for sound waves from fitting our data in the last section.
Overall, the most robust outcome of our simulations is that the best observable to benchmark is the integrated power spectrum. When normalized to the kinetic energy at late times in the simulation, $\wvv_{\rm 3d}$, the results become rather independent from the wall velocity and strength of the phase transition, see Fig.~\ref{fig:normalization_factor}. 

In comparison with former lattice results of the fluid interacting with a scalar field~\cite{Hindmarsh:2017gnf} we find the same qualitative features. The mean bubble separation and the sound shell thickness leave both traces in the power spectrum in the form of a double-broken power law. 
However, quantitatively we find some differences, see Fig.~\ref{fig:fitting_n_bigbox}. 
Overall, our GW signal is somewhat larger (by about a factor $\sim 2$ for $\alpha=0.0046$ and $\xi_w=0.8$). We also find a rather flat spectrum (a plateau) between the scale of the bubble separation and the shell thickness, while the lattice results and the sound shell model~\cite{Hindmarsh:2016lnk} seem to support a linear increase, $\Omega_{\rm GW} \sim k$. Some of these differences can be attributed to the fact that we nucleate bubbles according to an exponential increase in probability compared to simultaneous nucleations. Besides, the bubble count in our simulations is higher, which tends to reduce finite volume effects and slightly increases the result.

Moreover, our results indicate that the spectrum is rather independent of the wall velocity when normalized to the measured kinetic energy $\wvv_{\rm 3d}$, just as the (hydrodynamic) lattice simulations indicate. Normalizing to the energy of a single spherical bubble, $\kappa\alpha$, performs somewhat worse. This is especially true for hybrid solutions of weak to intermediate phase transitions. $\kappa\alpha$ shows a significant enhancement in this regime that is not seen in our simulations.

\section*{Acknowledgment}

The authors are grateful to Daniel Cutting and Mark Hindmarsh for helpful comments, especially for pointing out our poor modelling of the initial condition after the first collisions in the first version of the draft.
The work of RJ was supported by Grants-in-Aid for JSPS Overseas Research Fellow (No. 201960698).
This work is supported by the Deutsche Forschungsgemeinschaft 
under Germany's Excellence Strategy -- EXC 2121 ,,Quantum Universe`` -- 390833306.

\appendix

\section{Numerical Evolution of the 1d profile}
\label{app:1d_numerics}

\subsection{Equations}

We solve the hydrodynamic equations $\partial_\mu T^{\mu \nu} = 0$ and assume $d$-dimensional spherical symmetry
(e.g.~$d = 1$ is planar, $d = 2$ is cylindrical, and $d = 3$ is spherical).
We also assume a relativistic ideal gas $T_{\mu \nu} = w u_\mu u_\nu + p g_{\mu \nu}$ with $w = \rho + p$ and $p = \rho / 3$.
This leads to the evolution equations already stated in Eq.~(\ref{eq:1d_system}) 
\begin{align}
\partial_t u + \partial_r f + g
&= 0,
\end{align}
where
\begin{align}
u
&= 
\left(
\begin{matrix}
u_1 \\
u_2
\end{matrix}
\right)
=
\left(
\begin{matrix}
w \gamma^2 - p \\
w \gamma^2 v
\end{matrix}
\right),
~~
f
= 
\left(
\begin{matrix}
w \gamma^2 v \\
w \gamma^2 v^2 + p
\end{matrix}
\right),
~~
g
= 
\frac{d-1}{r}
\left(
\begin{matrix}
w \gamma^2 v \\
w \gamma^2 v^2
\end{matrix}
\right).
\end{align}
In terms of $\rho$ and $v$, we have
\begin{align}
\partial_t
\left(
\begin{matrix}
\rho \\
v
\end{matrix}
\right)
 + A~\partial_r
\left(
\begin{matrix}
\rho \\
v
\end{matrix}
\right)
+ h
&= 0,
\label{eq:delrhov}
\end{align}
where $A$ and $h$ are
\begin{align}
A
&= 
\frac{1}{1 - c_s^2 v^2}
\left(
\begin{matrix}
(1 - c_s^2) v & \rho + p
\\[0.4cm]
\displaystyle \frac{c_s^2 (1 - v^2)^2}{\rho + p} & (1 - c_s^2) v
\end{matrix}
\right),
~~~~
h
=
\frac{d - 1}{r}
\left(
\begin{matrix}
\displaystyle
\frac{(\rho + p) v}{1 - c_s^2 v^2}
\\[0.4cm]
\displaystyle
- \frac{c_s^2 v^2 (1 - v^2)}{1 - c_s^2 v^2}
\end{matrix}
\right).
\end{align}
For $p = \rho/3$, we can write $f$ and $g$ in terms of $u$ as
\begin{align}
u
&=
\left( 
\begin{matrix}
u_1 \\
u_2
\end{matrix}
\right),
~~
f
=
\left( 
\begin{matrix}
u_2 \\[1ex]
\displaystyle
\frac{5}{3} u_1
-
\frac{2}{3} \sqrt{4u_1^2 - 3u_2^2}, 
\end{matrix}
\right),
~~
g
=
\frac{d - 1}{r}
\left( 
\begin{matrix}
u_2 \\[1ex]
2u_1 - \sqrt{4u_1^2 - 3u_2^2}
\end{matrix}
\right).
\end{align}
Here we used 
\begin{align}
u_1
&= 
\frac{1+v^2/3}{1-v^2}\rho,
~~~~
u_2
= 
\frac{4v/3}{1-v^2}\rho,
~~~~
v
=
2\left( \frac{u_1}{u_2} \right)
- \sqrt{4\left(\frac{u_1}{u_2}\right)^2 - 3}
=
\frac{3u_2}{2u_1 + \sqrt{4u_1^2 - 3u_2^2}}.
\end{align}

\subsection{Numerical schemes}

In this appendix, we explain the numerical scheme adopted to evolve the 1d fluid dynamics, described by the equation system above. 
This kind of convective-diffusive systems admits a class of finite-difference solution schemes that are independent of the eigenvalues of the function $f$, called central schemes. One of the simplest and widely-used schemes is the so-called Lax-Friedrichs (LF) scheme \cite{lax, friedrichs}
\begin{eqnarray} \label{eq:lf_scheme}
u^{n+1}_j = \frac{u^{n}_{j+1}+u^{n}_{j-1}}{2} - \frac{\lambda}{2}\left[ f(u^{n}_{j+1}) - f(u^{n}_{j-1})\right] - g_j^n \Delta t \,,
\end{eqnarray}
with $\lambda = \Delta t/ \Delta x$ \footnote{It is important to guarantee that $\lambda \lesssim 0.5$, such that $\Delta t$ is smaller than $\Delta x$ and one can solve the shockwave elements.}. The upper and bottom indices denote the time and space lattice index, respectively. But the LF scheme, as will be shown below, has a large numerical viscosity, that demands a huge resolution to solve the shockwave fronts. The Kurganov-Tadmor (KT) discretization scheme, introduced in \cite{KURGANOV2000241}, reduces the numerical viscosity. The evolution of a lattice site is given by
\begin{eqnarray} \label{eq:kt_scheme}
u^{n+1}_j = u^{n}_{j} - \lambda \left[ H_{j+1/2}(t^n) - H_{j-1/2}(t^n) \right] - g_j^n \Delta t \,,
\end{eqnarray}
where $H$ is defined between two lattice points as
\begin{eqnarray}
 H_{j+1/2} (t) = \frac{ f(u^{+}_{j+1/2}(t)) + f(u^{-}_{j+1/2}(t)) }{2} - \frac{a_{j+1/2}(t) }{2} \left[ u^{+}_{j+1/2}(t) - u^{-}_{j+1/2}(t) \right] \,.
\end{eqnarray}
The value of $u^{+}_{j+1/2}$ is calculated through a decrement of $u_{j+1}$ while $u^{-}_{j+1/2}$ is calculated as an increment of $u_{j}$
\begin{eqnarray}
u^{+}_{j+1/2}(t) &=& u_{j+1}(t) - \frac{\Delta x}{2} (u_x)_{j+1}(t)\,, \label{eq:u_plus} \\
u^{-}_{j+1/2}(t) &=& u_{j}(t) + \frac{\Delta x}{2}(u_x)_{j}(t) \label{eq:u_minus}\,.
\end{eqnarray}
The derivative of $u$ at the site $j$, $(u_x)_{j}$ is calculated through 
\begin{eqnarray}
(u_x)_{j} = \textrm{minmod}\left( \theta \frac{u_{j}-u_{j-1}}{\Delta x},\frac{u_{j+1}-u_{j-1}}{2\Delta x},\theta \frac{u_{j+1}-u_{j}}{\Delta x} \right) \,,
\end{eqnarray}
where \it minmod \rm is equal to the minimum of the elements if all of them are positive, the maximum of the elements if all of them are negative, and zero otherwise. The value of $a_{j+1/2}(t)$ is defined as the maximum local speed of the fluid element in that intermediate cell, thus one needs to calculate the maximum value of the fluid velocity using both Eqs.~(\ref{eq:u_plus}) and (\ref{eq:u_minus}) and boost it by the sound speed:
\begin{eqnarray}
a_{j+1/2}(t) = \textrm{max} \left(|\mu(v^+_{j+1/2}, c_s)|,|\mu(v^+_{j+1/2}, -c_s)|,|\mu(v^-_{j+1/2}, c_s)|,|\mu(v^-_{j+1/2}, -c_s)| \right)\,,
\end{eqnarray}
with $\mu (a,b) = (a-b)/(1-ab)$ designating the Lorentz-transformed fluid velocity.

\begin{figure}[h]
\centering
\includegraphics[width=0.49\textwidth]{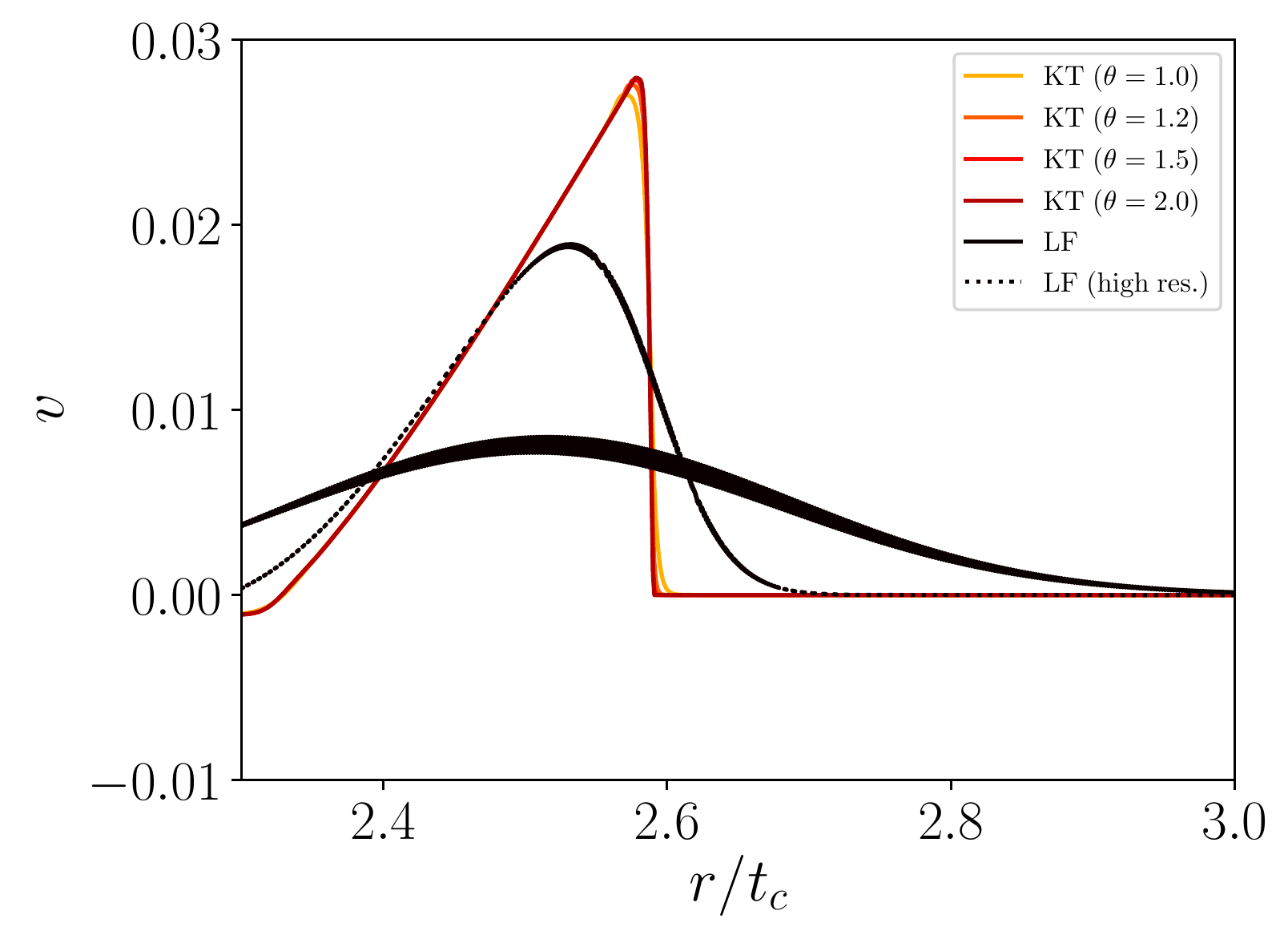}
\includegraphics[width=0.475\textwidth]{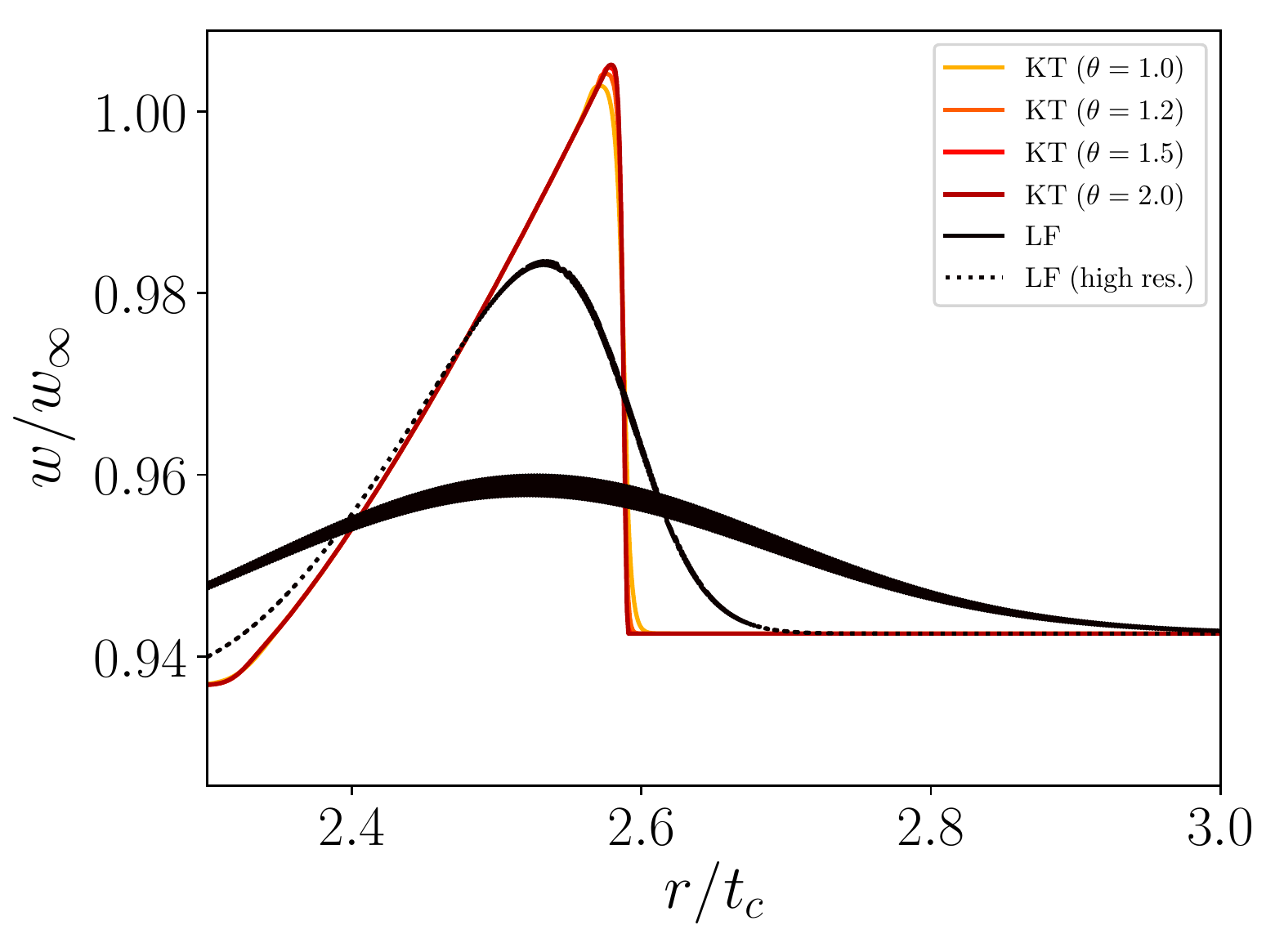}
\caption{\label{fig:theta_scheme_test} \small Velocity and enthalpy profiles a few time steps after collision. We used a initial detonation profile with $\xi_w = 0.8$. We compare the Kurganov-Tadmor (KT) scheme with different values of $\theta$ to the Lax-Friedrichs (LF) scheme. Notice that the LF scheme has some spurious oscillations (shown by the line thickness) and a huge numerical viscosity. For the KT scheme, all values of $\theta$ in the range shown have similar results. Decreasing $\theta$ leads to a larger viscosity and increasing it enhances numerical oscillations. 
}
\end{figure}

The \it minmod \rm derivative works as a flux limiter reducing spurious oscillations, which is recurrent in numerical schemes that aim to solve shocks and discontinuities. The free parameter $\theta$ characterizes the strength of the limiter. A small $\theta$ corresponds to smaller derivatives (in absolute value) and increases the numerical drag. Increasing $\theta$ makes the flux limiter weaker and introduces spurious numerical oscillations. In Figure~\ref{fig:theta_scheme_test} we display the velocity and enthalpy profile for the 1d simulation for a detonation with $\xi_w = 0.8$ a few times steps after the collision. We can see that the LF scheme has a huge numerical viscosity and the shockwave quickly develops into a smoother wave-packet. Also the thickness of the line corresponds to unstable numerical oscillations. We also show different simulations using the KT scheme for different values of $\theta$ inside the range $\left[1,2\right]$. For the same lattice parameters ($\Delta x = 10^{-3}$ and $\Delta t = 10^{-4}$), the KT scheme is much more stable which helps to maintain the shockwaves for longer in the 1d simulation, without any kind of spurious smoothing. In the same panels, in dotted, we display the results of a LF scheme with space and time grid 10 times smaller (high res.). Despite the higher resolution the fluid still does not develop a shock front. Also notice that there is no big difference among the KT lines in the chosen $\theta$ range, such that our result agrees with Ref.~\cite{KURGANOV2000241}.

The Kurganov-Tadmor method guarantees stability with respect to the spatial discretization. In order to guarantee the stability with respect to time discretization, one can use a Runge-Kutta method (of order $l$)
\begin{eqnarray}
u^{(1)}_{j} &=& u^n + \Delta t\, C[u^n]\,, \\
u^{(l+1)}_{j} &=& \eta _l u^n + (1-\eta _l)\left( u^{(l)} + \Delta t\, C[u^{(l)}] \right) \,,\\
u^{n+1} &=& u^{(l)}\,, 
\end{eqnarray}
with the conventional constants $\eta _l$. $C$ is calculated as
\begin{eqnarray}
C[u^\bullet] = - \left[ \frac{H_{j+1/2}(u^\bullet) - H_{j-1/2}(u^\bullet)}{\Delta x}\right]
\end{eqnarray}
In this work, we use the Runge-Kutta method of third order ($l=3$), i.e.~$\eta_1 = 3/4$ and $\eta_2 = 1/3$, which we found guarantees fast convergence in $\Delta t$. 

\begin{figure}[h]
\centering
\includegraphics[width=0.49\textwidth]{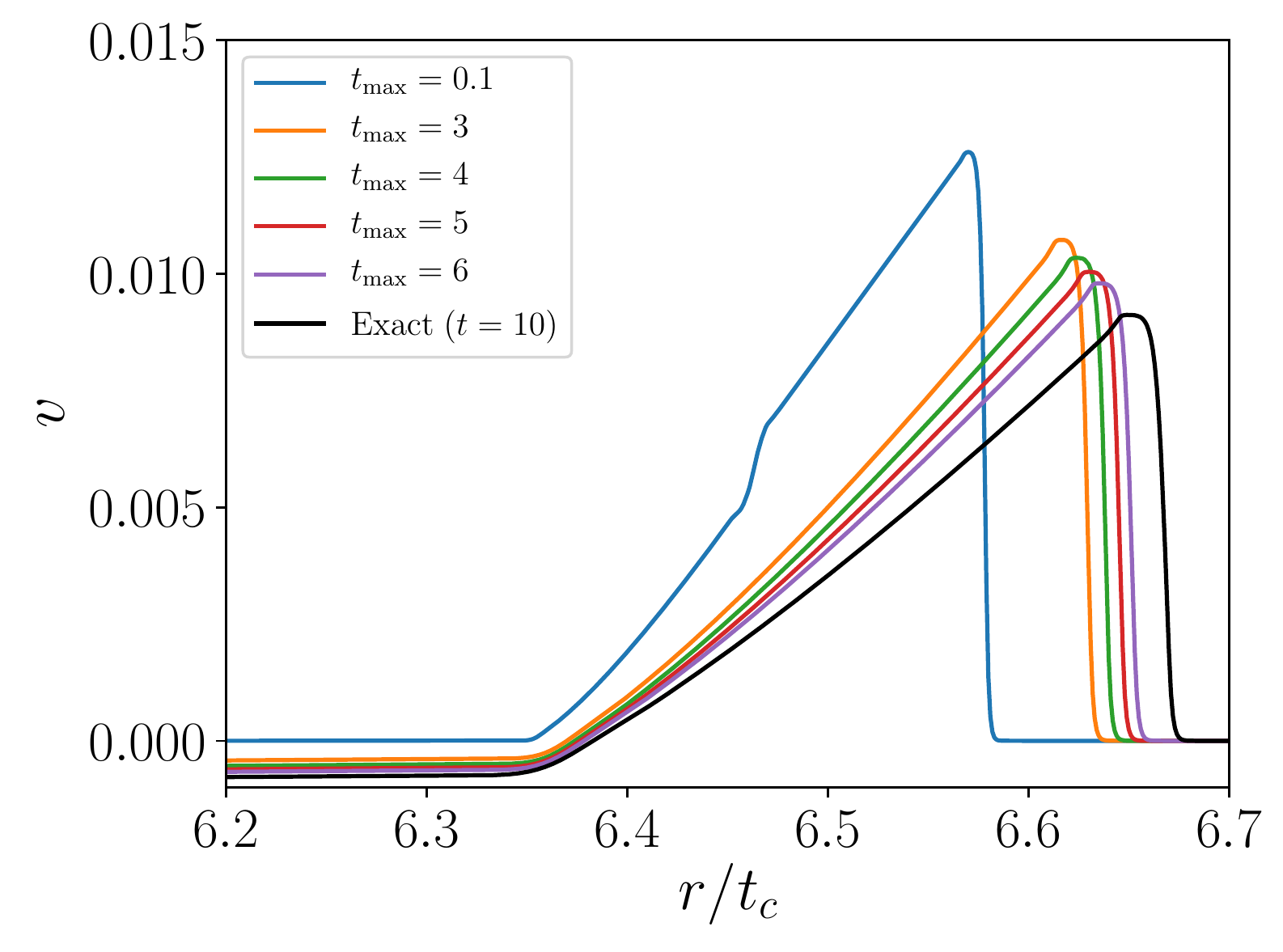}
\includegraphics[width=0.5\textwidth]{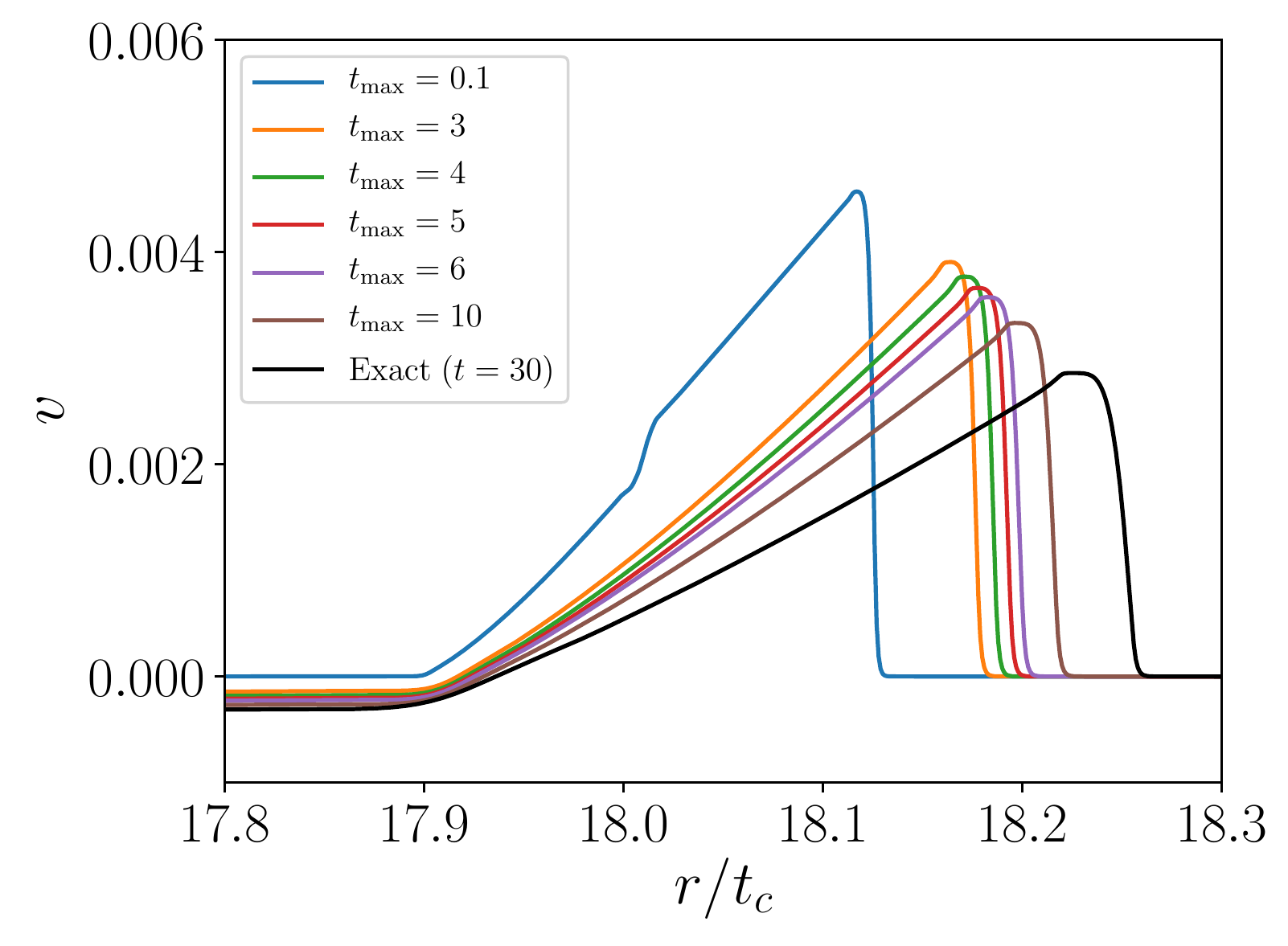}
\caption{\label{fig:extrap_test} \small Test of the 1d simulation extrapolation. On the left, $10$ time units after collision and, in the right, $30$ time units after collision. We compare the effect of considering different maximum extrapolation times used in Eq.~(\ref{eq:tmax_extrap}) with the exact numerical solution (black line). Notice that extrapolating the profile using any $t_{\rm max}$ in the interval $\left[ 3,6 \right]$ leads to similar results, which is rather close to the exact numerical result at short times. At later times (right), the extrapolation starts to fail, but the late time contribution to the final GW spectrum is smaller since it only affects very small bubbles. 
}
\end{figure}

Finally, we comment on the typical values of $\Delta t$ and $\Delta x$ used for the 1d simulations. The typical time scale $t$ and spatial scale $r$ in the 1d simulation are given respectively by
\begin{eqnarray}
\frac{t - t_n}{t_c - t_{n}} \,, \quad 
\frac{r}{t_c-t_{n}}\,,
\end{eqnarray}
with $t$ the simulation time (typically $\sim 10/\beta$) and $r$ the radial distance between the nucleating points and the grid cell (typically the simulation box size). The value $t_{c}$ is the collision time of the grid point and $t_{n}$ is the bubble nucleation time. Notice that since the surface element of the bubble might collide just after nucleating, $t_{c}-t_{n}$, can be arbitrarily small and consequently the 1d simulation time can be arbitrarily large. To overcome this issue, we extrapolate the profile after a maximum 1d simulation time $t_{\rm max}$ using 
\begin{equation} \label{eq:tmax_extrap}
f(r,t) = f(\bar r,t_{\rm max}) \times \,\frac{\bar r}{r} \qquad \textrm{with} \qquad \bar r = r - c_s \, (t-t_{\rm max}) \,.
\end{equation}
The final profile (and the GW outcome) has shown to be stable using $t_{\rm max}$ in the interval $\left[ 3,6 \right]$ after collision time (in $1/\beta$ units), see Fig.~\ref{fig:extrap_test}.\footnote{As stated in the main text, in this work we extrapolate after $t = 7t_c$, which corresponds to $t_{\rm max}=6$.} Increasing $t_{\rm max}$ too much implies not only running the 1d simulation for longer times but also reducing $\Delta t$ (and $\Delta x$) to overcome numerical friction. Having those scales in mind, the typical lattice size for which the 1d simulation is stable is $\Delta x = \mathcal{O}(10^{-3})$ and $\Delta t = \mathcal{O}(10^{-4})$ using KT scheme and Runge-Kutta of third order. 

\section{From the 3d velocity grid to the GW spectra}
\label{app:3d_scheme}

In this appendix, we describe the numerical procedure used in this work to calculate the GW spectra from the velocity grid. Whereas in Section~\ref{sec:Strategy} we described the overall set up and how to obtain the velocity grid, here the details of the scheme used to calculate $Q$ from the vector field $v^i(\vec{x})$ defined in the lattice site $\vec{x} = \left(x_i, y_j, z_k \right)$, where the indexes $x_i$, $y_i$ and $z_i$ run from 0 to $\GRIDSIZE - 1$. 

The first step is to construct the fluid stress-energy tensor in real space through
\begin{equation}
T^{ij}(\vec{x}) = w(\vec{x}) \gamma^2(\vec{x}) v^{i}(\vec{x})v^{j}(\vec{x}) \,.
\end{equation}
Note that we neglected the term proportional to $g_{ij}$, which does not contribute to the GW production.
Here we construct $w(\vec{x})$ and $v^{i}(\vec{x})$ from the superpositions of the different bubble contributions. 
Notice that in linearizing the Euler equations, terms of $O(v^2)$ and $O(v \, (w/w_0-1))$ are neglected but no terms of $O((w/w_0 -1)^2)$. Since the velocities point in different directions, they partially cancel out. If contributions of $n$ bubbles overlap, $v$ only grows as $\sqrt{n}$, while $(w/w_0 -1)$ grows as $n$. Hence, in the regime of many bubbles overlapping, it makes a parametric difference if one superimposes $(w/w_0 -1)$ or another quantity that is in leading order equivalent, e.g.~$\log(w/w_0)$. Therefore, superimposing $(w/w_0 -1)$ is the more consistent choice.

The second step involves taking the Fourier transform of each of the six components of $T^{ij}(\vec{x})$ using the FFTW3 package \cite{Frigo:2005zln}, obtaining the complex field $T^{ij}(\vec{k})$. The dual lattice sites of the compactified rectangular box are given by 
\begin{equation} \label{eq:kdef}
k_i = \sin\left(\frac{2\pi i}{\GRIDSIZE }\right) \,.
\end{equation}
Using this definition of $\vec{k}$, we calculate in the next step the projected components of the stress-energy tensor that appear in the equation of motion for $h$ (see Eq.~(\ref{eq:heq})). We define $T_{+}(\vec{k})$ and $T_{\times}(\vec{k})$ as
\begin{eqnarray}
T_{+}(\vec{k}) = \sum_{i,j} \frac{T^{ij}(\vec{k})}{\sqrt{2}}\left(\theta _i(\vec{k})\theta _j(\vec{k}) - \phi _i(\vec{k})\phi _j(\vec{k})\right)\,,
\\
T_{\times}(\vec{k}) = \sum_{i,j} \frac{T^{ij}(\vec{k})}{\sqrt{2}}\left( \theta _i(\vec{k}) \phi _j(\vec{k}) + \theta _i(\vec{k})\phi _j(\vec{k}) \right)\,.
\end{eqnarray}
Here $\theta_i$ and $\phi_i$ are normalized vectors that are orthogonal to the momentum defined in Eq.~(\ref{eq:kdef}). 

The fourth step involves calculating the Fourier transform with respect to time defined by Eq.~(\ref{eq:FT}). Here, instead of using FFTW, we stack past time slices as
\begin{eqnarray} 
T_{+,\times} (q, \vec{k},t) = \sum_{t'=t_{\rm init}}^t e^{iq t'}T_{+,\times} (t', \vec{k}) \, ,
\end{eqnarray}
in order to save memory during the simulation run. A few comments are in order concerning the choice of $t_{\rm init}$ and the definition of $k$. As mentioned in the main text, the typical $t_{\rm init}$ is taken to be $\sim 10/\beta$ time units after the nucleation of the first bubble, such that the GW from the sound shell does not get contaminated by IR effects related to GWs generated at the bubble size scale. The next point concerns the definition of $k$. Since $k$ is defined through the dispersion relation 
\begin{equation}
-k^2 = \triangle = \nabla_i\nabla^i \,,
\end{equation}
and the Laplacian operator needs to be invertible, we use
\begin{equation} \label{eq:omegadef}
\nabla_i = 2\frac{\GRIDSIZE}{\BOXSIZE} \sin\left(\frac{\pi i}{\GRIDSIZE}\right) \,,
\end{equation}
in contrast to (\ref{eq:kdef}).

The last step is to calculate the GW spectrum as defined in Eq.~(\ref{eq:Q2}), taking the mean of $T_{+}T_{+}^* + T_{\times}T_{\times}^*$ in the lattice sites with the correct norm
\begin{equation}
\Omega(q,t) = C \, q^3 \langle T_{+}T_{+}^* + T_{\times}T_{\times}^* \rangle|_{|\vec{k}| = q} \,.
\end{equation}
The constant $C$ is a normalization of the Fourier transform
\begin{equation}
C = \Delta t^2 \left(\frac{\BOXSIZE^3}{\GRIDSIZE^6}\right) \, .
\end{equation}

\section{The first collision of the fluid}
\label{app:first_collision}

In this appendix we discuss how much the fluid profile is deformed at the first crossing. One might wonder what happens after enthalpy injection from bubble collision stops and how the fluid profile is deformed by the collision.

In order to test this, we perform a 1d fluid simulation of the type described in Section~\ref{subsec:1d} and numerically evaluated according to Appendix~\ref{app:1d_numerics} for collisions of deflagrations, hybrid and detonation initial profiles. We stress that our special concern with the first collision is that it is also the collision of the scalar field solitons. In subsequent fluid shockwave collisions there is no bubble wall involved. In that case, our model considers the linear approximation, justified in the small fluid velocity limit, in which the fluid profiles simply superimpose and cross each other.

\begin{figure}
\centering
\includegraphics[width=0.42\textwidth]{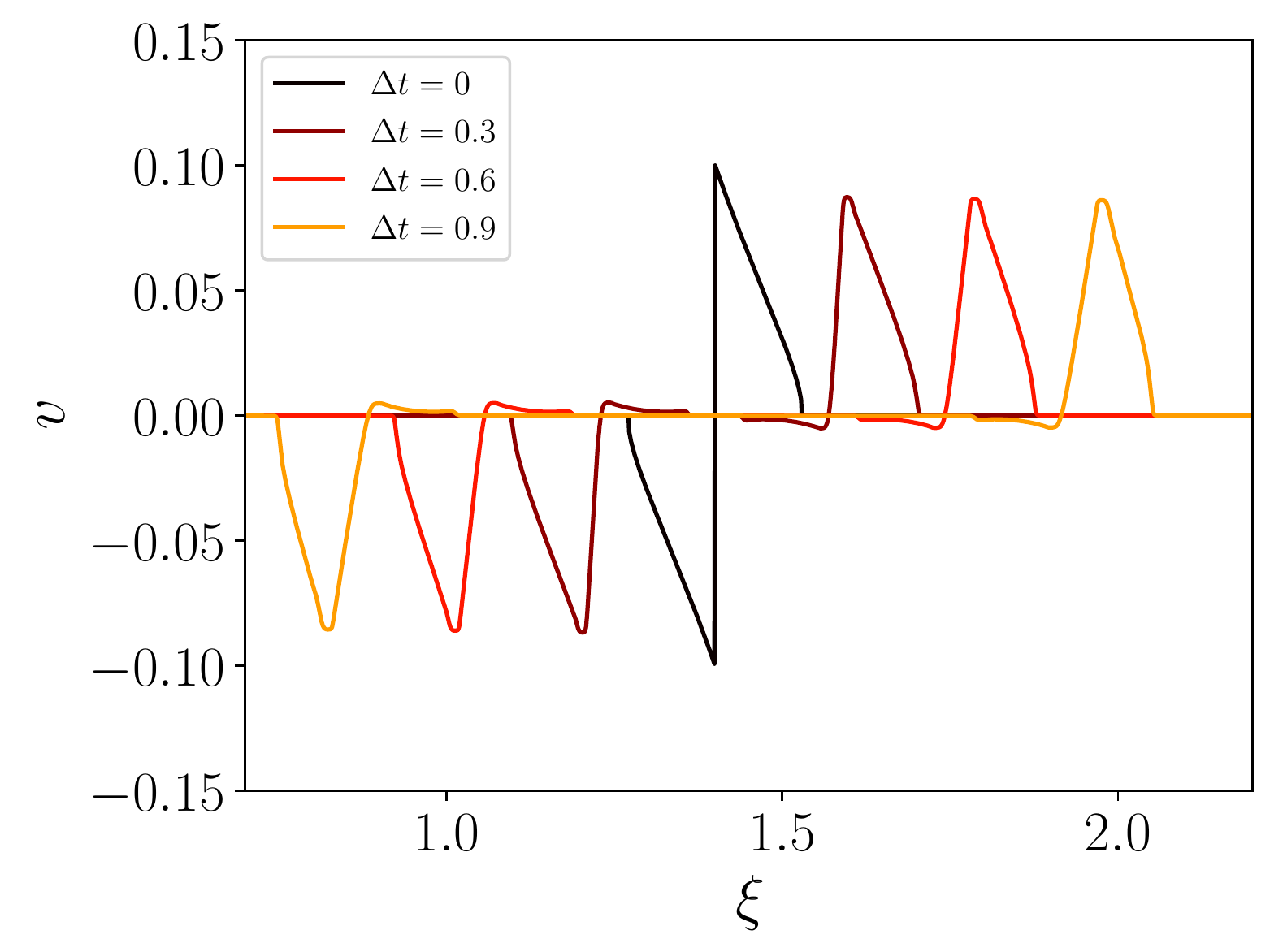} 
\includegraphics[width=0.4\textwidth]{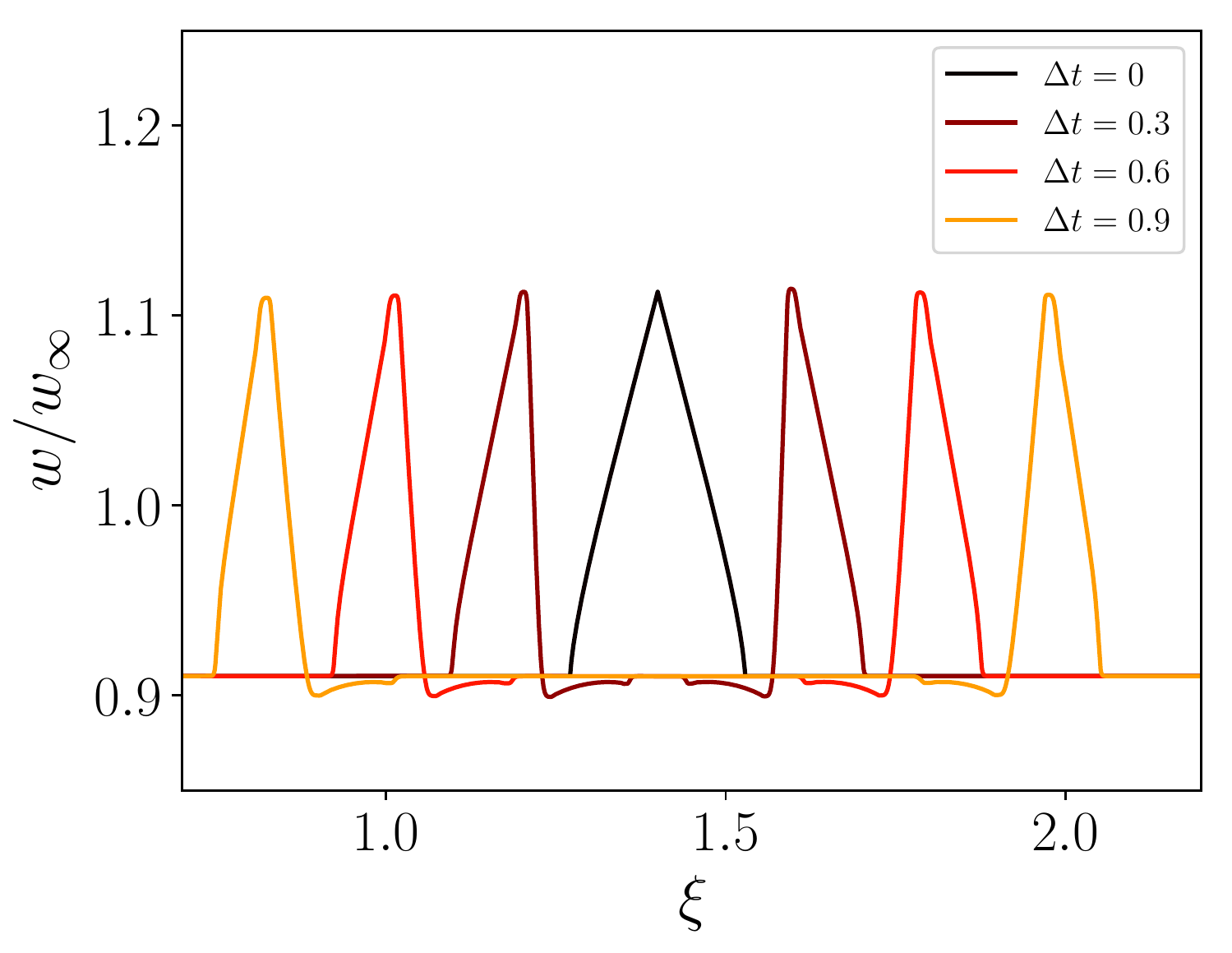} 
\includegraphics[width=0.42\textwidth]{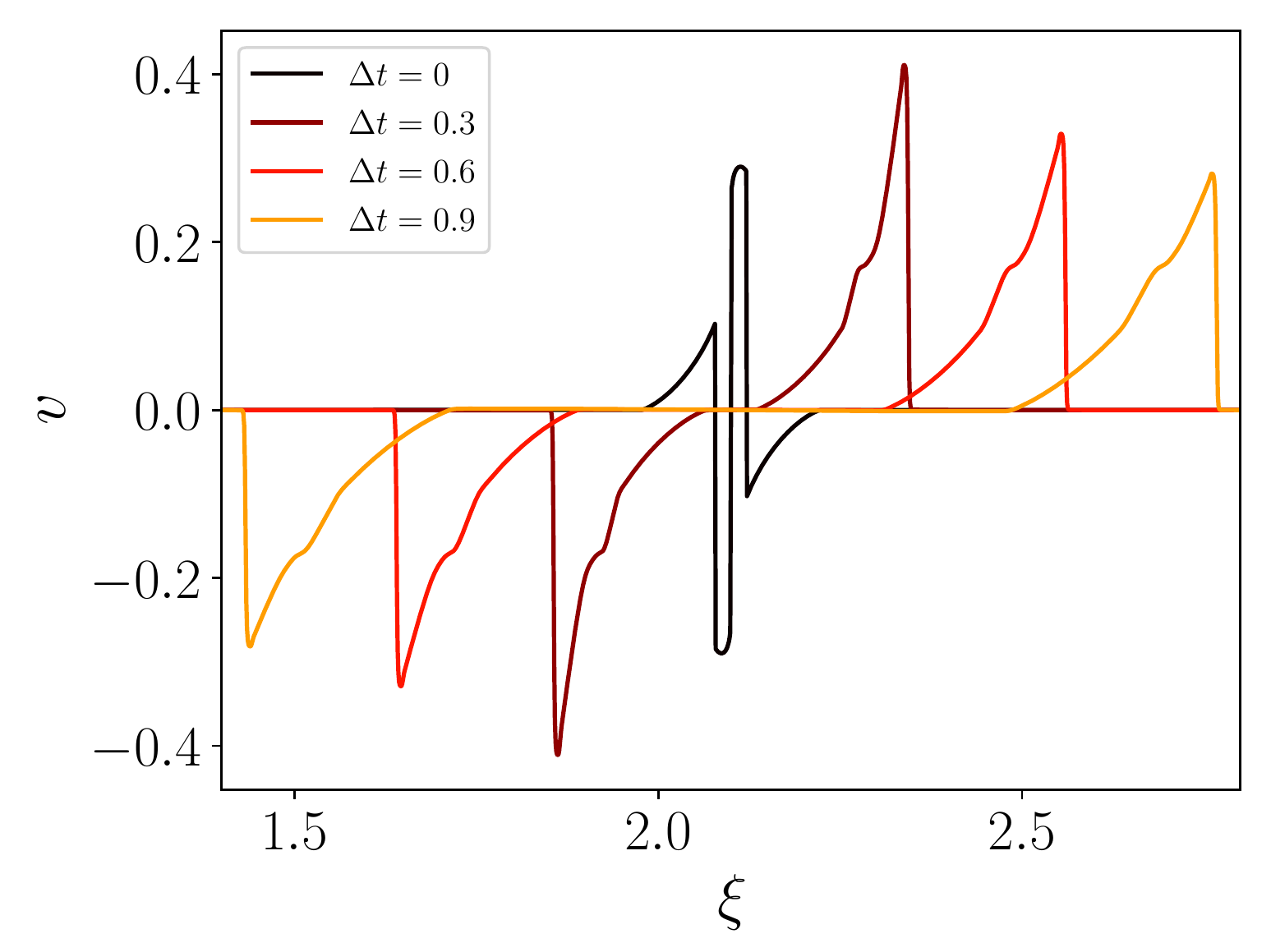} 
\includegraphics[width=0.4\textwidth]{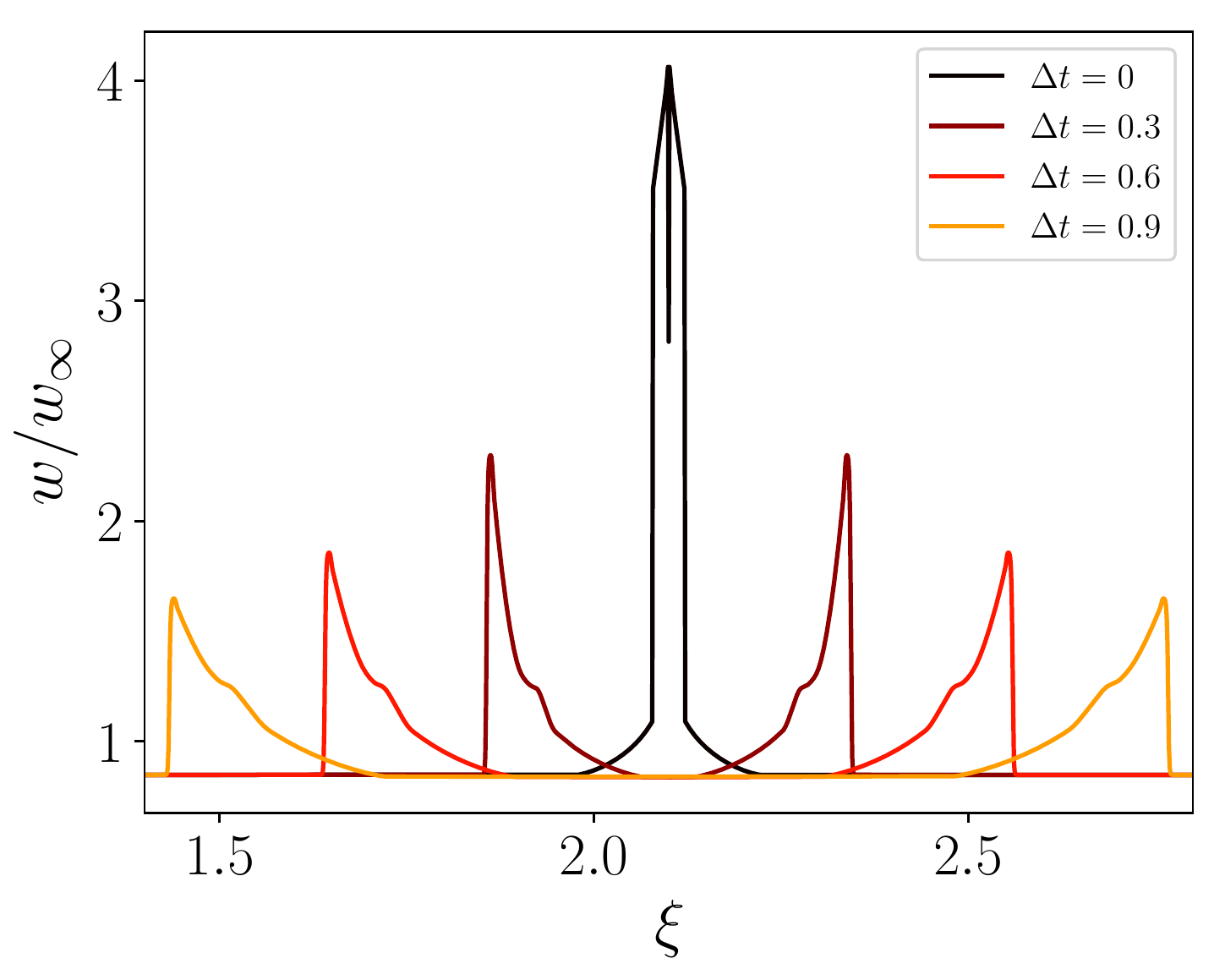} 
\includegraphics[width=0.42\textwidth]{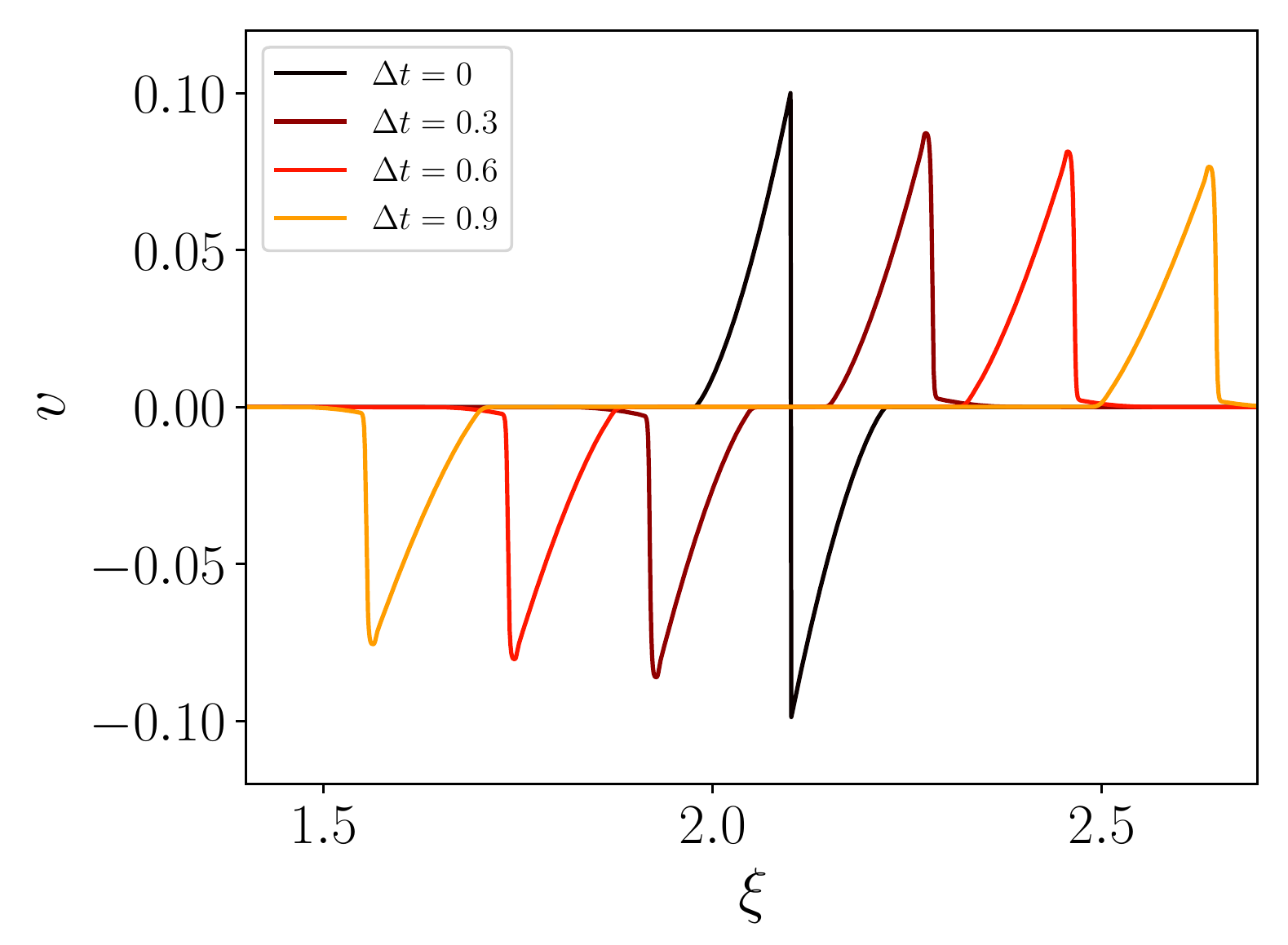} 
\includegraphics[width=0.4\textwidth]{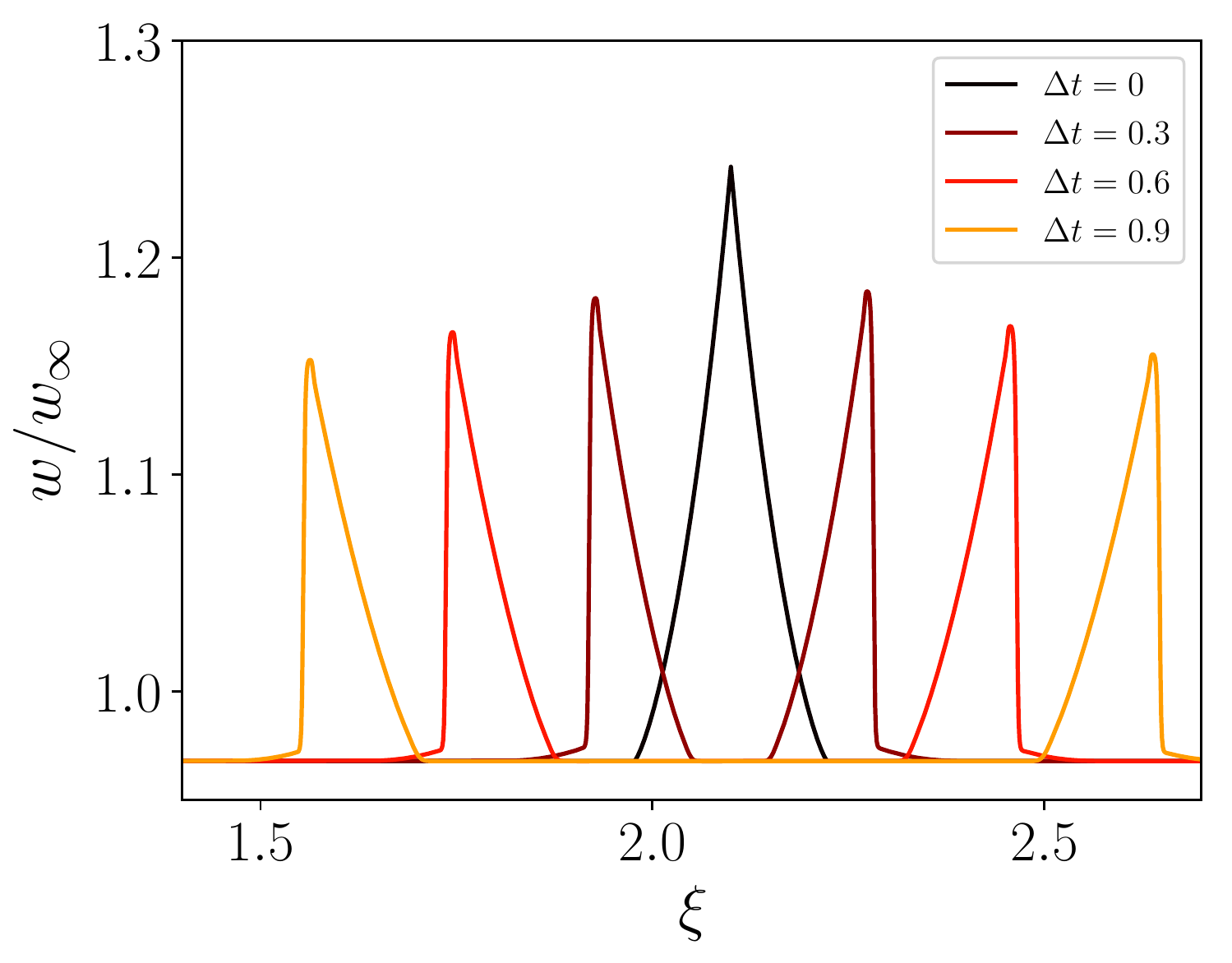} 
\caption{
Effect of first fluid collision for deflagration (top), hybrid (middle) and detonation (bottom), in the left for the velocity and in the right for the enthalpy profile. We set the initial profile at the moment of the bubble wall collision. For all three kind of profiles, after the collision, the enthalpy value in the inner part between the two bubbles is the same as in the innermost part of a single bubble.
}
\label{fig:collision}
\end{figure}

We take as initial conditions for the two frontally colliding fluid profiles the exact moment in which the two (incoming and outgoing) solitons collide. It is important to consider this exact moment since the 1d fluid simulation no longer absorbs the enthalpy injection from the scalar field. Starting the collision simulation earlier makes the system start to lose energy earlier. The initial condition for deflagration, hybrid and detonations are respectively described by the black lines in the top, middle and bottom panels of Figure~\ref{fig:collision}. In the left we show the velocity and in the right the enthalpy profile. Notice that in the case of deflagration and hybrid profiles, the fluid profiles (or part of them) already crossed each other when the scalar solitons collide. In that case, again, the linear approximation is used to justify that both waves pass each other without interacting.
After some time, the spectral shape of the velocity and enthalpy profiles (for all three types of initial conditions) resembles the superposition of the initial profiles. Remarkably, the fluid profiles are not affected much after crossing each other. 
We notice that a short time after collision and for all three profile types, the enthalpy in the region in between the two incoming and outgoing profiles settles down to the same value as the innermost part of the two bubbles. Ultimately, this is a consequence of energy-momentum conservation. We mimick this effect in the initial conditions of our 1D simulations with spherical symmetry by displacing the enthalpy in front of the bubble wall accordingly (i.e.~also in the shock front for deflagrations and hybrid solutions).

\section{Normalization of the spectrum}
\label{app:normalization}

In Sec.~\ref{subsec:DC} we presented the integrated spectrum $Q_{\rm int}$ normalized to $\kappa \alpha$, $\wvv_{\rm 1d}$, and $\wvv_{\rm 3d}$.
We found that the difference between $\kappa \alpha$ and $\wvv_{\rm 1d}$ is larger for wall velocities close to the speed of sound. In this appendix we discuss this behavior in more detail.

\begin{figure}
\centering
\includegraphics[width=\textwidth]{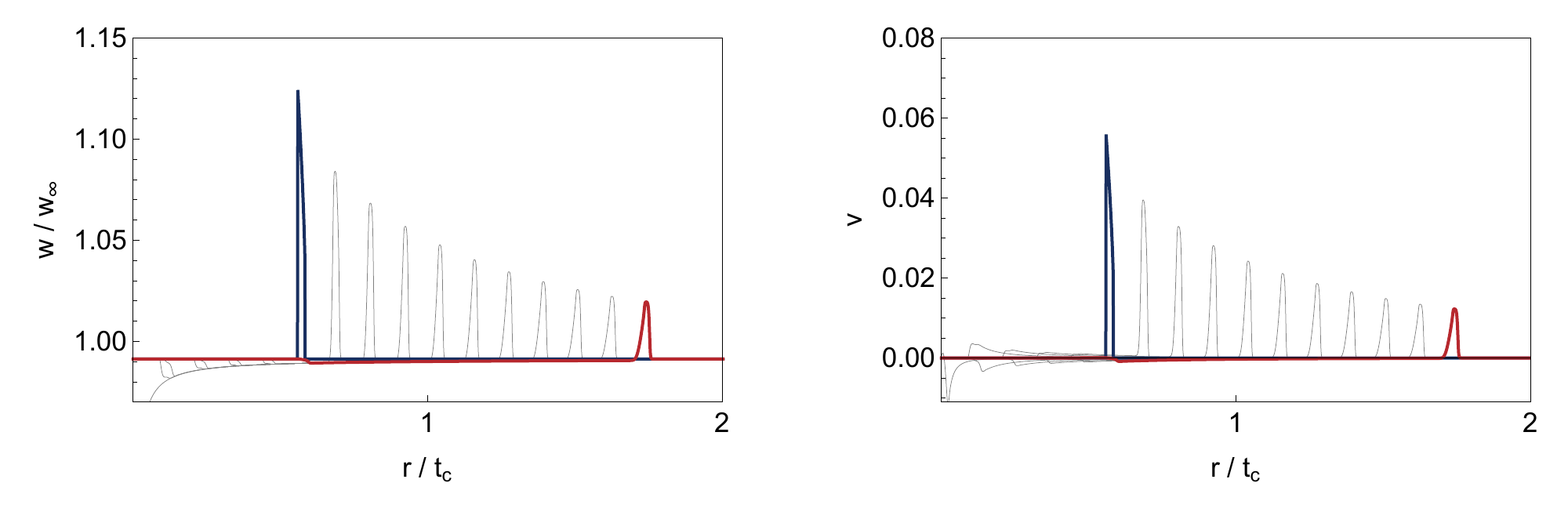} 
\includegraphics[width=\textwidth]{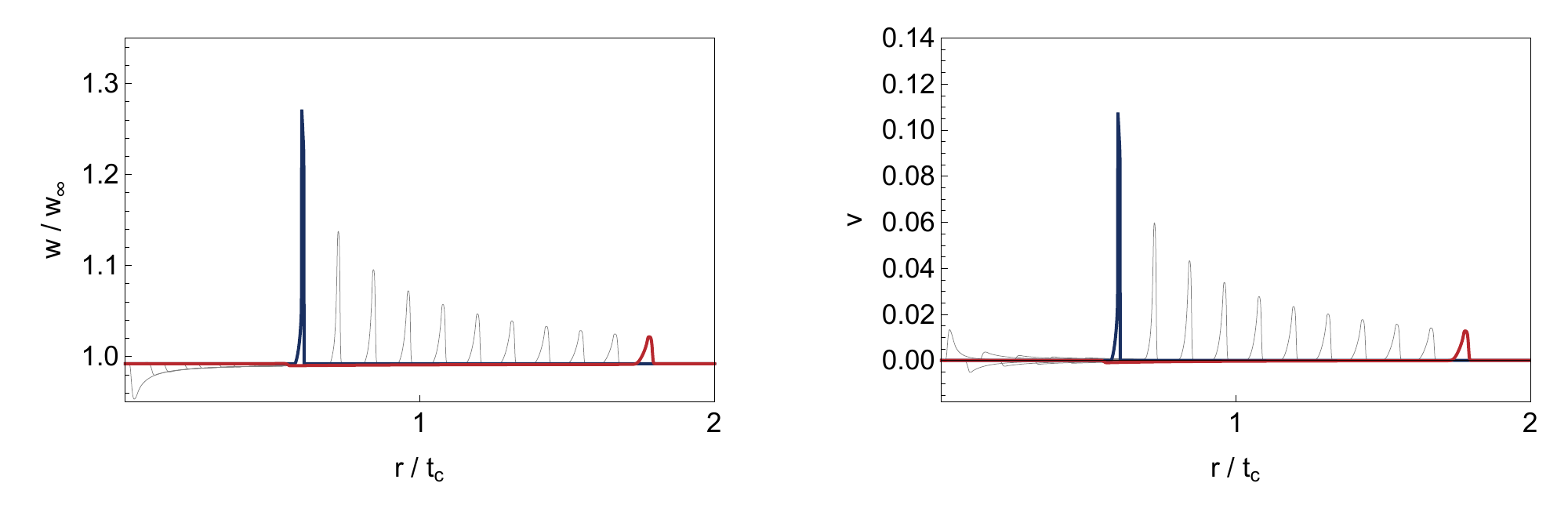} 
\caption{
Time evolution of the fluid profile for $\alpha = 0.0046$ and $\xi_w = 0.56$ (top) and $\xi_w = 0.6$ (bottom) from $t = t_c$ (blue) to $2t_c$ (red).
}
\label{fig:wv2_closer_look}
\end{figure}

Fig.~\ref{fig:wv2_closer_look} shows the time evolution of the fluid profile for $\alpha = 0.0046$ and $\xi_w = 0.56$ (top panels) and $\xi_w = 0.6$ (bottom panels).
Both $\Delta w = w - w_0$ and the maximal radial velocity $v$ decay relatively quickly while the shell volume increases as $\propto r^2$.
This tendency is particularly evident in the first few time slices for $\xi_w = 0.6$.

\begin{figure}
\centering
\includegraphics[width=0.7\textwidth]{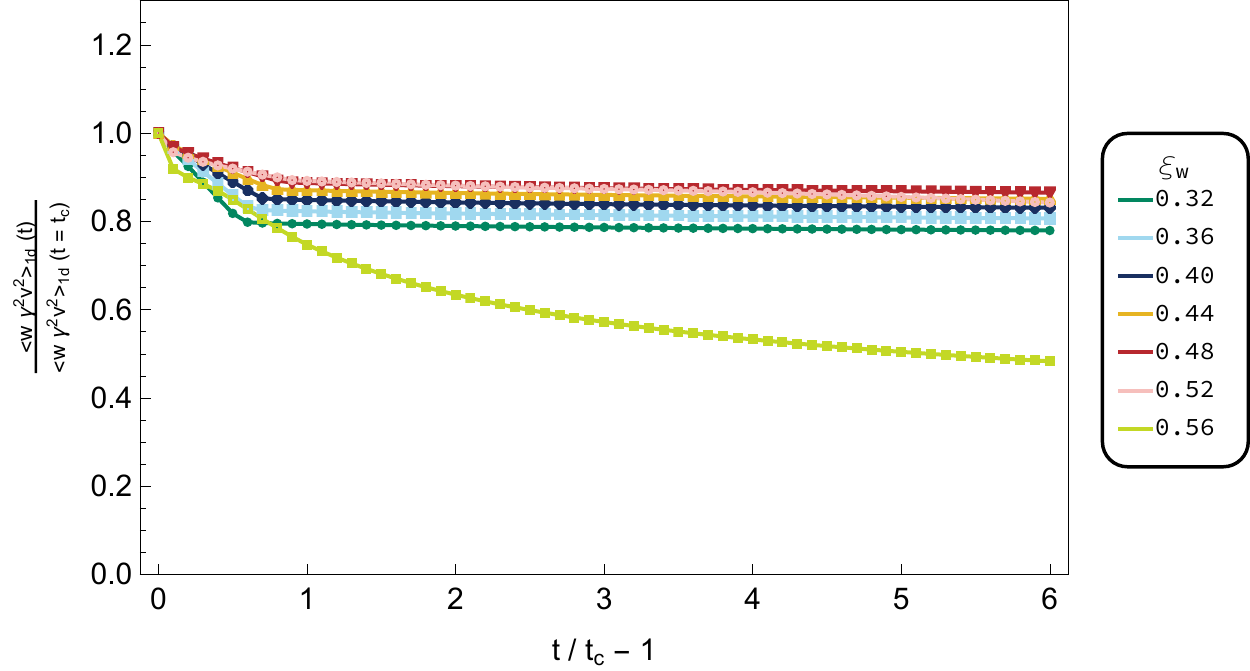} 
\vskip 0.5cm
\includegraphics[width=0.7\textwidth]{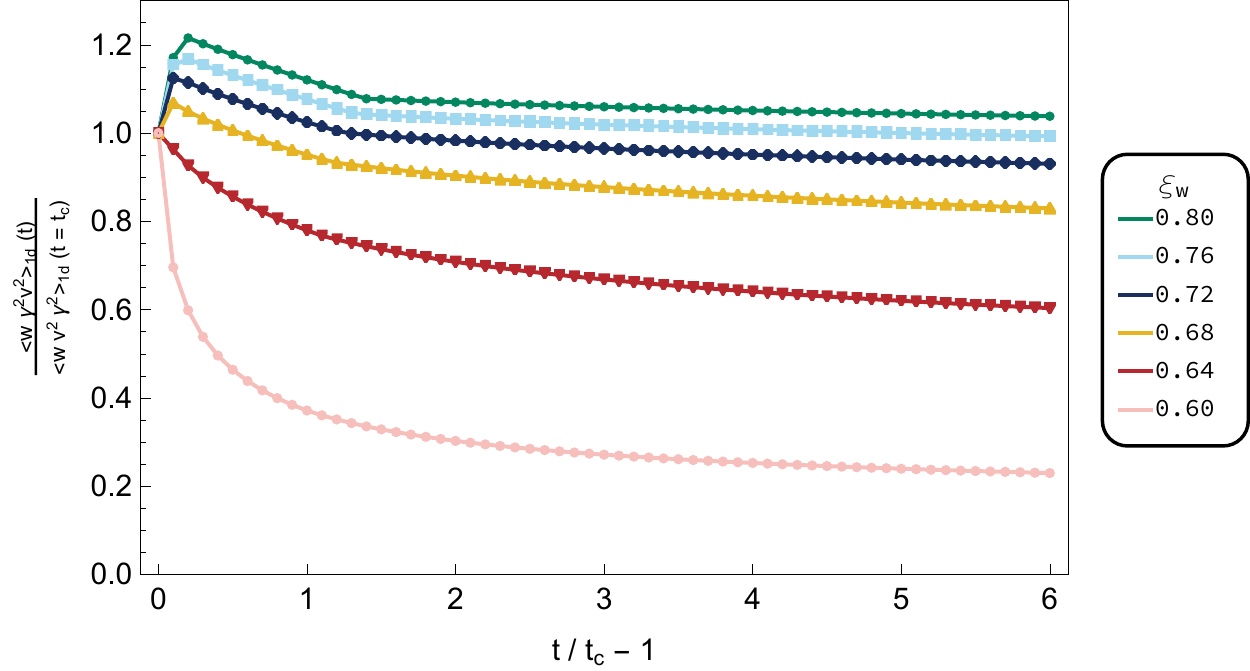} 
\caption{
Time evolution of $\wvv_{\rm 1d}$ for small (top) and large (bottom) $\xi_w$.
In this plot we fix $\alpha = 0.0046$.
Note that the normalization provided in the main text is $\wvv_{\rm 1d} (t = 7t_c)$ while in this plot we show its time evolution from $t = t_c$ to $7t_c$.
}
\label{fig:wv2_evolution_1d}
\end{figure}

\begin{figure}
\centering
\includegraphics[width=0.7\textwidth]{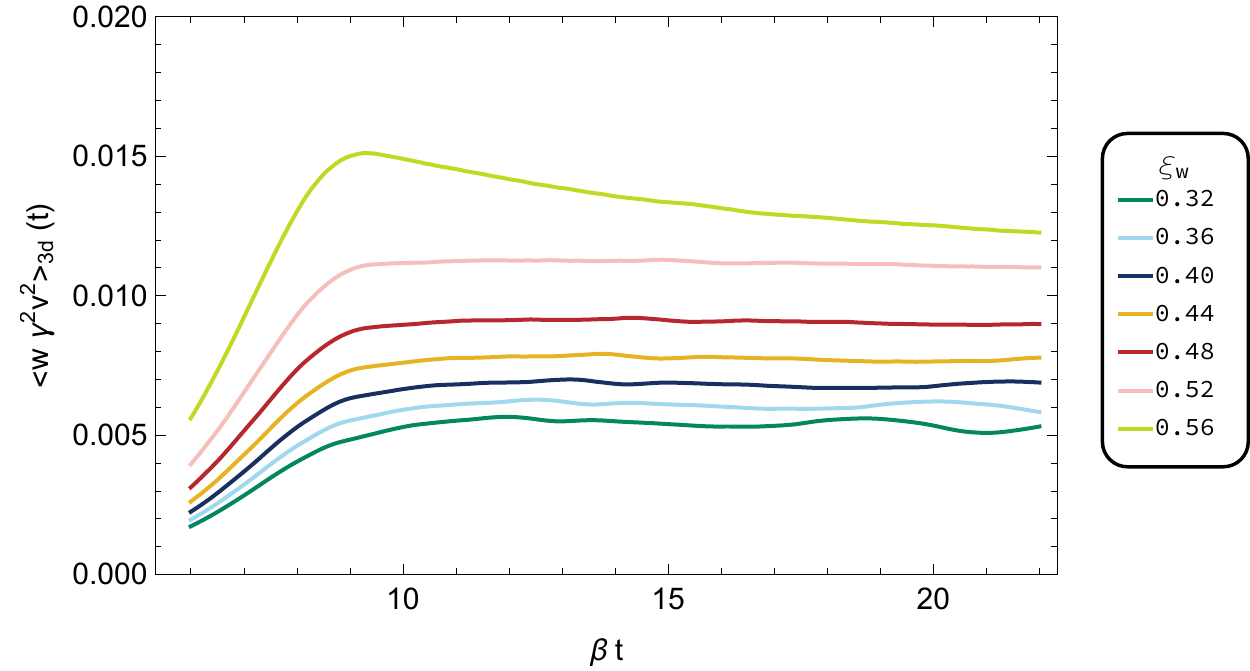} 
\vskip 0.5cm
\includegraphics[width=0.7\textwidth]{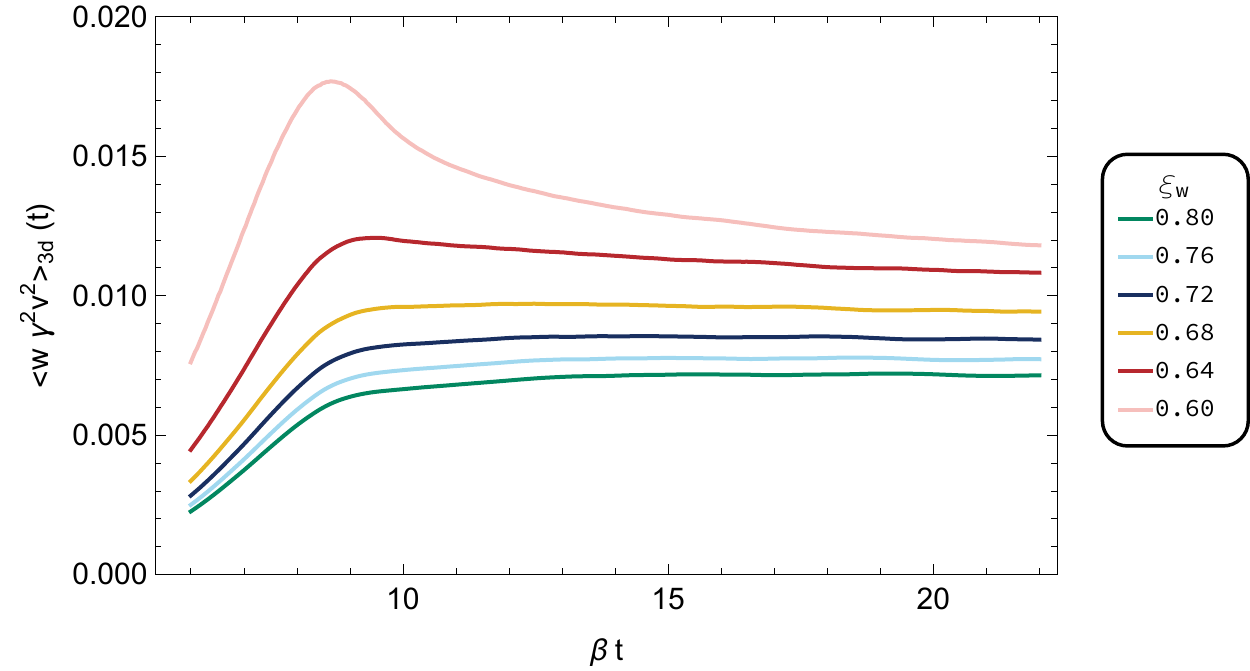} 
\caption{
Time evolution of $\wvv_{\rm 3d}$ for small (top) and large (bottom) $\xi_w$.
In this plot we fix $\alpha = 0.0046$.
}
\label{fig:wv2_evolution_3d}
\end{figure}

In Fig.~\ref{fig:wv2_evolution_1d} we plot the time evolution of $\wvv_{\rm 1d}$ from $t = t_c$ to $7t_c$. 
Note that $\wvv_{\rm 1d} (t = t_c)$ is equals to $(4/3) \kappa \alpha$, while we used $\wvv_{\rm 1d} (t = 7t_c)$ as the measure for the fluid kinetic energy in Sec.~\ref{subsec:DC}.
The rearrangement effect of the fluid profile after the energy injection from the Higgs wall ceases is stronger for $\xi_w$ close to the speed of sound as seen from the top and bottom panels of this figure. 
This effect {\it decreases} the kinetic energy for $\xi_w$ close to the sound speed.
As argued in Section~\ref{subsec:DC}, the total energy is still conserved and we checked this explicitly.
We also plot the time evolution of $\wvv_{\rm 3d}$ in Fig.~\ref{fig:wv2_evolution_3d}.

\section{Fitting}
\label{app:fitting}

\begin{figure}
\centering
\includegraphics[width=\textwidth]{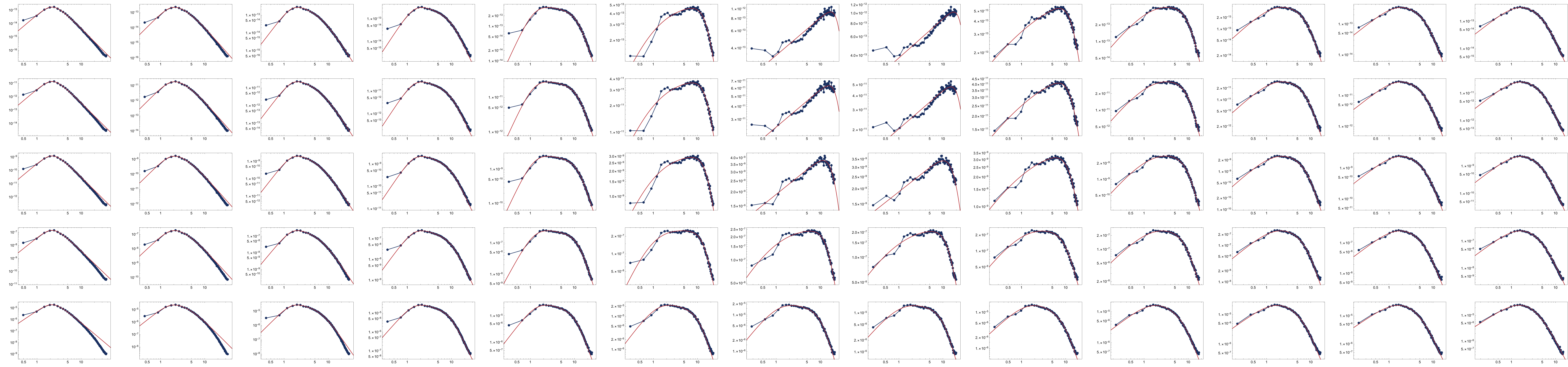} 
\caption{
Data (blue) and fitting function (red) for $\alpha = 0.0005$, $0.0015$, $0.0046$, $0.015$ and $0.05$ (from top to bottom) and $\xi_w = 0.32, 0.36, \cdots, 0.8$ (from left to right). 
The horizontal and vertical axes are 
$q/\beta$ and $Q'$, respectively.
We use box size $L = 40 \xi_w / \beta$ and grid $N = 256$.
}
\label{fig:fitting}
\vskip 0.5cm
\includegraphics[width=\textwidth]{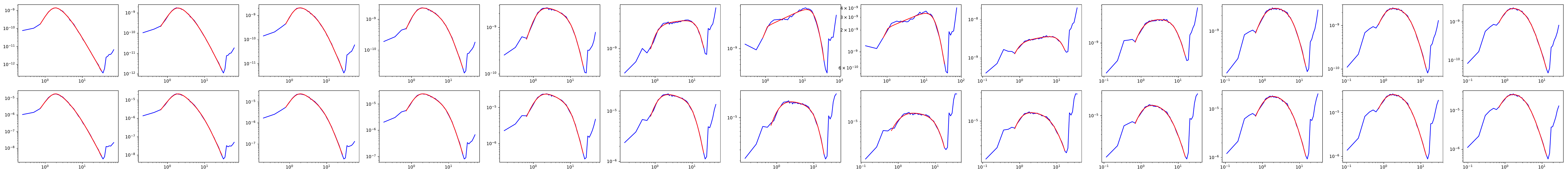} 
\caption{
Same as Fig.~\ref{fig:fitting} except that we use $L = 80 \xi_w / \beta$ and $N = 512$. 
}
\label{fig:fitting_bigbox}
\end{figure}

In this appendix we compare the data and the fitting function for the GW spectrum.
Fig.~\ref{fig:fitting} shows the data (blue) and fitting function (red) for $\alpha = 0.0005$, $0.0015$, $0.0046$, $0.015$, $0.05$ (from top to bottom) and $\xi_w = 0.32, 0.36, \cdots, 0.8$ (from left to right) for the box size $L = 40 \xi_w / \beta$ and grid $N = 256$.
The integration range is from $t = 14 / \beta$ to $22 / \beta$.
Fig.~\ref{fig:fitting_bigbox} is for $\alpha = 0.0046$ (top) and $0.05$ (bottom) and $\xi = 0.32, 0.36, \cdots, 0.8$ (from left to right) for the box size $L = 80 \xi_w / \beta$ and grid $N = 512$ with the same integration range.

\clearpage

\small
\bibliography{ref}

\end{document}